\documentclass{article}
\newtheorem{manualtheoreminner}{Proposition}
\newenvironment{manualtheorem}[1]{%
  \renewcommand\themanualtheoreminner{#1}%
  \manualtheoreminner
}{\endmanualtheoreminner}
 \newcommand{\ind}{\perp\!\!\!\!\perp}
 \newcommand{\notind}{\centernot{\ind}}
\usepackage{centernot}
\usepackage{xcolor}
\usepackage{arxiv}
\usepackage{tablefootnote}
\usepackage[para,online,flushleft]{threeparttable}
\usepackage{bm}
\usepackage{float}
\usepackage[title]{appendix}
\usepackage{dsfont}
\usepackage[authoryear]{natbib}
\usepackage[utf8]{inputenc} 
\usepackage[T1]{fontenc}    
\usepackage{hyperref}       
\usepackage{url}            
\usepackage{booktabs}       
\usepackage{amsfonts}       
\usepackage{nicefrac}       
\usepackage{microtype}      
\usepackage{lipsum}
\usepackage{graphicx}
\usepackage{amsmath}
\graphicspath{ {./images/} }

\title{Revisiting the Effects of Maternal Education on Adolescents' Academic Performance: Doubly Robust Estimation in a Network-Based Observational Study}
\vspace{-2cm}

\author{
 Vanessa McNealis\thanks{
VM is supported by doctoral fellowships from Natural Sciences and Engineering Research Council of Canada (NSERC) and the Fonds de Recherche du Québec (FRQ) - Nature et Technologie. EEMM acknowledges support from an NSERC Discovery Grant. EEMM is a Canada Research Chair (Tier 1) in Statistical Methods for Precision Medicine and acknowledges the support of a Chercheur-boursier de mérite career award from the FRQ - Santé. This research was enabled in part by support provided by Calcul Québec (https://www.calculquebec.ca/) and Compute Canada (www.computecanada.ca).
} \\
  Department of Epidemiology, Biostatistics and Occupational Health\\
  McGill University\\
  \texttt{vanessa.mcnealis@mail.mcgill.ca} \\
   \And
  Erica E. M. Moodie \\
  Department of Epidemiology, Biostatistics and Occupational Health\\
  McGill University\\
  \texttt{erica.moodie@mcgill.ca} \\
  \And
 Nema Dean \\
  School of Mathematics and Statistics\\
  University of Glasgow\\
  \texttt{Nema.Dean@glasgow.ac.uk} \\
  }

\begin{document}
\maketitle

\begin{abstract}
In many contexts, particularly when study subjects are adolescents, peer effects can invalidate typical statistical requirements in the data. For instance, it is plausible that a student’s academic performance is influenced both by their own mother’s educational level as well as that of their peers. Since the underlying social network is measured, the Add Health study provides a unique opportunity to examine the impact of maternal college education on adolescent school performance, both direct and indirect. However, causal inference on populations embedded in social networks poses technical challenges, since the typical no interference assumption no longer holds.  While inverse probability-of-treatment weighted (IPW) estimators have been developed for this setting, they are often highly unstable. Motivated by the question of maternal education, we propose doubly robust (DR) estimators combining models for treatment and outcome that are consistent and asymptotically normal if either model is correctly specified. We present empirical results that illustrate the DR property and the efficiency gain of DR over IPW estimators even when the treatment model is misspecified. Contrary to previous studies, our robust analysis does not provide evidence of an indirect effect of maternal education on academic performance within adolescents' social circles in Add Health. 
\end{abstract}
\section{Introduction}
\label{sec:intro}
Increasingly, there is a marked interest in estimating causal effects under interference, which arises when the potential outcome of an individual is affected by the treatment, exposure or intervention of another. In classical causal inference, interference is often assumed away in favour of the stable unit treatment value assumption (SUTVA) \citep{rubin1974estimating}. However, we can think of numerous settings where this assumption is likely violated and where interventions not only have direct effects on recipients but also \emph{spill over} onto their close contacts, hence generating indirect (spillover) effects. For instance, if we consider a vaccine trial on a population embedded in a network, chances are that the vaccination status of one individual will spill over onto a neighbour and reduce their risk of infection \citep{perez2014assessing, ogburn2017vaccines, liu2019doubly}. Outside the realm of infectious disease, another area where interference naturally manifests is education, where peer effects or \textit{spillover} effects have been extensively studied, be it in the context of an experiment \citep{hong2006evaluating, basse2018analyzing} or a nonrandomized exposure such as maternal education \citep{fletcher2020consequences, bifulco2011effect}. Indeed, it is this latter question that motivates this work: we wish to disentangle the direct and indirect effects of maternal education on adolescent school performance.

The bulk of the interference literature has focused on relaxing the SUTVA assumption to allow for arbitrary interference within disjoint clusters, a setting known as \textit{partial interference} \citep{sobel2006randomized, hudgens2008toward, tchetgen2012causal, perez2014assessing, liu2014large}. In this framework, independent groups of units are used to define the interference sets and information on shared connections within each cluster is typically not measured or utilized \citep{lee2021estimating}. \cite{liu2016inverse} developed an inverse probability-of-treatment weighted (IPW) estimator for causal effects under a general interference framework, which allowed for overlapping interference sets. This paved the way for methodology for observational network data under the counterfactual framework.

Recently, singly robust inference methods have been proposed for network-based observational studies where the interference graph is limited to first-order neighbours \citep{forastiere2021identification, lee2021estimating}. \cite{forastiere2021identification} rely on a propensity score regression approach, which requires modeling
assumptions regarding the dependence of the outcome on the covariates. Drawing upon the work of \citep{liu2016inverse} in the general interference setting, \cite{lee2021estimating} proposed inverse probability-of-treatment weighted (IPW) estimators for the direct and indirect effects of a treatment under network interference and derived a variance estimator that allows for overlapping interference sets. However, IPW estimators are known to be highly unstable and to have large variances, especially in the presence of extreme propensity scores (close to either 0 or 1) which are more likely to arise in the interference setting \citep{liu2019doubly}. 

Doubly robust (DR) estimators present certain advantages compared to IPW estimators. They combine an outcome model and a treatment model and produce a consistent estimator of the causal effect if either model has been correctly specified, hence providing some protection against model misspecification \citep{liu2019doubly}. Furthermore, DR estimators are known to have higher efficiency than IPW estimators, even when only the treatment nuisance model is correctly specified. Several estimators with the double-robust property have been proposed in the no interference setting \citep{scharfstein1999adjusting,bang2005doubly, bang2008correction}; see \cite{kang2007demystifying} for an extensive review. \cite{liu2019doubly} proposed extensions of these estimators in the partial interference framework. Extensions of the doubly robust estimators of \cite{liu2019doubly} have yet to be developed for observational network data under the counterfactual framework. While \cite{ogburn2022causal} proposed a targeted maximum likelihood estimator (TMLE) for observational social network data, their approach targets a single connected network component and is not specifically designed for multilevel network data, which is prevalent in educational research.  

An issue inherent to network-based observational studies is the presence of homophily, which is defined as the increased tendency of units with similar characteristics of forming ties. In our study of maternal education, we expect latent variable dependence due to latent traits that both drive the formation of friendships and predict an individual's socio-economic background. If these characteristics are observed, then observations can be rendered conditionally independent \citep{ogburn2022causal}. While it would be unrealistic to think that we could measure all traits driving homophily, latent variable dependence poses a threat to the identification of causal effects \citep{forastiere2018estimating}. Frameworks proposed by \cite{forastiere2021identification} and \cite{lee2021estimating}, which we draw upon, require neighbours' treatments to be conditionally independent given observed covariates. Given the possibility of latent treatment dependence, it appears even more important to develop inference methods that do not solely rely upon a correctly specified treatment model.

Our paper addresses an important methodological gap by extending the DR estimators proposed by \cite{liu2019doubly} in the partial interference setting to the more general context of network-based observational studies. We propose two novel estimators where the interference set is defined as the set of first-order neighbours in a sociometric network. By design, these estimators take into account the interconnections among individuals' treatments and outcomes. Under certain conditions, they can be shown to be consistent and asymptotically normal if either model, but not necessarily both, is correctly specified, assuming that the network can be written as the union of disjoint subgraphs. The first estimator, a regression estimator with residual bias correction, is endowed with the double robustness property whether or not the outcome has a multilevel structure. The second estimator, a regression estimator with inverse-propensity weighted coefficients, can be shown to be doubly robust if the outcome does not follow a hierarchical model. Using this framework, we apply M-estimation theory to propose appropriate asymptotic variance estimators. We also present empirical evidence of the double robustness property, as well as the efficiency gain of proposed estimators over IPW estimators, even when the treatment model is correctly specified. We make another contribution by illustrating the threat of latent treatment homophily to the identification of causal effects, and show how doubly robust estimation can guard against it. 

Studies suggest that an adolescent's academic outcomes (e.g., grade point average, retention) are positively affected not only by their own home environment but also by that of their friends \citep{bifulco2011effect, fletcher2020consequences}. For instance, the higher the proportion of friends or classmates with college-educated mothers, the greater the likelihood of finishing school and the higher the grades were according to their analyses. In our study leveraging data from Add Health, the direction of the effects found using our proposed causal inference methods supports conclusions drawn by previous authors, although the doubly robust analysis did not achieve statistical significance. Thus, our study does not provide evidence that the composition of friendship circles has an influence on learning outcomes. However, an important caveat is that we had to use a smaller subset of the dataset for computational reasons, potentially leading to an underpowered analysis. The validity of our causal analysis depends on, among other things, either a correctly specified model for the exposure (maternal college education) or a correctly specified model for the grade point average.

The paper is structured as follows. In Section \ref{section:motivating}, we introduce our motivating example using Add Health data which examines the direct and indirect effect of maternal education on school performance. 
In Section \ref{section:notation} we introduce the potential counterfactual framework under network interference and define relevant causal estimands. In Section \ref{section:inferential} we describe different inference strategies: inverse-probability-weighting estimation, regression estimation and doubly robust estimation. In Section \ref{section:simulations}, we assess finite sample properties of proposed estimators through simulation studies and illustrate the double robustness property of DR estimators. Having derived new, efficient and valid tools for estimation, we demonstrate the application of our proposed approach to estimate spillover effects of maternal college education on adolescent school performance using the Add Health data in Section \ref{section:addhealth}. We close with a general discussion in Section \ref{section:discussion}. 

\section{Motivating Context: Peer Effects in Education}
\label{section:motivating}

\begin{figure}[ht]
    \centering
    \begin{minipage}{.49\textwidth}
        \centering
        \includegraphics[width=0.85\linewidth, height=0.31\textheight]{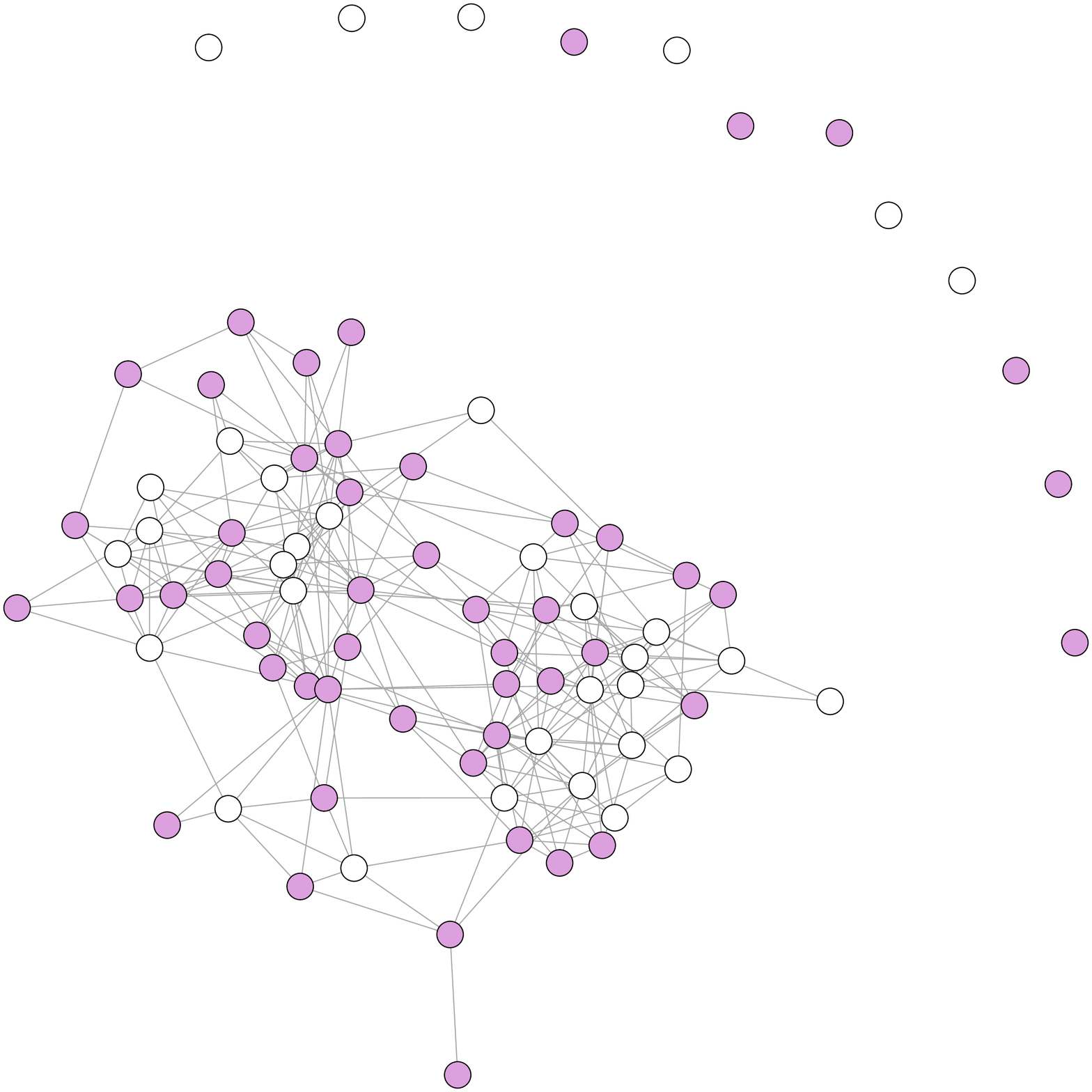}
    \end{minipage}%
    \hfill
    \begin{minipage}{0.49\textwidth}
        \centering
        \includegraphics[width=0.85\linewidth, height=0.3\textheight]{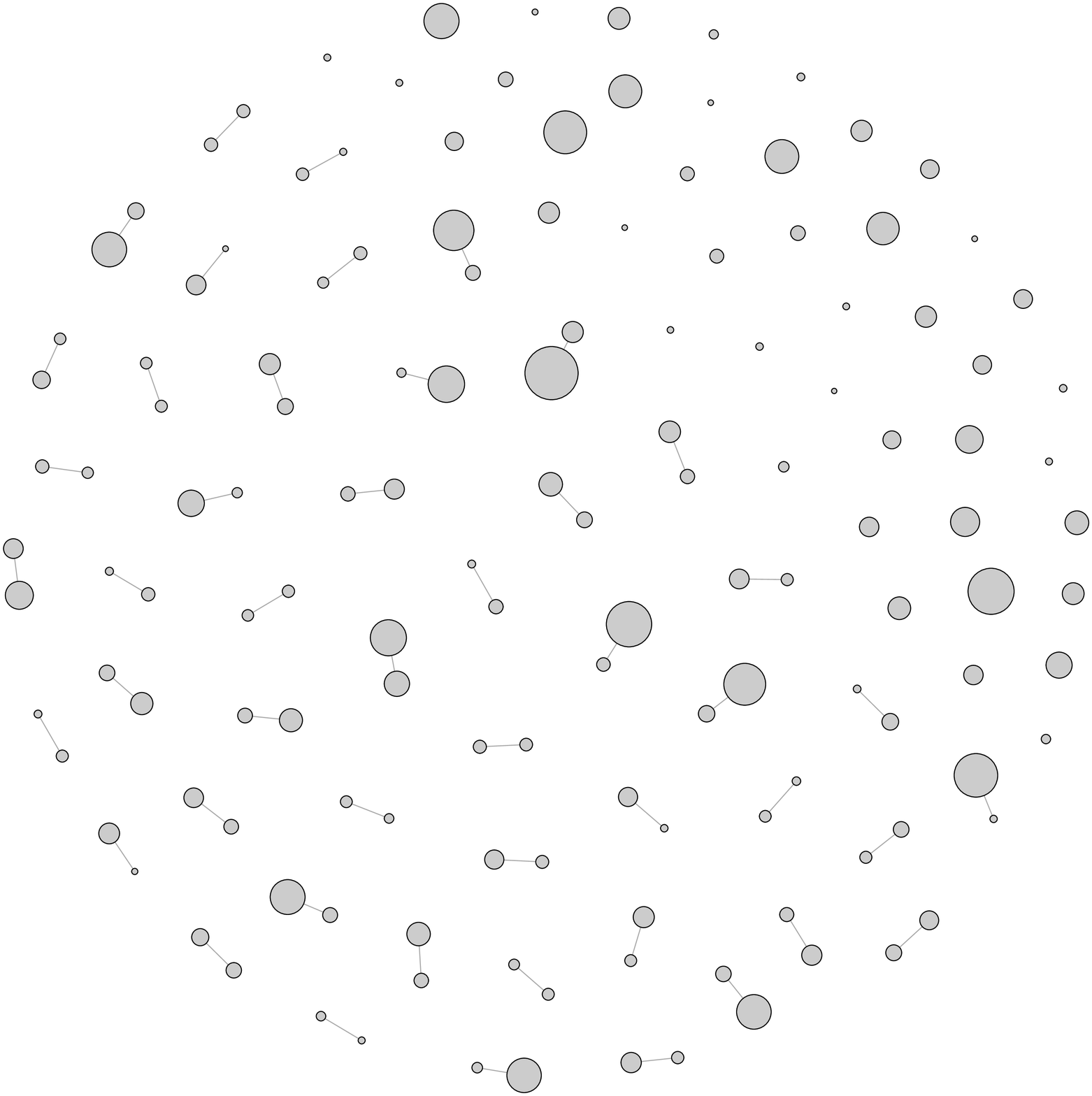}
    \end{minipage}
    \caption[Visualisations of the network formed by one school of 83 students in Add Health and of the network formed by 139 schools in Add Health.]{Left: One school of 83 students in the Add Health network, where an edge between two students indicates that either one nominated the other as a friend and plum-coloured nodes represent students with college-educated mothers. Right: Network formed by 139 schools in Add Health, where an edge is present if two schools share at least one friendship; and the size of each node is proportional to the size of the school. }
    \label{fig:plot11}
\end{figure}

The National Longitudinal Study of Adolescent Health (Add
Health) is a longitudinal study of a nationally representative sample of 90,118 adolescent students in grades 7-12 in the
United States in 1994-95 who were followed throughout adolescence. In addition to socio-demographic and academic achievement characteristics, the survey collected data on friendship nominations through in-school questionnaires
administered directly to students, from which a social network can be constructed. We considered an undirected network based on the data on friendship nominations from the first wave of data collection in 1994-95 of Add Health. We defined a friendship as a symmetric relationship, meaning that there was an edge between students $i$ and $j$ if student $i$ listed $j$ as a friend in the in-school survey, or $j$ listed $i$ as a friend, or both. The left side of Figure \ref{fig:plot11} represents one school in the Add Health network, whereas the right side displays relations among 139 schools, where an edge between two schools is indicative of at least one across-school friendship. A critical assumption is that the network is fully known and observed. Our methodology also assumes that the network can be decomposed into smaller disjoint subgraphs. This assumption is amenable to multilevel network data, which are prevalent in educational research, in which case the disjoint subgraphs correspond to different schools, classrooms, or simply network components \citep{van2018description,abacioglu2019exploring,stewart2019multilevel}.

As peer influence is an ongoing theme in educational research, the Add Health data provide a unique
opportunity to examine the causal effect of peers on academic success. Leveraging the Add Health data, \cite{bifulco2011effect} found that having a higher proportion of classmates with college-educated mothers was associated with a lower risk of dropping out and a higher likelihood of attending
college. Recently, \cite{fletcher2020consequences} made use of the friendship nominations data and found that having a greater proportion of friends with college-educated mothers increases Grade Point Average (GPA)
among female students, but not among male students. 
Recent methodological developments in causal inference with interference allow us to re-analyse this question under a counterfactual framework. Specifically, we aim to quantify the (direct) effect of having a college-educated mother on GPA, as well as the (indirect) effect of having a higher proportion of friends with college-educated mothers on GPA. Our proposed estimator combines a model for the propensity of maternal education among adolescents' social circles and a model for the outcome, GPA. Throughout the text, we will refer to maternal education as a (self-selected) treatment.

\section{Notation, Assumptions and Causal Estimands}
\label{section:notation}
Consider a binary undirected network $G=(\mathcal{N}, \mathbb{E})$, where $\mathcal{N} = \{1, 2, \ldots, N\}$ is a set of $N$ nodes and $\mathbb{E}$ is a set of edges with generic element $\{i,j\} = \{j,i\}$ denoting the presence of an edge between nodes $i$ and $j$. 
For convenience, we represent the network $G$ by the sociomatrix $A$, which has entries $A_{ij} = 1$ if $\{i,j\} \in \mathbb{E}$ and 0 otherwise. Since we assume that there are no self-loops, the diagonal elements of $A$ are identically $0$. We define the neighbourhood $\mathcal{N}_i$ of node $i$, $i=1,2,\ldots, N$, as the set of nodes $j$ which satisfy $A_{ij} = 1$ and the cardinality of $\mathcal{N}_i$ is denoted as $|\mathcal{N}_i| = d_i$. Additionally, we denote $\mathcal{N}_{i}^* = \mathcal{N}_{i} \cup {i}$ as the union of the node $i$ and the nodes in $\mathcal{N}_i$. Thus, as put by \cite{forastiere2021identification}, it follows that to each node $i$ is associated a partition of $\mathcal{N}$, that is $\{i, \mathcal{N}_i, \mathcal{N}_{-i}\}$, where  $\mathcal{N}_{-i} = \mathcal{N}\backslash \mathcal{N}_i^*$. 

A graph $H = (\mathcal{N}_H, \mathbb{E}_H)$ is a subgraph of $G$ if  $\mathcal{N}_H \subseteq \mathcal{N}$ and $\mathbb{E}_H \subseteq \mathbb{E}$. Further, a subgraph of a graph $G$ is said to be \textit{connected} if there exists a path for any two nodes in the subgraph \citep{kolaczyk2009}. We will assume that the network can be expressed as the union of $m$ components $C_1, C_2, \ldots, C_m$, where a component of a graph is defined as a maximally connected subgraph (i.e., a connected subgraph that is not part of any larger subgraph) \citep{kolaczyk2009}. Attributes measured on the nodes in disjoint subgraph are viewed as a random sample from a super-population and constitute independent data units. This assumption is necessary to introduce regression or model-based estimators of causal effects and to derive asymptotic properties of the proposed estimators. 

 For each node $i$, $i=1,2,\ldots, N$, we observe the node-level vector of attributes $O_i = (\bm{X}_i, Z_i, Y_i)$, where $\bm{X}_i=(X_{i1}, \ldots, X_{ip})'$ denotes a $p$-dimensional vector of pretreatment covariates, $Z_i$ denotes a binary (self-selected) treatment with $Z_i = 1$ if treated and $0$ otherwise, and $Y_i$ is a univariate outcome of interest. Let $\bm{X}_{\mathcal{N}_i} = (\bm{X}_{j_1}, \ldots, \bm{X}_{j_{d_i}})'$ and $\bm{Z}_{\mathcal{N}_i} = (Z_{j_1}, \ldots, Z_{j_{d_i}})'$ denote the neighbourhood matrix of pretreatment covariates and neighbourhood vector of treatments for node $i$, respectively, where $j_{k} \in \mathcal{N}_i, k = 1, \ldots, d_i$. Throughout this article, we will use the terms node and individual interchangeably.

In accordance with the Neyman-Rubin counterfactual framework \citep{rubin1974estimating}, let $y_i(z_i,\bm{z}_{\mathcal{N}_i}, \bm{z}_{\mathcal{N}_{-i}})$ denote the potential outcome for individual $i$ if, possibly contrary to fact, they receive treatment value $z_i$, their neighbourhood receives $\bm{z}_{\mathcal{N}_i}$ and the nodes outside of the neighbourhood receive $\bm{z}_{\mathcal{N}_{-i}}$. Later, we will assume spillover only from the nodes in $\mathcal{N}_i$ and write the $i$-th potential outcome as $y_i(z_i, \bm{z}_{\mathcal{N}_i})$. Note that $z \in \{0,1\}$ and $\bm{z}_{\mathcal{N}_i} \in \mathcal{Z}(d_i)$, where $\mathcal{Z}(d_i)$ denotes the set of vectors of possible treatment allocations of length $d_i$ with $|\mathcal{Z}(d_i)| = 2^{d_i}$. To ensure identifiability of causal effects, we first make the consistency assumption, which states that there cannot be multiple versions of a treatment. 

\begin{assumption}[Consistency]
For $i \in \mathcal{N}$, $Y_i = y_i(Z_i, \bm{Z}_{\mathcal{N}_i}, \bm{Z}_{\mathcal{N}_{-i}})$.
\label{assum1}
\end{assumption}
We further assume the potential outcome of individual $i$ depends on their first-order neighbours' treatments only through a function of the neighbourhood treatment, which is referred to as the neighbourhood \citep{forastiere2021identification} or stratified interference assumption \citep{lee2021estimating}.
\begin{assumption}[Stratified interference]
For $i \in \mathcal{N}$, $\forall \ \bm{z}_{\mathcal{N}_{-i}}, \bm{z}_{\mathcal{N}_{-i}}'$, and \\ $\forall$ $\bm{z}_{\mathcal{N}_{i}}, \bm{z}_{\mathcal{N}_{i}}'$ such that $\phi(\bm{z}_{\mathcal{N}_{i}}) = \phi(\bm{z}_{\mathcal{N}_{i}}')$ given a function $\phi: \{0,1\}^{d_i} \rightarrow \Phi_i$, then, $$y_i(z_i, \bm{z}_{\mathcal{N}_{i}}, \bm{z}_{\mathcal{N}_{-i}})  =  y_i(z_i, \bm{z}_{\mathcal{N}_{i}}, \bm{z}_{\mathcal{N}_{-i}}').$$
\end{assumption}

In other words, we assume that the potential outcome of an individual depends on their neighbourhood through a summary of neighbourhood treatment vector, $\phi(\bm{z}_{\mathcal{N}_i})$. In this paper, we will focus on the transformation $\phi: \{0,1\}^{d_i} \rightarrow \Phi_i$ defined by $\phi(\bm{z}_{\mathcal{N}_i}) = \sum_{j \in \mathcal{N}_i} z_j$, where $\Phi_i \in \{0, 1, \ldots, d_i\}$, which corresponds to the number of treated neighbours. We will denote this quantity $\Sigma\bm{z}_{\mathcal{N}_{i}}$ from this point onward. In the context of this case study, this implies that we assume that the GPA of a student is only affected by the home environment (as captured by maternal education) of their peers through the number of friends with a college-educated mother. This device is introduced to reduce the number of potential outcomes from $2^{d_i + 1}$ (if $\phi$ was the identity function) to $2d_i$.
 Because of Assumption 2, we can drop the dependence of $y_i(\cdot)$ on the nodes in $\mathcal{N}_{-i}$ and write more simply $y_i(z_i, z_{\mathcal{N}_i})$. 
 
 We let $f(z_i, \bm{z}_{\mathcal{N}_i} | \bm{X}_i, \bm{X}_{\mathcal{N}_i}) = P(Z_i=z_i, \bm{Z}_{\mathcal{N}_i} = \bm{z}_{\mathcal{N}_i}| \bm{X}_i, \bm{X}_{\mathcal{N}_i})$ denote the probability of observing the joint treatment $(z_i, \bm{z}_{\mathcal{N}_i})$ conditional on individual and neighbourhood covariates, which is also referred to as the propensity score. The following assumptions are extensions of the typical no unmeasured confounding and positivity assumptions. The no unmeasured confounding assumption states that for all $i \in \mathcal{N}$, the covariates $\bm{X}_i$ are a sufficient conditioning set, such that groups defined by levels of the joint exposure $(Z_i, \bm{Z}_{\mathcal{N}_i})$ are exchangeable conditionally on $\bm{X}_i$. The positivity assumption ensures that there is appropriate covariate overlap across different treatment groups.
 \begin{assumption}[Conditional exchangeability] For all $i \in \mathcal{N}$, $z_i \in \{0, 1\}$, $\bm{z}_{\mathcal{N}_i} \in \{0,1\}^{d_i},$  $y_i(z_i, \bm{z}_{\mathcal{N}_i}) \ind Z_i, \bm{Z}_{\mathcal{N}_i} | \bm{X}_i.$
 \label{assum3}
\end{assumption}
 \begin{assumption}[Positivity] For all $i \in \mathcal{N}$ and for all $z_i, \bm{z}_{\mathcal{N}_i}, \bm{X}_i, \bm{X}_{\mathcal{N}_i}$, $$f(z_i, \bm{z}_{\mathcal{N}_i} | \bm{X}_i, \bm{X}_{\mathcal{N}_i}) > 0.$$
  \label{assum4}
\end{assumption}
To define the causal estimands of interest, we adopt the Bernoulli treatment allocation standardization proposed by \cite{tchetgen2012causal} in the context of partial interference and \cite{liu2016inverse} in the general interference setting. That is, average potential outcomes are standardized according to a reference population in which individuals independently receive the exposure (e.g., maternal education) with probability $\alpha$. Our framework only differs in the fact that we marginalize the potential outcomes over the number of treated neighbours $\Sigma \bm{z}_{\mathcal{N}_i}$ instead of the neighbourhood treatment vector $\bm{z}_{\mathcal{N}_i}$ to remain consistent with Assumption 2. Let $\pi(\Sigma \bm{z}_{\mathcal{N}_i}; \alpha) = \binom{d_i}{\Sigma \bm{z}_{\mathcal{N}_i}}\alpha^{\Sigma\bm{z}_{\mathcal{N}_i}}(1-\alpha)^{d_i -\Sigma \bm{z}_{\mathcal{N}_i}}$ denote the probability of node $i$ receiving neighbourhood treatment $\Sigma \bm{z}_{\mathcal{N}_i}$ under Bernoulli allocation strategy $\alpha$. Similarly, under counterfactual scenario $\alpha$, the probability of individual $i$ and their neighbourhood jointly receiving exposure $(z_i, \Sigma \bm{z}_{\mathcal{N}_i})$ is given by  $\pi(z_i, \Sigma \bm{z}_{\mathcal{N}_i}; \alpha) = \alpha^{z_i}(1-\alpha)^{1-z_i} \binom{d_i}{\Sigma \bm{z}_{\mathcal{N}_i}}\alpha^{\Sigma\bm{z}_{\mathcal{N}_i}}(1-\alpha)^{d_i -\Sigma \bm{z}_{\mathcal{N}_i}} $. Define the $i$-th individual's potential outcome conditional on individual exposure $z$ and allocation strategy $\alpha$ as $$\bar{y}_{i}(z;\alpha) = \sum_{\Sigma \bm{z}_{\mathcal{N}_i} = 0 }^{d_i} y_{i}(z, \Sigma \bm{z}_{\mathcal{N}_i}) \pi( \Sigma \bm{z}_{\mathcal{N}_i}; \alpha),$$
and the marginal individual potential outcome as 
$$\bar{y}_{i}(\alpha) = \sum_{\substack{z_i \in \{0, 1\}\\ \Sigma \bm{z}_{\mathcal{N}_i} \in \{0, 1, \ldots, d_i\} }} y_{i}(z_i, \Sigma \bm{z}_{\mathcal{N}_i}) \pi(z_i, \Sigma \bm{z}_{\mathcal{N}_i}; \alpha),$$
for $i=1, 2, \ldots, N$. Similar to \cite{liu2019doubly}, we define the population average potential outcome as $\bar{y}(z;\alpha) = m^{-1} \sum_{\nu=1}^m N_{\nu}^{-1} \sum_{i \in C_{\nu}} \bar{y}_{i}(z;\alpha)$ and the marginal average potential outcome as $\bar{y}(\alpha)= m^{-1} \sum_{\nu=1}^m N_{\nu}^{-1} \sum_{i\in C_{\nu}} \bar{y}_{i}(\alpha)$, where $N_{\nu} = |C_{\nu}|$ is the number of nodes in component $C_{\nu}$. These definitions of the population average potential outcomes differ from those used by \cite{liu2016inverse} and \cite{lee2021estimating}, who considered a global average of individual potential outcomes rather than an average of group-level averages. Finally, let $\mu_{z,\alpha} = \mathbb{E}[\bar{y}(z;\alpha)]$  and $\mu_{\alpha} = \mathbb{E}[\bar{y}(\alpha)]$ denote the average potential outcome and the marginal average potential outcome, respectively, where $\mathbb{E}[\cdot]$ denotes the expected value in the super-population. The average potential outcome is an expectation over the marginal distribution of the covariates $\bm{X}$ whereas the marginal average potential outcome takes an additional expectation over the individual treatment allocation $Z$.

Following \cite{hudgens2008toward} and \cite{liu2016inverse}, we consider different contrasts of the average potential outcomes to define the causal estimands. The direct exposure effect is defined as the difference between the average potential outcomes of untreated and treated individuals for a fixed allocation strategy $\alpha$, that is, $DE(\alpha) = \mu_{1,\alpha} - \mu_{0,\alpha}$. The spillover effect, also referred to as the \textit{indirect effect}, is defined as the difference between the average potential outcomes under counterfactual scenarios $\alpha$ and $\alpha'$ among the untreated, i.e. $IE(\alpha, \alpha') = \mu_{0,\alpha}- \mu_{0,\alpha'}$. For instance, in the context of our motivating question, the spillover effect corresponds to the expected increment in GPA when changing the counterfactual probability of having a college-educated mother from $\alpha'$ to $\alpha$ among students whose mother has not completed a 4-year college degree. It is possible to define a spillover effect for the treated individuals as well. For allocation strategies $\alpha$ and $\alpha'$, we also define the total effect $TE(\alpha, \alpha') = \mu_{1,\alpha}  - \mu_{0,\alpha'}$ as well as the overall effect $OE(\alpha, \alpha') = \mu_{\alpha} - \mu_{\alpha'}$. As the name suggests, $TE(\alpha, \alpha')$ can be interpreted as the total effect of treatment, which encompasses both the direct and the spillover effect, since $TE(\alpha, \alpha') = DE(\alpha) + IE(\alpha, \alpha')$.

\section{Inferential Procedures}
\label{section:inferential}
\subsection{IPW and Regression Estimators}
IPW is commonly used to address confounding in observational studies. By weighting each observation by the inverse probability-of-treatment, we create a pseudo-population in which the covariates are no longer predictive of the treatment, and thus from which a causal effect can be consistently estimated. In the present case study, the procedure consists of weighting each observation by the joint propensity score so as to balance the covariates across groups defined by the maternal education assignment vector in a given neighbourhood, $(Z_i, Z_{\mathcal{N}_i})$.  We consider the following IPW estimators of $\mu_{z, \alpha}$ and $\mu_{\alpha}$ \citep{liu2019doubly,lee2021estimating}:
\begin{align} \hat{Y}^{\text{IPW}}(z, \alpha) = \frac{1}{m} \sum_{\nu=1}^m \frac{1}{N_{\nu}} \sum_{i\in C_{\nu}}\frac{y_i(Z_i, \Sigma \bm{Z}_{\mathcal{N}_i}) \mathds{1}(Z_i = z) \pi(\Sigma\bm{Z}_{\mathcal{N}_i}; \alpha)}{\binom{d_i}{\Sigma \bm{Z}_{\mathcal{N}_i}} f(Z_i, \bm{Z}_{\mathcal{N}_i} \ \rvert \ \bm{X}_i, \bm{X}_{\mathcal{N}_i})},\label{eq:ipw1}
\end{align}
\begin{align}
\hat{Y}^{\text{IPW}}(\alpha) = \frac{1}{m}\sum_{\nu=1}^m \frac{1}{N_{\nu}} \sum_{i\in C_{\nu}}\frac{y_i(Z_i, \Sigma \bm{Z}_{\mathcal{N}_i})  \pi(Z_i, \Sigma\bm{Z}_{\mathcal{N}_i}; \alpha)}{\binom{d_i}{\Sigma \bm{Z}_{\mathcal{N}_i}} f(Z_i,  \bm{Z}_{\mathcal{N}_i} \ \rvert \ \bm{X}_i, \bm{X}_{\mathcal{N}_i})}\label{eq:ipw2}.
\end{align}
The estimators in (\ref{eq:ipw1}) and (\ref{eq:ipw2}) require a model for the propensity score. We introduce the following assumption, which states that all variables driving homophily in the treatment are included in the vector of pre-treatment covariates $\bm{X}_i, \ \forall i \in \mathcal{N}$. This assumption was made by previous authors \citep{forastiere2021identification,lee2021estimating}, although not explicitly.
\begin{assumption}[No latent treatment homophily] For all $i \in \mathcal{N}$, conditional on a component-level random effect $b_{\nu}$, $\nu =1, \ldots, m$,
$$P(Z_i = z_i, \bm{Z}_{\mathcal{N}_i} =  \bm{z}_{\mathcal{N}_i} | \bm{X}_i, \bm{X}_{\mathcal{N}_i}, b_{\nu}) = \prod_{j \in \{i, \mathcal{N}_i\}}  P(Z_j = z_j | \bm{X}_j, b_{\nu}).$$
\label{assump5}
\end{assumption}
Thus, if we adopt a mixed effects logistic regression model for the treatment, the propensity score is assumed to take the form
\begin{align}f(Z_i, \bm{Z}_{\mathcal{N}_i}\ \rvert \ \bm{X}_i, \bm{X}_{\mathcal{N}_i}; \bm{\gamma}) = \int_{-\infty}^{\infty} \prod_{j \in \{i, \mathcal{N}_i\}} p_j^{Z_j}(1-p_j)^{1-Z_j}f(b_{\nu}) db_{\nu},
\label{eqn:propensity}
\end{align}
where $b_{\nu} \sim N(0, \phi_b)$ is a component-level random effect, $\bm{\gamma}=(\gamma_0, \gamma_1, \ldots, \gamma_p)^{\top}$ is the $(p+1)$-dimensional parameter vector of the propensity score model, and $p_j = P(Z_j = 1 | \bm{X}_j, b_{\nu}) = \mathrm{logit}^{-1}(\gamma_0 + \bm{\gamma}_{1:p}^{\top}\bm{X}_j + b_{\nu})$, where $\bm{\gamma}_{1:p} = (\gamma_1, \ldots, \gamma_p)^{\top}$. The component-level random effect accounts for potential dependence among the maternal education indicators of students within each component. The IPW estimators can thus be computed using $f(Z_i, \bm{Z}_{\mathcal{N}_i}\ \rvert \ \bm{X}_i, \bm{X}_{\mathcal{N}_i}; \hat{\bm{\gamma}}, \hat{\phi}_{b}),$ where $(\hat{\bm{\gamma}}, \hat{\phi}_{b})$ is the maximum likelihood estimator of $(\bm{\gamma}, {\phi}_{b})$. Estimators in (\ref{eq:ipw1}) and (\ref{eq:ipw2}) differ from those of \cite{lee2021estimating} in that they assume a priori that the potential outcomes of individual $i$ only depends on the exposures of other individuals through the number of exposed nearest neighbours $\Sigma \bm{z}_{\mathcal{N}_i}$, so that $Y_i = y_i(Z_i, \bm{Z}_{\mathcal{N}_i}) = y_i( Z_i, \Sigma \bm{Z}_{\mathcal{N}_i})$. Estimators in (\ref{eq:ipw1}) and (\ref{eq:ipw2}) can be shown to be unbiased for the true average potential outcomes conditional on $z$ and allocation strategy $\alpha$ when the propensity scores are known (see Appendix \ref{section:appendixA}). When $(\bm{\gamma}, \phi_b)$ is estimated from the observed data, \cite{lee2021estimating} showed that the IPW estimators are consistent and asymptotically normal under Assumptions \ref{assum1}-\ref{assump5} and assuming a correctly specified mixed effects logistic regression model for $f(Z_i, \bm{Z}_{\mathcal{N}_i} | \bm{X}_i, \bm{X}_{\mathcal{N}_i}; \bm{\gamma}, {\phi}_{b})$.

We could alternatively estimate the average potential outcomes via outcome modeling, thereby extending the regression estimators introduced by \cite{liu2019doubly} to the network setting. Direct standardization may be used to identify causal effects, since Assumptions \ref{assum1} and \ref{assum3} imply
$$\mathbb{E}\left[y_i(z_i, \bm{z}_{\mathcal{N}_i}) | \bm{X}_i\right] = \mathbb{E}\left[y_i(z_i, \bm{z}_{\mathcal{N}_i}) | Z_i = z_i, \bm{Z}_{\mathcal{N}_i} = \bm{z}_{\mathcal{N}_i}, \bm{X}_i\right] =  \mathbb{E}\left[Y_i | Z_i = z_i, \bm{Z}_{\mathcal{N}_i} = \bm{z}_{\mathcal{N}_i}, \bm{X}_i\right].$$
Thus, this suggests fitting a parametric outcome regression model for GPA such as \\$m_i(Z_i, \Sigma \bm{Z}_{\mathcal{N}_i}, \bm{X}_i; \bm{\beta})=\mathbb{E}\left[Y_i | Z_i, \bm{Z}_{\mathcal{N}_i}, \bm{X}_i\right]= \beta_0 + \beta_{Z} Z_i + \beta_{\bm{Z}_\mathcal{N}} h(\Sigma \bm{Z}_{\mathcal{N}_i}) + \sum_{k=1}^p \beta_k X_{ik}$. The parameter estimate $\hat{\bm{\beta}}$ may then be used to compute the standardized mean outcome from the predicted outcome values, leading to the following regression estimators of $\mu_{z, \alpha}$ and $\mu_{\alpha}$:
\begin{align} 
\hat{Y}^{\text{REG}}(z, \alpha) = \frac{1}{m} \sum_{\nu=1}^m \frac{1}{N_{\nu}} \sum_{i\in C_{\nu} }\sum_{ \Sigma{\bm{z}_{\mathcal{N}_i}} = 0}^{d_i} m_i(z, \Sigma \bm{z}_{\mathcal{N}_i}, \bm{X}_i; \hat{\bm{\beta}}) \pi(\Sigma{\bm{z}_{\mathcal{N}_i}};\alpha),\label{eq:reg1}\\
\hat{Y}^{\text{REG}}(\alpha) = \frac{1}{m} \sum_{\nu=1}^m \frac{1}{N_{\nu}} \sum_{i\in C_{\nu} } \sum_{\substack{z_i \in \{0, 1\}\\ \Sigma \bm{z}_{\mathcal{N}_i} \in \{0, 1, \ldots, d_i\} }} m_i(z_i, \Sigma \bm{z}_{\mathcal{N}_i}, \bm{X}_i; \hat{\bm{\beta}}) \pi(z_i, \Sigma{\bm{z}_{\mathcal{N}_i}};\alpha)\label{eq:reg2}.
\end{align}
It can be shown that the estimators in \ref{eq:reg1} and \ref{eq:reg2} are consistent and asymptotically normal using standard estimating equation theory if the outcome regression model is correctly specified.

Estimators of the direct, indirect, total and overall effect are then defined as $\widehat{DE}(\alpha) = \hat{Y}(1, \alpha) - \hat{Y}(0, \alpha)$, $\widehat{IE}(\alpha, \alpha') = \hat{Y}(0, \alpha) - \hat{Y}(0, \alpha')$, $\widehat{TE}(\alpha, \alpha') = \hat{Y}(1, \alpha) - \hat{Y}(0, \alpha')$ and $OE(\alpha, \alpha') = \hat{Y}(\alpha) - \hat{Y}( \alpha')$, respectively, where $\hat{Y}(\cdot)$ is either the IPW or the regression estimator of the average potential outcome. The IPW estimators of the causal effects are consistent if the model for the treatment is correctly specified. Likewise, the regression estimators are consistent if the outcome model is correctly specified. In the next section, we introduce doubly robust estimators, which have the appealing property of being consistent if either the treatment or the outcome model is correctly specified.

\subsection{Regression Estimation with Residual Bias Correction}

Following the definitions of inverse-probability weighting and regression estimators, we propose the bias corrected DR (DR-BC) estimators for the average potential outcomes $\mu_{z, \alpha}$ and $\mu_{\alpha}$, which are given by $\hat{Y}^{\text{DR-BC}}(z,\alpha) = m^{-1} \sum_{\nu=1}^m \hat{Y}_{\nu}^{\text{DR-BC}}(z, \alpha)$ and \\$\hat{Y}^{\text{DR-BC}}(\alpha) = m^{-1} \sum_{\nu=1}^m \hat{Y}_{\nu}^{\text{DR-BC}}( \alpha)$, respectively, where
\begin{multline}
\hat{Y}^{\text{DR-BC}}_{\nu}(z, \alpha)  = \frac{1}{N_{\nu}} \sum_{i\in C_{\nu} }  \biggr\{ \sum_{\Sigma \bm{z}_{\mathcal{N}_i} = 0}^{d_i} m_i(z, \Sigma \bm{z}_{\mathcal{N}_i}, \bm{X}_i; \hat{\bm{\beta}}) \pi(\Sigma\bm{z}_{\mathcal{N}_i};\alpha) +\\ \frac{\mathds{1}(Z_i = z)}{\binom{d_i}{\Sigma \bm{Z}_{\mathcal{N}_i}}f(Z_i, \bm{Z}_{\mathcal{N}_i}\ \rvert \ \bm{X}_i, \bm{X}_{\mathcal{N}_i}; \hat{\bm{\gamma}}, \hat{\phi}_{b}) } \left\{y_i(Z_i, \Sigma \bm{Z}_{\mathcal{N}_i}) - m_i(Z_i, \Sigma \bm{Z}_{\mathcal{N}_i}, \bm{X}_i; \hat{\bm{\beta}})\right\}\pi(\Sigma\bm{Z}_{\mathcal{N}_i}; \alpha) \biggr\},
\label{eqn:dr-bc1}
\end{multline}
and
\begin{multline}
\hat{Y}^{\text{DR-BC}}_{\nu}(\alpha) 
 =\frac{1}{N_{\nu}} \sum_{i\in C_{\nu} } \biggr\{ \sum_{\substack{z_i \in \{0, 1\}\\ \Sigma \bm{z}_{\mathcal{N}_i} \in \{0, 1, \ldots, d_i\} }} m_i(z_i, \Sigma \bm{z}_{\mathcal{N}_i}, \bm{X}_i; \hat{\bm{\beta}}) \pi(z_i,\Sigma \bm{z}_{\mathcal{N}_i};\alpha) +\\ \frac{y_i(Z_i, \Sigma \bm{Z}_{\mathcal{N}_i}) - m_i(Z_i, \Sigma \bm{Z}_{\mathcal{N}_i}, \bm{X}_i; \hat{\bm{\beta}})}{\binom{d_i}{\Sigma \bm{Z}_{\mathcal{N}_i}}f(Z_i, \bm{Z}_{\mathcal{N}_i}\ \rvert \ \bm{X}_i, \bm{X}_{\mathcal{N}_i}; \hat{\bm{\gamma}}, \hat{\phi}_{b}) } \pi(Z_i, \Sigma \bm{Z}_{\mathcal{N}_i};\alpha) \biggr\}.
 \label{eqn:dr-bc2}
\end{multline} 

Estimators in (\ref{eqn:dr-bc1}) and (\ref{eqn:dr-bc2}) are extensions of the DR estimators proposed by \cite{liu2019doubly} in the partial interference setting, which are themselves motivated by the estimators proposed by \cite{scharfstein1999adjusting} for the setting where there is no interference. The DR estimators with residual bias correction combine the regression estimator and the inverse weighted residuals of the regression estimator. They differ from those of \cite{liu2019doubly} in that they make use of network information by allowing each individual to have their own interference set. The DR property can be grasped informally as follows. When the outcome model is correctly specified, the regression estimator converges to $\mu_{z,\alpha}$ since the expected value of the inverse weighted residuals is 0. On the other hand, when the treatment model is correctly specified while the outcome model is not, the second term consistently estimates the bias of the first term \citep{kang2007demystifying}. 

To characterize the large-sample distribution of the DR estimators, we leverage the asymptotic framework proposed by \cite{liu2016inverse} in the general interference setting, of which network interference is a special case. Recall that we assumed that the observed network $G$ could be divided up into $m$ components or disjoint subgraphs $C_1, C_2, \ldots, C_m$, where the observable random variables $\bm{O}_{\nu}$ from different components are independent and identically distributed. 
We write the observed outcome of the $i$-th node in component $\nu$ as $Y_{\nu i}$ and the vector of observed outcomes as $\tilde{\bm{Y}}_{\nu}= (Y_{\nu 1}, \ldots, Y_{\nu N_{\nu}})^{\top}$. Similarly, let $\bm{X}_{\nu i}$ and $Z_{\nu i}$ denote the observed covariates and treatment for individual $i$ in component $\nu$ and define the corresponding component-level vectors $\tilde{\bm{X}}_{\nu}$ and $\tilde{\bm{Z}}_{\nu}$. Under the above assumptions, we have that $\bm{O}_{\nu} = (\tilde{\bm{X}}_{\nu}, \tilde{\bm{Z}}_{\nu}, \tilde{\bm{Y}}_{\nu})$ are independently and identically distributed for $\nu=1, \ldots, m$. Let $F$ denote the cumulative distribution function of $\bm{O}_{{\nu}}$. 
Again, we make the consistency assumption, i.e., $Y_{\nu i} = y_{\nu i}(Z_{\nu i}, \bm{Z}_{\nu \mathcal{N}_i})$, where $\bm{Z}_{\nu \mathcal{N}_i}$ denotes the neighbourhood vector of treatments for node $i$ in component $\nu$. Finally, we also make the exchangeability assumption, that is, $y_{\nu i}(z_i, \bm{z}_{\mathcal{N}_i}) \ind Z_{\nu i}, \bm{Z}_{\nu \mathcal{N}_i} | \bm{X}_i.$ 
 Let $\psi_{z\alpha}^{\text{DR-BC}}(\bm{O}_{\nu}; \mu_{z\alpha}, \bm{\beta}, \bm{\gamma}, \phi_b) = \hat{Y}^{\text{DR-BC}}_{\nu}(z,\alpha) - \mu_{z\alpha}$ and $\psi_{\beta}(\bm{O}_{\nu}; \bm{\beta})$, $\psi_{\gamma}(\bm{O}_{\nu}; \bm{\gamma}, \phi_b)$, and $\psi_{\phi_b}(\bm{O}_{\nu}; \bm{\gamma}, \phi_b)$ denote the estimating functions corresponding to $\hat{\bm{\beta}}$, $\hat{\bm{\gamma}}$, and $\hat{\phi}_b$ such that $ \hat{\bm{\theta}}^{\text{DR-BC}} = (\hat{Y}^{\text{DR-BC}}(0, \alpha), \hat{Y}^{\text{DR-BC}}(1, \alpha), \hat{\bm{\beta}}, \hat{\bm{\gamma}}, \hat{\phi}_b)^{\top}$
 solves the estimating equation
$$\bm{\psi}^{\text{DR-BC}}(\bm{\theta}) = \sum_{\nu=1}^m \bm{\psi}^{\text{DR-BC}}(\bm{O}_{\nu}; \bm{\theta}) = \bm{0} $$
where $\bm{\theta}=(\mu_{0\alpha}, \mu_{1\alpha}, \bm{\beta},\bm{\gamma}, {\phi}_b)^{\top}$ and \begin{multline*}
\bm{\psi}^{\text{DR-BC}}(\bm{O}; \bm{\theta})=( \psi_{0\alpha}^{\text{DR-BC}}(\bm{O}; \mu_{0\alpha}, \bm{\beta}, \bm{\gamma}, \phi_b), \psi_{1\alpha}^{\text{DR-BC}}(\bm{O}; \mu_{1\alpha}, \bm{\beta}, \bm{\gamma}, \phi_b),\\ \psi_{\beta}(\bm{O}; \bm{\beta}), \psi_{\gamma}(\bm{O}; \bm{\gamma}, \phi_b),\psi_{\phi_b}(\bm{O}; \bm{\gamma}, \phi_b))^{\top}.
\end{multline*}
The following proposition establishes the double robustness property and asymptotic normality of the bias corrected estimator.


\begin{proposition}
\label{proposition:dr}
If either $f(Z_i, \bm{Z}_{\mathcal{N}_i}| \bm{X}_i, \bm{X}_{\mathcal{N}_i}; \bm{\gamma}, \phi_b)$ or $m(Z_i, \Sigma \bm{Z}_{\mathcal{N}_i}, \bm{X}_i; \bm{\beta})$ is correctly specified, then $\hat{\bm{\theta}}^{\text{DR-BC}} \xrightarrow{p} \bm{\theta}$ and $$\sqrt{m}(\hat{\bm{\theta}}^{\text{DR-BC}} -  \bm{\theta}) \xrightarrow{d}  N(\bm{0}, \bm{U}(\bm{\theta})^{-1}\bm{V}(\bm{\theta})(\bm{U}(\bm{\theta})^{-1})^{\top}),$$
where $\bm{U}(\bm{\theta})=\mathbb{E}\left[-\partial \bm{\psi}^{\text{DR-BC}}(\bm{O}_{\nu}; \bm{\theta})/ \partial \bm{\theta}^{\top} \right]$ and $\bm{V}(\bm{\theta}) = \mathbb{E}\left[\bm{\psi}^{\text{DR-BC}}(\bm{O}_{\nu}; \bm{\theta}) \bm{\psi}^{\text{DR-BC}}(\bm{O}_{\nu}; \bm{\theta})^{\top} \right].$
\end{proposition}
The proof of Proposition \ref{proposition:dr} can be found in Appendix \ref{A.B} and follows from standard estimating equation theory. The empirical sandwich variance estimator of $\hat{\bm{\theta}}^{\text{DR-BC}}$ can be obtained by plugging the sample analogues of $\bm{U}(\bm{\theta})$ and $\bm{V}(\bm{\theta})$ in the sandwich matrix:
$$\bm{\Sigma}_m = \bm{U}_m(\hat{\theta})^{-1} \bm{V}_m(\bm{\hat{\theta}}) \left(\bm{U}_m(\hat{\theta})^{-1}\right)^{\top},  $$
where $\bm{U}_m(\hat{\bm{\theta}}) = m^{-1} \sum_{\nu=1}^m -\partial \bm{\psi}^{\text{DR-BC}}(\bm{O}_{\nu}; \hat{\bm{\theta}})/ \partial \bm{\theta}^{\top}$ and \\$\bm{V}(\hat{\bm{\theta}}) = m^{-1} \sum_{\nu=1}^m \bm{\psi}^{\text{DR-BC}}(\bm{O}_{\nu}; \hat{\bm{\theta}}) \bm{\psi}^{\text{DR-BC}}(\bm{O}_{\nu}; \hat{\bm{\theta}})^{\top} $.
Asymptotic normality and asymptotic variance of causal effect estimators such as $\widehat{DE}(\alpha)=\tau^{\top}\bm{\theta}$ can be derived using the delta method, where $\tau = \begin{pmatrix} -1 & 1  & 0 & \cdots & 0 \end{pmatrix}^{\top}$. For instance, the sandwich variance estimator of $\widehat{DE}(\alpha)$ is given by $\tau^{\top} \bm{\Sigma}_m \tau$, which can be used to construct a Wald-type confidence interval for $DE(\alpha)$. The computation of the sandwich variance estimator can entail error-prone derivative calculations, owing to the presence of a component-level random effect in the treatment model (see the appendix of \cite{perez2014assessing}). Tedious calculations can be avoided using the \texttt{R} package \textbf{geex}, which computes the roots and the sandwich variance estimator for any set of unbiased estimating equations \citep{saul2020calculus}.

\subsection{Other Extensions}
In Appendix \ref{section:AppendixA}, we extend the regression estimator with inverse-propensity weighted coefficients (IP-WLS) considered by \cite{liu2019doubly} to the network interference setting. Instead of incorporating a correction term as in (\ref{eqn:dr-bc1}) and (\ref{eqn:dr-bc2}), inverse-propensity weighted regression is used remove confounding and hence eliminate the bias in the regression estimators of average potential outcomes \citep{kang2007demystifying}. Contrary to what its name suggests, this estimator actually uses inverse probability-of-treatment weights, i.e., $\mathds{1}(Z_i=z)/f(Z_i, \bm{Z}_{\mathcal{N}_i}|\bm{X}_i, \bm{X}_{\mathcal{N}_i})$, as opposed to inverse propensity weights, i.e., $f(1, \bm{Z}_{\mathcal{N}_i}|\bm{X}_i, \bm{X}_{\mathcal{N}_i})^{-1}$. The IP-WLS estimators of $\mu_{z\alpha}$ and $\mu_{\alpha}$ are given by equations (1) and (2) in Appendix \ref{section:AppendixA}, respectively. We show that the IP-WLS estimator has the double robustness property if errors in the outcome model are conditionally independent.

So far, the regression and doubly robust estimators discussed have assumed independent errors, in that
the outcome nuisance model does not assume a multilevel structure. Given that the Add Health data is clustered, we aimed to extend the methods to account for potential dependence in the GPAs received by students within each component. Generalizations of the above estimators to multilevel outcomes under the random intercept linear mixed model are discussed in Appendix \ref{section:AppendixB}. Although these generalizations rely upon a different model to produce predicted outcome values, the form of the estimators remains unchanged. In other words, equations (\ref{eq:reg1}), (\ref{eqn:dr-bc1}), and equation (\ref{eq:drwls1}) of Appendix \ref{section:AppendixA} can still be used to respectively compute the regression estimator, the regression estimator with residual bias correction, and the regression estimator with inverse-propensity weighted coefficients of average potential outcomes by leveraging predictions from a linear mixed model. The same is true for outcome-based estimators of marginal average potential outcomes in equations (\ref{eq:reg2}), (\ref{eqn:dr-bc2}), and equation (\ref{eq:drwls2}) of Appendix \ref{section:AppendixA}.

For the most part, M-estimation of parameters is straightforward except for those involved in the IP-WLS estimator. In this case, the implementation requires weighting fixed effect coefficients by the inverse of the joint propensity score. Therefore, inference for the outcome nuisance and treatment nuisance models is performed jointly and one must be able to specify a likelihood that also accounts for the hierarchical nature of the data. More details on the implementation of weighted mixed linear models are provided in Appendix \ref{section:appendixb.4}. Contrary to the case of conditionally independent outcomes discussed above, the IP-WLS estimator does not appear to retain the double robustness property in the presence of multilevel outcomes as explained in Appendix \ref{section:appendixb.4}.

\section{Simulations}
\label{section:simulations}
We designed simulation studies to assess finite sample bias of the IPW, regression (REG), and DR estimators described in Section \ref{section:inferential} as well as to compare their efficiency and robustness to misspecification. The first simulation scheme assumed equal component sizes and conditionally independent errors in the outcome model, while the second scheme aimed to mimic certain aspects of the Add Health study, namely the unequal network component sizes and the multilevel structure of the outcome data.

Five scenarios based on correct or incorrect specification of the outcome or treatment models, and one setting in which both models were ``correctly specified'' but there was latent treatment dependence that was not captured by either model (see Steps 5-6 below). This first simulation study was conducted according to the following steps with $m = 30$ network components:
\begin{enumerate}
\item \textbf{Generation of the network}: For $\nu=1,\ldots, m$, we considered a fixed number of nodes $N_{\nu} = 30$. For each component, we generated the random vector $\bm{H}_{\nu}$ whose elements were a $\mathrm{Bernoulli}(0.5)$ distribution. The $m$ components were generated according to an exponential random graph model (ERGM) in which matching $H$ values increased the probability of tie between two nodes, inducing network homophily \citep{kolaczyk2009}. Specifically, the ERGM used to simulate each network component was such that the conditional log-odds of two nodes having a tie while holding the rest of the network fixed was $1.5 \delta_1 - 2.5 \delta_2,$
where $\delta_1$ is the indicator for matching $H$ values and $\delta_2$ represents the increase in the number of ties. The final network $G$ was taken as the union of the $m$ components. 
\item For $i = 1, \ldots, N_{\nu}$, two covariates were generated as $X_{1i} \sim \mathcal{N}(0,1)$ and $X_{2i} \sim \mathrm{Bernoulli}(0.5)$. Denote the proportion of treated nodes in a neighbourhood as $p(\bm{z}_{\mathcal{N}_i}) = \Sigma \bm{z}_{\mathcal{N}_i}/d_i$. Potential outcomes for all $z_i \in \{0,1\}$, $\Sigma \bm{z}_{\mathcal{N}_i} \in \{0,1,\ldots, d_i\}$ are given by
$$y_i(z_i, \bm{z}_{\mathcal{N}_i}) = 2 + 2 z_i + p(\bm{z}_{\mathcal{N}_i}) + z_i p(\bm{z}_{\mathcal{N}_i}) -1.5 |X_{1i}| + 2X_{2i} - 3 |X_{1i}|X_{2i} + \varepsilon_i, \ \varepsilon_i \sim N(0,1).$$
\item  The treatment was generated with homophily as $Z_i = \mathrm{Bernoulli}(p_i)$, where $$p_i = \mathrm{logit}^{-1}\left[0.1 +0.2 |X_{1i}| + 0.2X_{2i}|X_{1i}| - H_i+ b_{\nu}\right], b_{\nu} \sim N(0, 1).$$ 
\item Based on the potential outcomes defined in Step 2, the observed outcomes were set equal to $Y_i = y_i\left(Z_i,  \Sigma \bm{Z}_j\right)$.
\item The following models were fitted to the simulated data:
\begin{itemize}
    \item Correct treatment model: $$\mathbb{E}[Z_i|\bm{X}_i] = \mathrm{logit}^{-1}(\gamma_0 + \gamma_1 |X_{1i}| + \gamma_2 |X_{1i}| X_{2i} + \gamma_3 H_{i} + b_{\nu} )$$
    \item Incorrect treatment model: $$\mathbb{E}[Z_i|\bm{X}_i] = \mathrm{logit}^{-1}\left(\gamma_0 + \gamma_1 X_{1i} + \gamma_2 H_{i} + b_{\nu} \right)$$
    \item Correct outcome model: $$\mathbb{E}[Y_i|\bm{X}_i, Z_i, \bm{Z}_{\mathcal{N}_i}] = \beta_0 + \beta_1 Z_i + \beta_2 p(\bm{Z}_{\mathcal{N}_i}) + \beta_3 Z_i p(\bm{Z}_{\mathcal{N}_i}) + \beta_4 |X_{1i}| + \beta_5 X_{2i} + \beta_6 |X_{1i}|X_{2i} $$
    \item Incorrect outcome model $$\mathbb{E}[Y_i|\bm{X}_i, Z_i, \bm{Z}_{\mathcal{N}_i}] = \beta_0 + \beta_1 Z_i + \beta_2 p(\bm{Z}_{\mathcal{N}_i}) + \beta_3 Z_ip(\bm{Z}_{\mathcal{N}_i}) + \beta_4 X_{1i} + \beta_5 X_{2i} $$
\end{itemize}
\item Additionally, a treatment model excluding the variable driving homophily was also fitted to illustrate the threat of latent treatment homophily:
$$\mathbb{E}[Z_i|\bm{X}_i] = \mathrm{logit}^{-1}\left(\gamma_0 + \gamma_1 |X_{1i}| + \gamma_2 |X_{1i}| X_{2i}  + b_{\nu} \right).$$
\item Estimators in (\ref{eq:ipw1}), (\ref{eq:reg1}), (\ref{eqn:dr-bc1}) and as well as IP-WLS (equation (1) of Appendix \ref{section:AppendixA}) were computed from the simulated sample for each scenario with $z \in \{0,1\}$ and $\alpha \in \{0, 0.1, 0.2, \ldots, 1\}$. The propensity score in (\ref{eqn:propensity}) was computed using Gauss-Hermite quadrature with 10 points. 
\end{enumerate}
Simulations were carried out 1,000 times by repeating steps 2-7. The true parameters were calculated by averaging the potential outcomes generated in Step 2. Empirical expectations of estimators were computed by averaging the point estimates over the 1,000 replicates. The estimated standard errors were obtained from the sandwich variance matrix. Empirical coverage was defined as the proportion of simulation replicates for which the true parameters were contained in the Wald-type 95\% confidence intervals. Lack of convergence in the computation of the sandwich variance estimators led to the exclusion of a few samples depending on the scenario under investigation. One sample was excluded for the IPW estimator in scenarios (a) and (b), while 3 samples were excluded for the IPW estimator in scenarios (c) and (d). Between 5 and 8 samples were discarded when assessing the DR-BC and IP-WLS estimators across all scenarios. No sample was excluded for the REG method. 

\begin{figure}[!htb]
    \centering
    \begin{minipage}{.50\textwidth}
        \centering
\includegraphics[width=0.9\linewidth, height=0.30\textheight]{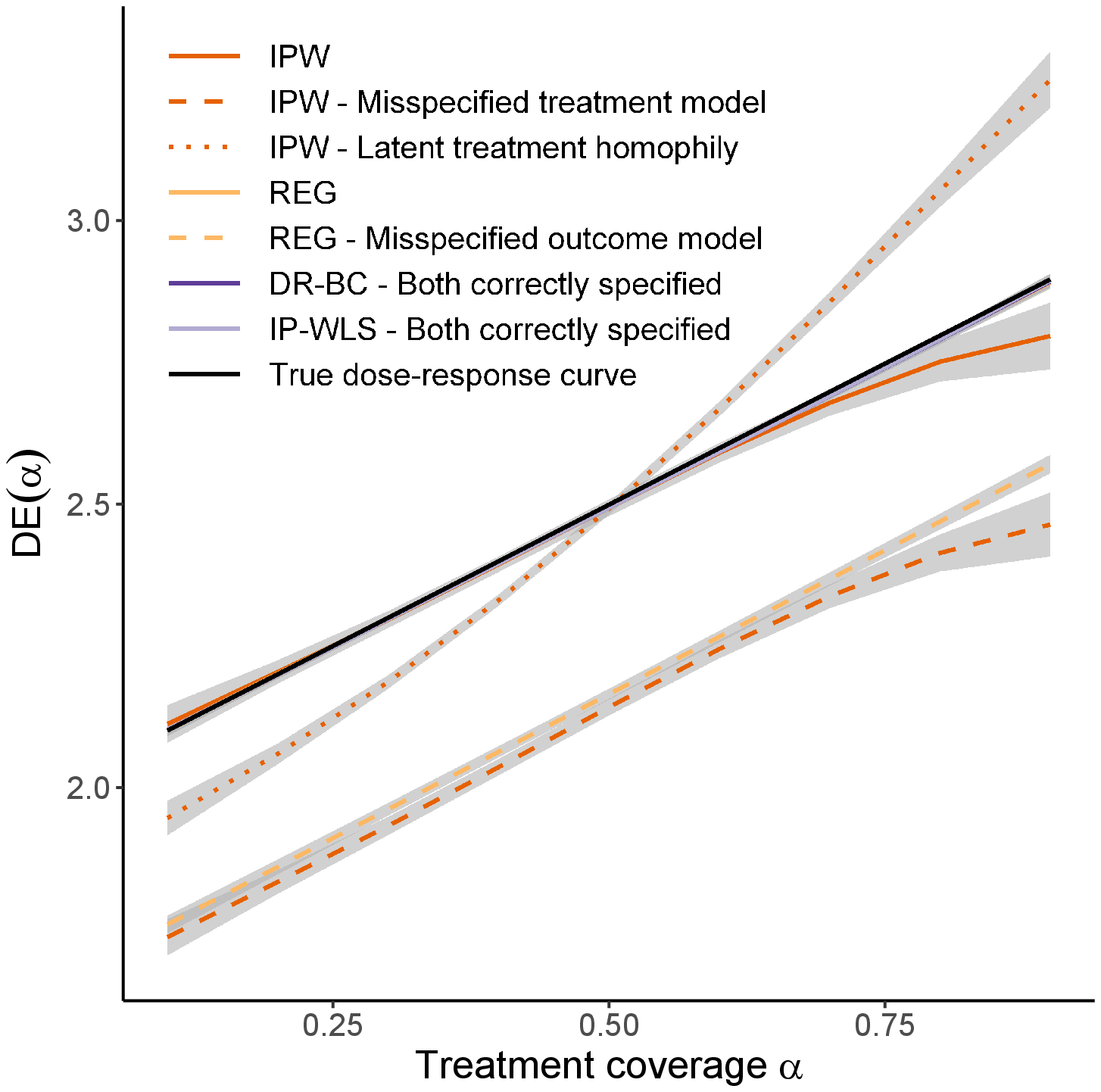}
    \end{minipage}%
    \hspace{0.05\textwidth}%
    \begin{minipage}{0.45\textwidth}
        \centering
                \vspace*{-0.3cm}
                \includegraphics[width=0.95\linewidth, height=0.31\textheight]{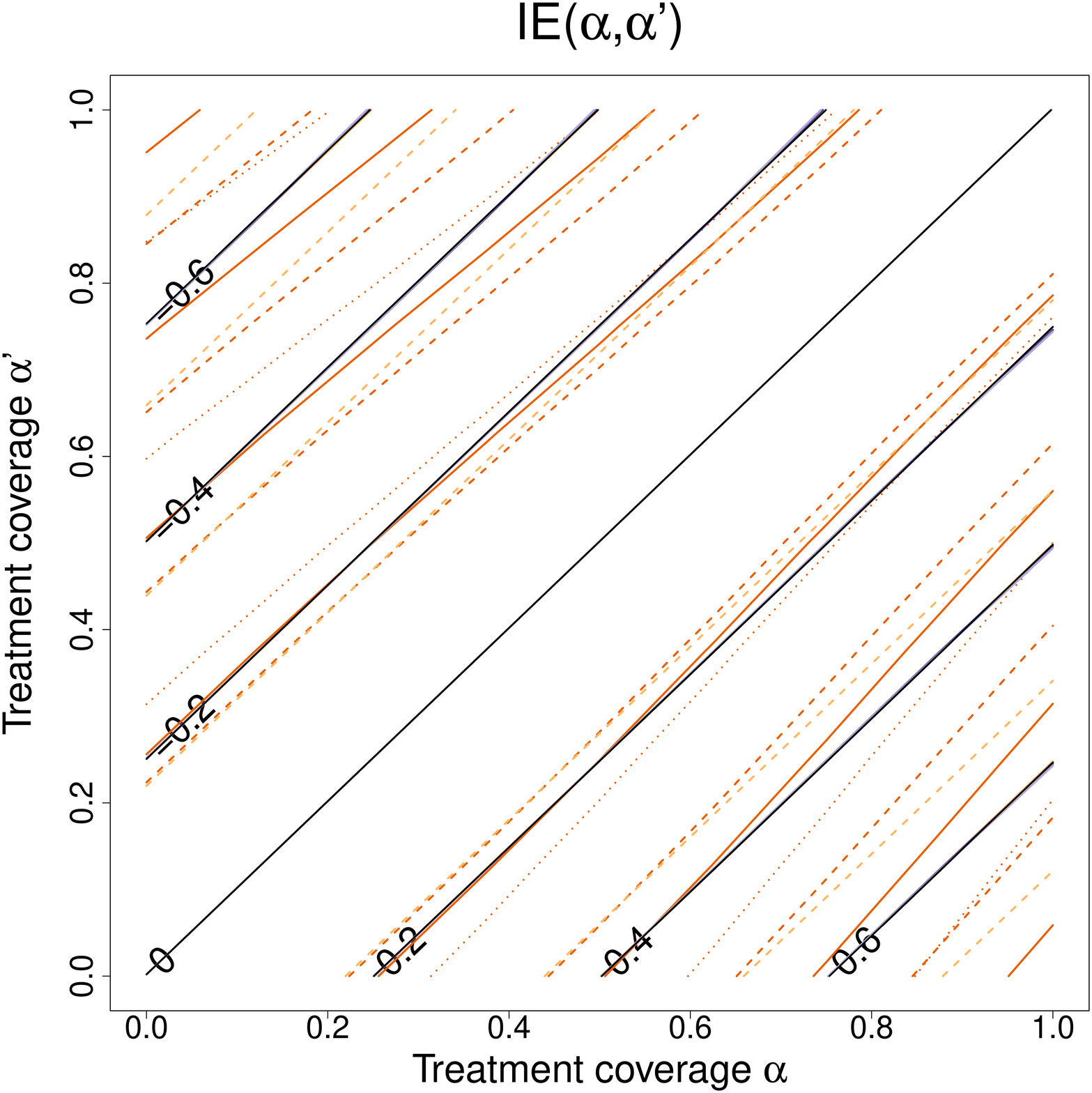}
    \end{minipage}
    \caption[Average estimates of the dose response curve $DE(\alpha)$ and the surface $IE(\alpha, \alpha')$ based on 1,000 simulation replicates.]{Left: Estimates of the dose response curve $DE(\alpha)$. Right: Estimates of the surface $IE(\alpha, \alpha')$. The black solid line represents the true effect. The light gray area represents 95\% pointwise confidence intervals based on $S = 1,000$ simulation replicates.} 
    \label{fig:plot1}
\end{figure}

\begin{figure}[!htb]
\centering
    \includegraphics[scale=0.40]{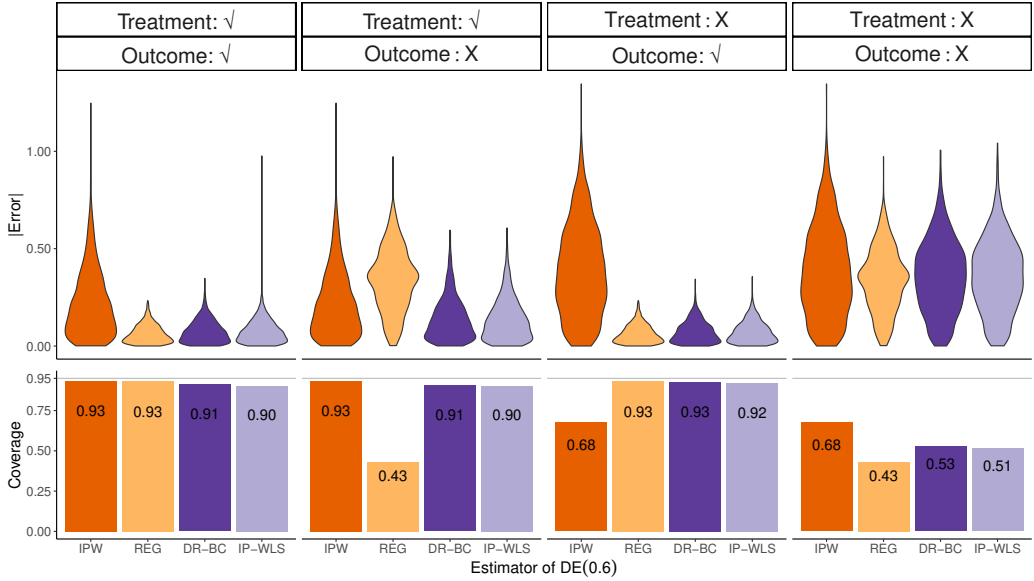}
    \caption[Absolute bias and confidence interval coverage for the IPW, REG, DR-BC, and IP-WLS estimators of $DE(0.6)$ under the first simulation scheme.]{Absolute bias and confidence interval coverage for the IPW, REG, DR-BC, and IP-WLS estimators of $DE(0.6)$ under the first simulation scheme. In scenarios (a) and (b), results for IPW, REG, DR-BC, and IP-WLS are based on 999, 1,000, 992, and 992 simulated datasets, respectively. In scenarios (c) and (d), results for IPW, REG, DR-BC, and IP-WLS are based on 997, 1,000, 995, and 994 simulated datasets, respectively.} 
    \label{fig:absbias}
\end{figure}

Figure \ref{fig:plot1} displays the average of the 1,000 IPW, regression (REG), the regression estimator with residual bias correction (DR-BC) and the regression estimator with inverse-propensity weighted coefficients (IP-WLS) discussed in Section \ref{section:inferential} of the direct effect $DE(\alpha)$ and indirect effect $IE(\alpha, \alpha')$, respectively, as well as the corresponding true effects. A different version of the plot in the right panel can be found in Appendix \ref{section:AppendixC} for increased readability. Unless otherwise stated, estimators are based on the correctly specified models for either treatment or outcome. As expected, the true dose response curve and the REG, DR-BC, and IP-WLS estimators almost perfectly overlap, while the IPW estimator exhibits some instability for $\alpha$ values closer to 1. Furthermore, the IPW estimator of $IE(\alpha, \alpha')$ departs from the true surface for larger gaps between $\alpha$ and $\alpha$'. The IPW and REG estimators of the dose-response curve based on misspecified models show systematic bias. The DR-BC and IP-WLS estimators of $DE(\alpha)$ and $IE(\alpha, \alpha')$ based on either a misspecified treatment or a misspecified outcome model behaved as expected (results not shown). Finally, unmeasured homophily violates Assumption \ref{assump5} and leads to a biased IPW estimate (orange dotted curve).
 \indent Figure \ref{fig:absbias} shows the absolute bias of the IPW, REG, DR-BC, and IP-WLS estimators of $DE(0.6)$ under different misspecification scenarios. We see that the bias of the DR-BC and IP-WLS estimators is negligible whenever one of the two models is correctly specified. For instance, for scenario (b), the bias of IPW, REG, DR-BC, and IP-WLS are -0.011, -0.332, -0.014, and -0.024, respectively, whereas for scenario (c), these values are -0.358, -0.004, -0.007, and -0.007. In addition to exhibiting the double robustness property, the DR estimator proves to be more efficient than the IPW estimator whenever at least one model of the two is correctly specified. For example, when only the model for the propensity score is correctly specified, the Monte Carlo mean squared error of the IPW estimator is 0.163 while that of the DR-BC and IP-WLS estimators are 0.032 and 0.030, respectively. However, when the outcome model is correctly specified, the REG estimator exhibits the smallest mean squared error (0.006 compared to 0.008 for both DR-BC and IP-WLS). 
 
 Empirical coverages of Wald-type 95\% confidence intervals for $DE(0.6)$ based on the sandwich variance estimates can be found at the bottom of Figure \ref{fig:absbias}. When either model is correctly specified, the empirical coverages of DR-BC and IP-WLS are approximately 0.95. The empirical coverages for the singly robust methods are well below 0.95 when the working model is not correctly specified. The same is true for the DR-BC and IP-WLS methods when both models are incorrect. For instance, when the treatment and outcome models are incorrect, the empirical coverages of the IPW, REG, DR-BC and IP-WLS are 0.68, 0.43, 0.53, and 0.51, respectively. These comparisons align with what \cite{liu2019doubly} found in their simulation studies, albeit in a setting with partial interference only.

Table \ref{tab:tab1} in Appendix \ref{section:AppendixC} shows the bias and the mean squared error (MSE) of the IPW, REG, DR-BC, and IP-WLS estimators of the indirect/spillover effect using the correctly specified treatment model versus the treatment model deprived of the homophilous variable, which, it should be noted, is not a confounder in the classical sense of directly impacting both treatment and outcome. In a classical causal inference setting, where units are independently and identically distributed, the omission of a predictor of the treatment that is not a confounder in the treatment nuisance model can improve the estimation (reducing the mean squared error of the estimator). However, in the network setting, these empirical results show that the IPW estimator in (\ref{eq:ipw1}) is biased if traits shared by adjacent nodes that affect treatment assignment are unobserved or not conditioned upon. 

Appendix \ref{section:AppendixC} also contains additional simulations in which we assessed sensitivity to failure of the stratified interference assumption. In one scenario considered, the potential outcomes do not merely depend on the sum of treated neighbours but through a more complex function of the neighbourhood treatment vector and show that for this particular violation and data generating scheme, the performance of the estimators is mildly affected. In an additional scenario, we allow the potential outcomes to depend on second-order neighbours (friends of friends) and show that a misspecification of the order of the spillover effect in the outcome model leads to  slight bias and undercoverage for the data generating mechanism considered.

In Appendix \ref{section:B}, we present results for a second simulation scheme in which network components were unbalanced and the outcomes were generated from a random intercept linear model in order to investigate the performance of the proposed estimators under conditions similar to those encountered in the data analysis in the following section. We investigated properties of estimators for an increasingly large number of network components $m$, where $m$ is from the set \{50, 100, 150, 200\}. This gave further empirical evidence of consistency and asymptotic normality of the singly robust methods (IPW and REG) when the correct model is used. We also gained empirical evidence of the double robustness of the DR-BC estimator for multilevel outcomes when at least one of the two models is correctly specified. Finally, when examining point estimates for the IP-WLS method, we observed a non-vanishing bias as the number of network components increased. Since there is no theoretical guarantee for the double robustness of the IP-WLS method in the presence of multilevel outcomes, we excluded this estimator from the analysis of Add Health.
\section{Effect of Maternal Education on School Performance}
\label{section:addhealth}

We re-examine a question considered by previous authors about the impact of maternal college education on adolescent school performance \citep{bifulco2011effect, fletcher2020consequences}. 
In the following, the outcome is a student’s GPA and the exposure is defined as the binary
indicator of whether their mother has completed four years of college. The students who did not report
living with their biological mother, stepmother, foster mother, or adoptive mother at the time of the survey were excluded from the analytic sample. Furthermore, we decided to exclude three schools from the analysis at outset because more than 98\% of their nodes were isolates, which correspond to nodes with degree 0. This initial preparation resulted in an undirected network of 139 schools and 73,580 nodes in total. 

Further pre-processing of the full sample detailed in Appendix \ref{section:AppendixE} was performed to arrive at our analytic sample. The final network structure $G=(\mathcal{N}, E)$ had 48 components (observed subnetworks) with $N=$ 5,724 students and 19,776 reciprocal friendships. The size of the 48 components ranged from 2 to 508 students, with a median of 31 (1st to 3rd quartile (Q1--Q3) : 2--201). Characteristics of the network and descriptive statistics for this subsample are presented in Tables S2 and S3 of Appendix \ref{section:AppendixE}, respectively. Using this analytic subsample, we fitted a linear mixed effects model for GPA and a distinct mixed effects logistic model for the propensity of maternal education at the student level. A random intercept for the network component was incorporated into each model. Given that the in-school questionnaire collected limited information about the student's mother, the model for the propensity of maternal education included only four covariates: student's race (as a proxy for the race of the mother), whether the mother was born in the US, whether the father is at home, and household size. Similarly to \cite{egami2021identification}, in addition to those included in the model for the exposure, the covariates in the outcome model included gender, age, whether the student was adopted, whether the mother works for pay, perception of mother care, club participation, motivation at school, trouble at school, sense of belonging, attendance, health status, physical fitness, and screen time.
\begin{figure}[ht]
        \centering
        \includegraphics[width=1\linewidth, height=0.35\textheight]{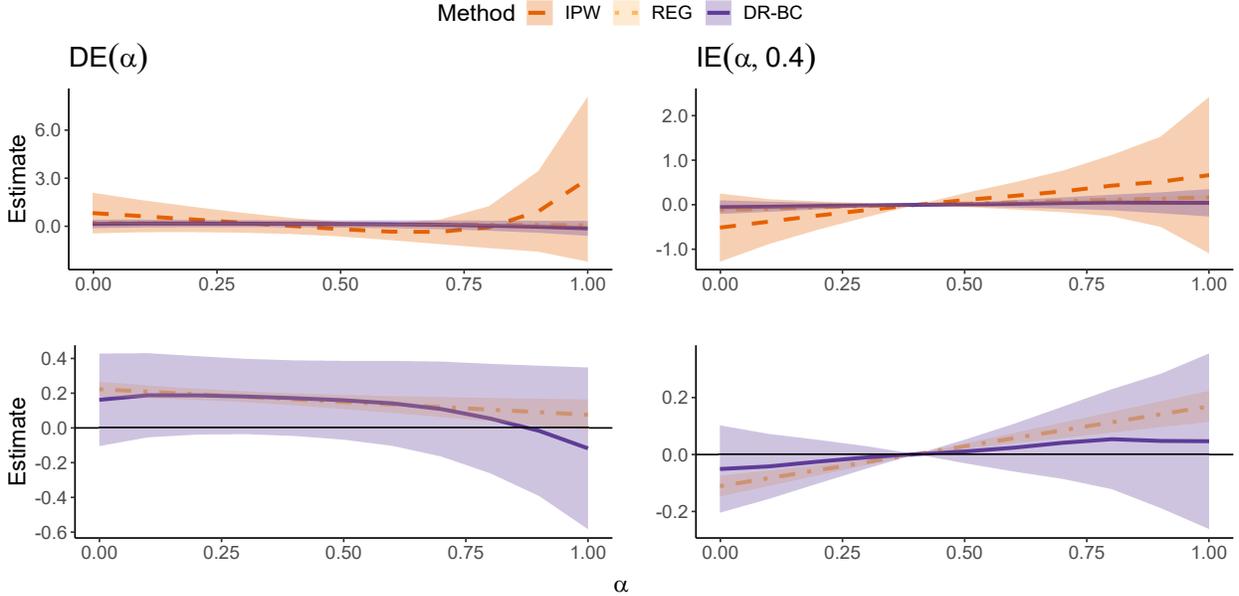}
        \caption[IPW, REG, and DR-BC estimates of ${DE}(\alpha)$ and ${IE}(\alpha, 0.4)$ effects with 95\% confidence bands.]{IPW, REG, and DR-BC estimates of ${DE}(\alpha)$ and ${IE}(\alpha, 0.4)$ effects with 95\% confidence bands. The bottom panel only shows the REG and DR-BC estimates. 
        }
        \label{fig:plotDEIE}
\end{figure}
\begin{figure}[ht]
        \centering
        \includegraphics[width=1\linewidth, height=0.35\textheight]{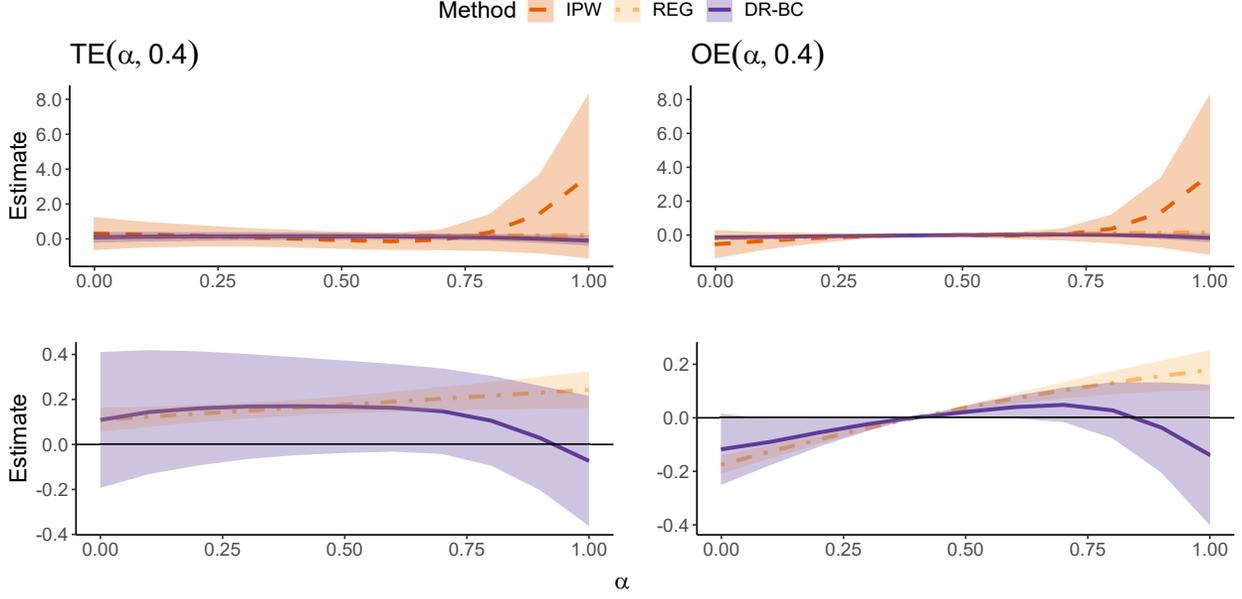}
        \caption[IPW, REG, and DR-BC estimates of ${TE}(\alpha, 0.4)$ and ${OE}(\alpha, 0.4)$ with 95\% confidence bands.]{IPW, REG, and DR-BC estimates of ${TE}(\alpha, 0.4)$ and ${OE}(\alpha, 0.4)$ with 95\% confidence bands. The bottom panel only shows the REG and DR-BC estimates.}
        \label{fig:plotTEOE}
\end{figure}

Figures \ref{fig:plotDEIE} and \ref{fig:plotTEOE} compare the estimators of the direct effect ${DE}(\alpha)$, indirect effect ${IE}(\alpha, 0.4)$, total effect ${TE}(\alpha, 0.4)$, and overall effect ${OE}(\alpha, 0.4)$. For all estimands, the REG and DR-BC estimators generally give similar conclusions, in that the directions of the dose-response curves align. In comparison, the dose-response curves estimated by IPW can be non-monotonic and attain relatively high values for large $\alpha$ values. For instance, the REG and DR-BC estimates of $DE(\alpha)$ range between 0.22 and -0.12 and decrease with increasing coverage $\alpha$, whereas the IPW estimate first decreases and increases again past $\alpha = 0.7$ to reach the value 2.95 at $\alpha=1.0$, which is a rather extreme value considering the range of GPA (between 0 and 4). Recall that the model for maternal education was very limited compared to the outcome model, with only 4 covariates. Also, as shown by the histogram in Figure \ref{fig:proptreated} of Appendix \ref{section:AppendixE}, the distribution of observed proportion of friends with a college-educated mother was right skewed. The reduced amount of information for higher $\alpha$ values is reflected in the width of the confidence bands of the IPW estimator, and in that of the DR-BC estimator to a lesser extent.
Table \ref{tab:estimatesAddHealth} displays the IPW, REG and DR-BC estimates along with 95\% confidence intervals for ${DE}(\alpha)$, ${IE}(\alpha, 0.4)$, ${TE}(\alpha, 0.4)$ and ${OE}(\alpha, 0.4)$. We provide an interpretation of $DE(0.1)$ for the DR-BC point estimate. If every student in this network were to be ``assigned'' a college-educated mother versus none, we would expect the average GPA difference to be 0.18 (95\% CI: [-0.06, 0.43]) points under a counterfactual probability of receiving treatment $\alpha=10$\%. The interpretation of $DE(0.10)$ is nuanced, since it is not feasible for each person to be exposed and simultaneously have only 10\% of their friends exposed. Recall that the target estimand $DE(\alpha)$ is defined as the expected average of individual-level direct effects standardized according to a reference population in which individuals independently receive the exposure with probability $\alpha$. Both the REG and DR-BC methods yield a decreasing $DE(\alpha)$ curve, although the point estimates given by DR-BC never reach statistical significance. This could mean that there is little to no added benefit for a student to have a college-educated mother, and even less so if there is already a high proportion of ``exposed'' friends in their social circle.

\begin{table}
\caption{\label{tab:estimatesAddHealth} IPW, REG, and DR-BC estimates for the Add Health study along with 95\% Wald-type confidence intervals for $\mathrm{DE}(\alpha)$, $\mathrm{IE}(\alpha, 0.4)$, $\mathrm{TE}(\alpha, 0.4)$ and $\mathrm{OE}(\alpha, 0.4)$.}
\centering
\begin{tabular}{lccc}
  \toprule
 Estimand & IPW & REG & DR-BC \\ 
  \midrule
  $DE(\alpha)$\\
 $\alpha=0.1$ & 0.64 [-0.36,1.64] & 0.20 [0.17,0.24] & 0.18 [-0.06,0.43] \\
$\alpha=0.3$ & 0.24 [-0.39,0.86] & 0.18 [0.14,0.21] & 0.18 [-0.04,0.39] \\
 $\alpha=0.5$ & -0.17 [-0.62,0.29] & 0.15 [0.11,0.19] & 0.16 [-0.07,0.38] \\
  $\alpha=0.7$ & -0.34 [-1.10,0.42] & 0.12 [0.06,0.17] & 0.11 [-0.17,0.38] \\ 
 \midrule
 $IE(\alpha, 0.4)$\\
   $\alpha=0.1$ & -0.38 [-0.88,0.13] & -0.08 [-0.11,-0.06] & -0.04 [-0.16,0.07] \\
   $\alpha=0.3$ & -0.12 [-0.27,0.04] & -0.03 [-0.04,-0.02] & -0.01 [-0.05,0.03] \\
   $\alpha=0.5$ & 0.11 [-0.05,0.26] & 0.03 [0.02,0.04] & 0.01 [-0.03,0.05] \\ 
      $\alpha=0.7$ & 0.30 [-0.16,0.76] & 0.08 [0.06,0.11] & 0.04 [-0.09,0.17] \\ 
    \midrule
     $TE(\alpha, 0.4)$\\
     $\alpha=0.1$ & 0.26 [-0.48,1.01] & 0.12 [0.08,0.17] & 0.14 [-0.13,0.42] \\ 
   $\alpha=0.3$ & 0.12 [-0.42,0.66] & 0.15 [0.11,0.18] & 0.17 [-0.07,0.4] \\ 
   $\alpha=0.5$  & -0.06 [-0.55,0.43] & 0.17 [0.14,0.21] & 0.17 [-0.04,0.37] \\ 
      $\alpha=0.7$  & -0.04 [-0.62,0.55] & 0.20 [0.15,0.25] & 0.15 [-0.04,0.33] \\ 
   \midrule
   $OE(\alpha, 0.4)$\\
    $\alpha=0.1$ & -0.32 [-0.85,0.20] & -0.13 [-0.15,-0.10] & -0.09 [-0.18,-0.01] \\ 
   $\alpha=0.3$ & -0.06 [-0.20,0.09] & -0.04 [-0.05,-0.03] & -0.03 [-0.05,0.01] \\ 
   $\alpha=0.5$  & 0.01 [-0.10,0.13] & 0.04 [0.03,0.05] & 0.02 [-0.01,0.04] \\
      $\alpha=0.7$ & 0.05 [-0.30,0.40] & 0.10 [0.07,0.13] & 0.05 [-0.02,0.11] \\ 
   \bottomrule
\end{tabular}
\end{table}

Using the DR-BC method, we obtain an estimate of -0.04 (95\% CI: [-0.16, 0.07]) for $IE(0.1, 0.4)$. This implies that if we held the maternal education indicator to 0, the GPA of students with 10\% of friends with college-educated mothers would be on average lower by -0.04 points compared to those with 40\% of friends exposed, assuming both groups' covariate distribution matched that of the overall population. The REG method suggests a small positive spillover effect of maternal education on GPA, as $IE(\alpha, 0.4)$ increases with $\alpha$. This aligns with some findings of \cite{fletcher2020consequences}. However, the DR-BC method does not suggest such an effect, as the estimates do not significantly differ from 0. Since the total effect is the sum of the direct and indirect effects, it remains relatively constant across different $\alpha$ values. Similarly to the indirect effect, the point estimate of the overall effect, which marginalizes over the individual exposure/treatment, increases with the proportion of treated friends for values up to $\alpha=0.7$. For instance, the DR-BC estimates are -0.09 (95\% CI: [-0.18, -0.01]), -0.03 (95\% CI: [-0.05, 0.01]), and 0.05 (95\% CI: [-0.02, 0.11]) for $OE(0.1, 0.4)$, $OE(0.3, 0.4)$, and $OE(0.7, 0.4)$, respectively.
\section{Discussion}
\label{section:discussion}
In this paper, we propose doubly robust estimators for causal effects in the presence of clustered network interference. They are shown to be consistent and asymptotically normal if either the treatment model or the outcome regression model is correctly specified. In simulations, we generated networks with equal and unequal component sizes and outcome data with and without a multilevel structure. For conditionally independent outcomes, empirical results illustrate the DR property and the potential efficiency gain of the DR-BC and IP-WLS estimators over previously proposed IPW estimators when the propensity score model is correctly specified. Results also underscore the importance of the no latent treatment homophily assumption for the identification of causal effects, an assumption that we have made explicit in our framework. If the outcome is assumed to follow a hierarchical model, then the DR-BC estimator remains doubly robust, while evidence suggests that IP-WLS is only consistent if the outcome model is correctly specified. It is worth noting that the proposed methods can be applied to observational network data that do not have a clear multilevel structure from the outset. Indeed, community detection algorithms can be used to identify clusters beforehand, as was done under partial interference \citep{perez2014assessing, liu2019doubly} and networked interference \citep{lee2021estimating}. Using causal structural equation models, \cite{ogburn2022causal} proposed a TMLE approach for a single social network. However, they use a bounded dependence central limit theorem to establish asymptotic properties, which precludes the presence of nodes with high degree centrality or so-called ``hubs'' in the social network. That is, conditions for asymptotic normality will not hold if the degree of one or more nodes grows faster than $\sqrt{N}$ \citep{ogburn2022causal}. Our methods are free of such assumptions, as our asymptotic framework views each network component as a datum and holds regardless of the network topology within each component.

Our study of peer characteristics in education leveraging the Add Health dataset did not provide evidence of a spillover effect of maternal education on academic performance within adolescents' social circles. However, when outcome regression was used, we did find that a higher proportion of friends with a college-educated mother led to slightly higher GPA among students whose mothers had not completed a 4-year college degree, corroborating some substantive conclusions of \cite{fletcher2020consequences}. This study is nonetheless subject to important limitations. First of all, the exposure of interest, maternal education, was not well defined and did not correspond to a standard intervention, potentially violating Assumption \ref{assum1}. Also, the exposure and outcome were contemporaneous, making it essentially impossible to distinguish effects due to interference and homophily or environmental confounding \citep{vanderweele2013social, egami2021identification}. In their causal peer effect analysis of Add Health, \cite{egami2021identification} proposed a double negative control approach to adjust for potential network confounding and introduced a time element by using GPA outcomes reported in the in-school and the in-home interviews, which were separated by a period of at least 90 days. However, it is known that the Add Health network at baseline changed overtime. Methods such as stochastic actor-oriented models could be better adapted to this longitudinal setting, as they model social influence and selection of ties jointly \citep{ snijders2005models,vanderweele2013social}. Two other limitations worth mentioning were the need to select a subsample of the network for computational reasons, and the use of single imputation. The first of these was largely practical, but did lead to some meaningful differences in the covariate distribution of the subsample compared with that of the full sample, affecting generalizability of the results.

While our analysis assumes first-order spillover effects only, this assumption could be relaxed by widening the interference set to include second-order neighbours (friends of friends). In some settings it could be desirable to allow the treatment assignments received by friends of friends to have a smaller impact than those received by immediate neighbours. One avenue for future work would be to allow spillover effects to decay with the neighbourhood order, which would require a more substantial extension. Nonetheless, we assessed finite sample properties of the estimators under a violation of first-order spillover effects via simulation and observed that bias and coverage were not severely affected even when the data were generated under a second-order interference mechanism.

Methodological work on causal inference in the presence of network interference is still in its infancy, and there are several areas of future research related to the counterfactual framework presented in this paper. One natural route would be to extend this framework to longitudinal network data, which would allow to address network endogeneity by disentangling effects from social influence and homophily \citep{tchetgen2012causal}. Future work could entail formulating stochastic actor-oriented models within a counterfactual framework. Furthermore, in this paper, the observed social network is assumed to be undirected and considered as known and fixed, such that inference is entirely driven by random treatment assignment and random potential outcomes \citep{li2020random}. However, it is possible that subjects misreport their connections and that social ties are missing not at random \citep{vanderweele2013social}. While the general problem of imputation of network data has received some attention (see \cite{huisman2009imputation, krause2018missing}), to our knowledge, the only existing imputation procedure for causal inference under network interference was proposed in the context of experiments \citep{kao2017causal}. This could be another important avenue for future work in the area of observational network-based studies. Existing causal inference methods could also possibly be extended to the classes of directed networks and weighted networks, wherein the direction (respectively strength) of edges would be of importance. 

A partially observed network, that is, a network with uncertain or missing edges, would imply that the interference sets are not known with certainty which would have an impact on the estimation of spillover effects. In the experimental setting, \cite{li2021causal} formulated the impact of network error on the bias of estimators of average causal effects. The impact of such errors in observational network-based studies has yet to be studied. In addition, asymptotic results are derived under the assumption that the known interference graph can be partitioned into several components \citep{liu2016inverse, lee2021estimating}. \cite{forastiere2021identification} assumed that the social network is a random sample from a larger underlying network and devised bootstrapping procedures based on egocentric and cluster sampling. Future research could involve relaxing the fixed network assumption and assume that the interference graph is a random draw from an (unknown) graphon as was done in experimental settings \citep{li2020random}. Such an inference framework would point to a clearer use of resampling methods for network data to derive finite-sample variance estimators for causal effects.

 In the no interference setting, many types of doubly robust estimators have been proposed. Notably, the class of DR estimators with propensity-based covariates remains to be investigated in the network interference setting. For instance, \cite{liu2019doubly} formulated a generalization of the estimator proposed by \cite{scharfstein1999adjusting} to the partial interference setting, but other estimators such as the one proposed by \cite{bang2005doubly, bang2008correction} could also be considered. Lastly, while \cite{park2022efficient} showed that the bias-corrected doubly robust estimator of \cite{liu2019doubly} was locally efficient under partial interference \citep{zhang2023propensity}, whether the proposed estimators are semiparametric efficient has yet to be investigated in the network interference setting.

\bibliographystyle{plainnat}  

\bibliography{references}

\begin{thebibliography}{46}
\providecommand{\natexlab}[1]{#1}
\providecommand{\url}[1]{\texttt{#1}}
\expandafter\ifx\csname urlstyle\endcsname\relax
  \providecommand{\doi}[1]{doi: #1}\else
  \providecommand{\doi}{doi: \begingroup \urlstyle{rm}\Url}\fi

\bibitem[Abacioglu et~al.(2019)Abacioglu, Isvoranu, Verkuyten, Thijs, and Epskamp]{abacioglu2019exploring}
Ceren~Su Abacioglu, Adela-Maria Isvoranu, Maykel Verkuyten, Jochem Thijs, and Sacha Epskamp.
\newblock Exploring multicultural classroom dynamics: {A} network analysis.
\newblock \emph{Journal of School Psychology}, 74:\penalty0 90--105, 2019.

\bibitem[Bailey et~al.(2021)Bailey, Kelley, Nguyen, and Huo]{bailey2020package}
Paul Bailey, Claire Kelley, Trang Nguyen, and Huade Huo.
\newblock \emph{{W}e{M}ix: {W}eighted {M}ixed-{E}ffects {M}odels {U}sing {M}ultilevel {P}seudo {M}aximum {L}ikelihood {E}stimation}, 2021.
\newblock URL \url{https://CRAN.R-project.org/package=WeMix}.
\newblock R package version 3.2.1.

\bibitem[Bang and Robins(2005)]{bang2005doubly}
Heejung Bang and James~M Robins.
\newblock Doubly robust estimation in missing data and causal inference models.
\newblock \emph{Biometrics}, 61\penalty0 (4):\penalty0 962--973, 2005.

\bibitem[Bang and Robins(2008)]{bang2008correction}
Heejung Bang and James~M Robins.
\newblock Correction to “{D}oubly {R}obust {E}stimation in {M}issing {D}ata and {C}ausal {I}nference {M}odels,” by {H.} {B}ang and {JM} {R}obins; 61, 962--972, {D}ecember 2005.
\newblock \emph{Biometrics}, 64\penalty0 (2):\penalty0 650--650, 2008.

\bibitem[Basse and Feller(2018)]{basse2018analyzing}
Guillaume Basse and Avi Feller.
\newblock Analyzing two-stage experiments in the presence of interference.
\newblock \emph{Journal of the American Statistical Association}, 113\penalty0 (521):\penalty0 41--55, 2018.

\bibitem[Bates et~al.(2014)Bates, M{\"a}chler, Bolker, and Walker]{bates2014fitting}
Douglas Bates, Martin M{\"a}chler, Ben Bolker, and Steve Walker.
\newblock Fitting linear mixed-effects models using lme4.
\newblock \emph{arXiv preprint arXiv:1406.5823}, 2014.

\bibitem[Bifulco et~al.(2011)Bifulco, Fletcher, and Ross]{bifulco2011effect}
Robert Bifulco, Jason~M Fletcher, and Stephen~L Ross.
\newblock The effect of classmate characteristics on post-secondary outcomes: Evidence from the {A}dd {H}ealth.
\newblock \emph{American Economic Journal: Economic Policy}, 3\penalty0 (1):\penalty0 25--53, 2011.

\bibitem[Csardi and Nepusz(2006)]{igraph}
Gabor Csardi and Tamas Nepusz.
\newblock The igraph software package for complex network research.
\newblock \emph{InterJournal}, Complex Systems:\penalty0 1695, 2006.
\newblock URL \url{https://igraph.org}.

\bibitem[Egami and Tchetgen(2023)]{egami2021identification}
Naoki Egami and Eric J~Tchetgen Tchetgen.
\newblock Identification and estimation of causal peer effects using double negative controls for unmeasured network confounding.
\newblock \emph{Journal of the Royal Statistical Society Series B: Statistical Methodology}, page qkad132, 2023.

\bibitem[Fletcher et~al.(2020)Fletcher, Ross, and Zhang]{fletcher2020consequences}
Jason~M Fletcher, Stephen~L Ross, and Yuxiu Zhang.
\newblock The consequences of friendships: Evidence on the effect of social relationships in school on academic achievement.
\newblock \emph{Journal of Urban Economics}, 116:\penalty0 103241, 2020.

\bibitem[Forastiere et~al.(2018)Forastiere, Mealli, Wu, and Airoldi]{forastiere2018estimating}
Laura Forastiere, Fabrizia Mealli, Albert Wu, and Edoardo~M Airoldi.
\newblock Estimating causal effects on social networks.
\newblock In \emph{2018 IEEE 5th International Conference on Data Science and Advanced Analytics (DSAA)}, pages 60--69. IEEE, 2018.

\bibitem[Forastiere et~al.(2021)Forastiere, Airoldi, and Mealli]{forastiere2021identification}
Laura Forastiere, Edoardo~M Airoldi, and Fabrizia Mealli.
\newblock Identification and estimation of treatment and interference effects in observational studies on networks.
\newblock \emph{Journal of the American Statistical Association}, 116\penalty0 (534):\penalty0 901--918, 2021.

\bibitem[Hong and Raudenbush(2006)]{hong2006evaluating}
Guanglei Hong and Stephen~W Raudenbush.
\newblock Evaluating kindergarten retention policy: {A} case study of causal inference for multilevel observational data.
\newblock \emph{Journal of the American Statistical Association}, 101\penalty0 (475):\penalty0 901--910, 2006.

\bibitem[Hudgens and Halloran(2008)]{hudgens2008toward}
Michael~G Hudgens and M~Elizabeth Halloran.
\newblock Toward causal inference with interference.
\newblock \emph{Journal of the American Statistical Association}, 103\penalty0 (482):\penalty0 832--842, 2008.

\bibitem[Huisman(2009)]{huisman2009imputation}
Mark Huisman.
\newblock Imputation of missing network data: {S}ome simple procedures.
\newblock \emph{Journal of Social Structure}, 10\penalty0 (1):\penalty0 1--29, 2009.

\bibitem[Hunter et~al.(2008)Hunter, Handcock, Butts, Goodreau, and Morris]{hunter2008ergm}
David~R Hunter, Mark~S Handcock, Carter~T Butts, Steven~M Goodreau, and Martina Morris.
\newblock ergm: A package to fit, simulate and diagnose exponential-family models for networks.
\newblock \emph{Journal of Statistical Software}, 24\penalty0 (3):\penalty0 nihpa54860, 2008.

\bibitem[Kang and Schafer(2007)]{kang2007demystifying}
Joseph~DY Kang and Joseph~L Schafer.
\newblock Demystifying double robustness: A comparison of alternative strategies for estimating a population mean from incomplete data.
\newblock \emph{Statistical {S}cience}, 22\penalty0 (4):\penalty0 523--539, 2007.

\bibitem[Kao(2017)]{kao2017causal}
Edward~K Kao.
\newblock \emph{Causal inference under network interference: A framework for experiments on social networks}.
\newblock PhD thesis, Harvard University, 2017.

\bibitem[Kolaczyk(2009)]{kolaczyk2009}
Eric~D Kolaczyk.
\newblock \emph{Statistical Analysis of Network Data}.
\newblock Springer New York, NY, 2009.

\bibitem[Krause et~al.(2018)Krause, Huisman, Steglich, and Snijders]{krause2018missing}
Robert~W Krause, Mark Huisman, Christian Steglich, and Tom~AB Snijders.
\newblock Missing network data: {A} comparison of different imputation methods.
\newblock In \emph{2018 IEEE/ACM International Conference on Advances in Social Networks Analysis and Mining (ASONAM)}, pages 159--163. IEEE, 2018.

\bibitem[Lee et~al.(2023)Lee, Buchanan, Katenka, Forastiere, Halloran, Friedman, and Nikolopoulos]{lee2021estimating}
TingFang Lee, Ashley~L Buchanan, Natallia~V Katenka, Laura Forastiere, M~Elizabeth Halloran, Samuel~R Friedman, and Georgios Nikolopoulos.
\newblock Estimating causal effects of non-randomized {HIV} prevention interventions with interference in network-based studies among people who inject drugs.
\newblock \emph{Annals of Applied Statistics}, 17\penalty0 (3):\penalty0 2165--2191, 2023.

\bibitem[Li and Wager(2022)]{li2020random}
Shuangning Li and Stefan Wager.
\newblock Random graph asymptotics for treatment effect estimation under network interference.
\newblock \emph{The Annals of Statistics}, 50\penalty0 (4):\penalty0 2334--2358, 2022.

\bibitem[Li et~al.(2021)Li, Sussman, and Kolaczyk]{li2021causal}
Wenrui Li, Daniel~L Sussman, and Eric~D Kolaczyk.
\newblock Causal inference under network interference with noise.
\newblock \emph{arXiv preprint arXiv:2105.04518}, 2021.

\bibitem[Liu and Hudgens(2014)]{liu2014large}
Lan Liu and Michael~G Hudgens.
\newblock Large sample randomization inference of causal effects in the presence of interference.
\newblock \emph{Journal of the American Statistical Association}, 109\penalty0 (505):\penalty0 288--301, 2014.

\bibitem[Liu et~al.(2016)Liu, Hudgens, and Becker-Dreps]{liu2016inverse}
Lan Liu, Michael~G Hudgens, and Sylvia Becker-Dreps.
\newblock On inverse probability-weighted estimators in the presence of interference.
\newblock \emph{Biometrika}, 103\penalty0 (4):\penalty0 829--842, 2016.

\bibitem[Liu et~al.(2019)Liu, Hudgens, Saul, Clemens, Ali, and Emch]{liu2019doubly}
Lan Liu, Michael~G Hudgens, Bradley Saul, John~D Clemens, Mohammad Ali, and Michael~E Emch.
\newblock Doubly robust estimation in observational studies with partial interference.
\newblock \emph{Stat}, 8\penalty0 (1):\penalty0 e214, 2019.

\bibitem[Manski(1993)]{manski1993identification}
Charles~F Manski.
\newblock Identification of endogenous social effects: The reflection problem.
\newblock \emph{The Review of Economic Studies}, 60\penalty0 (3):\penalty0 531--542, 1993.

\bibitem[Ogburn and VanderWeele(2017)]{ogburn2017vaccines}
Elizabeth~L Ogburn and Tyler~J VanderWeele.
\newblock Vaccines, contagion, and social networks.
\newblock \emph{The Annals of Applied Statistics}, 11\penalty0 (2):\penalty0 919--948, 2017.

\bibitem[Ogburn et~al.(2022)Ogburn, Sofrygin, Diaz, and Van~der Laan]{ogburn2022causal}
Elizabeth~L Ogburn, Oleg Sofrygin, Ivan Diaz, and Mark~J Van~der Laan.
\newblock Causal inference for social network data.
\newblock \emph{Journal of the American Statistical Association}, pages 1--15, 2022.

\bibitem[Park and Kang(2022)]{park2022efficient}
Chan Park and Hyunseung Kang.
\newblock Efficient semiparametric estimation of network treatment effects under partial interference.
\newblock \emph{Biometrika}, 109\penalty0 (4):\penalty0 1015--1031, 2022.

\bibitem[Perez-Heydrich et~al.(2014)Perez-Heydrich, Hudgens, Halloran, Clemens, Ali, and Emch]{perez2014assessing}
Carolina Perez-Heydrich, Michael~G Hudgens, M~Elizabeth Halloran, John~D Clemens, Mohammad Ali, and Michael~E Emch.
\newblock Assessing effects of cholera vaccination in the presence of interference.
\newblock \emph{Biometrics}, 70\penalty0 (3):\penalty0 731--741, 2014.

\bibitem[Rabe-Hesketh and Skrondal(2006)]{rabe2006multilevel}
Sophia Rabe-Hesketh and Anders Skrondal.
\newblock Multilevel modelling of complex survey data.
\newblock \emph{Journal of the {R}oyal {S}tatistical {S}ociety: {S}eries {A} ({S}tatistics in {S}ociety)}, 169\penalty0 (4):\penalty0 805--827, 2006.

\bibitem[Rubin(1974)]{rubin1974estimating}
Donald~B Rubin.
\newblock Estimating causal effects of treatments in randomized and nonrandomized studies.
\newblock \emph{Journal of Educational Psychology}, 66\penalty0 (5):\penalty0 688--701, 1974.

\bibitem[Saul and Hudgens(2020)]{saul2020calculus}
Bradley~C Saul and Michael~G Hudgens.
\newblock The calculus of {M}-estimation in {R} with geex.
\newblock \emph{Journal of Statistical Software}, 92\penalty0 (2), 2020.

\bibitem[Scharfstein et~al.(1999)Scharfstein, Rotnitzky, and Robins]{scharfstein1999adjusting}
Daniel~O Scharfstein, Andrea Rotnitzky, and James~M Robins.
\newblock Adjusting for nonignorable drop-out using semiparametric nonresponse models.
\newblock \emph{Journal of the American Statistical Association}, 94\penalty0 (448):\penalty0 1096--1120, 1999.

\bibitem[Shardell and Ferrucci(2018)]{shardell2018joint}
Michelle Shardell and Luigi Ferrucci.
\newblock Joint mixed-effects models for causal inference with longitudinal data.
\newblock \emph{Statistics in {M}edicine}, 37\penalty0 (5):\penalty0 829--846, 2018.

\bibitem[Snijders(2005)]{snijders2005models}
Tom~AB Snijders.
\newblock Models for longitudinal network data.
\newblock \emph{Models and Methods in Social Network Analysis}, 1:\penalty0 215--247, 2005.

\bibitem[Sobel(2006)]{sobel2006randomized}
Michael~E Sobel.
\newblock What do randomized studies of housing mobility demonstrate? {C}ausal inference in the face of interference.
\newblock \emph{Journal of the American Statistical Association}, 101\penalty0 (476):\penalty0 1398--1407, 2006.

\bibitem[Stefanski and Boos(2002)]{stefanski2002calculus}
Leonard~A Stefanski and Dennis~D Boos.
\newblock The calculus of {M}-estimation.
\newblock \emph{The American Statistician}, 56\penalty0 (1):\penalty0 29--38, 2002.

\bibitem[Stewart et~al.(2019)Stewart, Schweinberger, Bojanowski, and Morris]{stewart2019multilevel}
Jonathan Stewart, Michael Schweinberger, Michal Bojanowski, and Martina Morris.
\newblock Multilevel network data facilitate statistical inference for curved ergms with geometrically weighted terms.
\newblock \emph{Social Networks}, 59:\penalty0 98--119, 2019.

\bibitem[Tchetgen and VanderWeele(2012)]{tchetgen2012causal}
Eric J~Tchetgen Tchetgen and Tyler~J VanderWeele.
\newblock On causal inference in the presence of interference.
\newblock \emph{Statistical Methods in Medical Research}, 21\penalty0 (1):\penalty0 55--75, 2012.

\bibitem[Van~der Vaart(2000)]{van2000asymptotic}
Aad~W Van~der Vaart.
\newblock \emph{Asymptotic Statistics}, volume~3.
\newblock Cambridge university press, 2000.

\bibitem[van Rijsewijk et~al.(2018)van Rijsewijk, Oldenburg, Snijders, Dijkstra, and Veenstra]{van2018description}
Louise Gerharda~Maria van Rijsewijk, Beau Oldenburg, Tom Augustinus~Benedictus Snijders, Jan~Kornelis Dijkstra, and Ren{\'e} Veenstra.
\newblock A description of classroom help networks, individual network position, and their associations with academic achievement.
\newblock \emph{PLoS One}, 13\penalty0 (12):\penalty0 e0208173, 2018.

\bibitem[VanderWeele and An(2013)]{vanderweele2013social}
Tyler~J VanderWeele and Weihua An.
\newblock Social networks and causal inference.
\newblock \emph{Handbook of Causal Analysis for Social Research}, pages 353--374, 2013.

\bibitem[Wakefield(2013)]{wakefield2013bayesian}
Jon Wakefield.
\newblock \emph{Bayesian and {F}requentist {R}egression {M}ethods}, volume~23.
\newblock Springer, 2013.

\bibitem[Zhang et~al.(2023)Zhang, Hudgens, and Halloran]{zhang2023propensity}
Bo~Zhang, Michael~G Hudgens, and M~Elizabeth Halloran.
\newblock Propensity score in the face of interference: Discussion of {R}osenbaum and {R}ubin (1983).
\newblock \emph{Observational Studies}, 9\penalty0 (1):\penalty0 125--131, 2023.

\end{thebibliography}

\newpage
\setcounter{table}{0}
\setcounter{figure}{0}
\setcounter{equation}{0}
\renewcommand{\thefigure}{S\arabic{figure}}
\renewcommand{\thetable}{S\arabic{table}}
\renewcommand{\theequation}{S\arabic{equation}}
\begin{center}
{\Large\bf Online Appendix}
\end{center}
\begin{appendices}
\section{Regression Estimation with Inverse-Propensity Weighted Coefficients}
\label{section:AppendixA}
In this section, we extend a second estimation method considered by \cite{liu2019doubly} to the network interference setting. With the propensities estimated from the treatment model, we can compute a weighted least-squares estimate of $\bm{\beta}$ by regressing $Y_i$ on the covariate vector $\bm{L}_i = (1, \Sigma \bm{Z}_{\mathcal{N}_i},\bm{X}_i)^{\top}$ of dimension $p+2$ for each treatment group $z \in \{0, 1\}$ with individual weights $\omega_i^{z,\alpha} = \mathds{1}(Z_i = z) \pi(\Sigma \bm{Z}_{\mathcal{N}_i};\alpha)/[\binom{d_i}{\Sigma \bm{Z}_{\mathcal{N}_i}}f(z, \bm{Z}_{\mathcal{N}_i}| \bm{X}_i, \bm{X}_{\mathcal{N}_i}; \hat{\bm{\gamma}},\hat{\phi}_{b})]$, $i=1,\ldots, N,$ where $d_i$ is the degree of the $i$-th individual, $\pi(\Sigma \bm{z}_{\mathcal{N}_i}; \alpha) = \binom{d_i}{\Sigma \bm{z}_{\mathcal{N}_i}}\alpha^{\bm{z}_{\mathcal{N}_i}}(1-\alpha)^{d_i -\Sigma \bm{z}_{\mathcal{N}_i}}$ and $f(z, \bm{Z}_{\mathcal{N}_i}| \bm{X}_i, \bm{X}_{\mathcal{N}_i}; \hat{\bm{\gamma}},\hat{\phi}_{b})$ is the model-based estimator of the joint probability  $P(Z_i=z, \bm{Z}_{\mathcal{N}_i} = \bm{z}_{\mathcal{N}_i}| \bm{X}_i, \bm{X}_{\mathcal{N}_i})$ described in the main text. We propose the following inverse-propensity weighted (IP-WLS) estimators for the average potential outcomes $\mu_{z, \alpha}$ and $\mu_{\alpha}$ under network interference:
\begin{align} \hat{Y}^{\text{IP-WLS}}(z, \alpha) =\frac{1}{m} \sum_{\nu=1}^m \frac{1}{N_{\nu}} \sum_{i\in C_{\nu} } \sum_{ \Sigma{\bm{z}_{\mathcal{N}_i}} = 0}^{d_i} m_i(z, \Sigma \bm{z}_{\mathcal{N}_i}, \bm{X}_i; \hat{\bm{\beta}}_{z, \alpha}^{\text{WLS}}) \pi(\Sigma{\bm{z}_{\mathcal{N}_i}};\alpha), \label{eq:drwls1}\\
\hat{Y}^{\text{IP-WLS}}(\alpha) = \frac{1}{m} \sum_{\nu=1}^m \frac{1}{N_{\nu}} \sum_{i\in C_{\nu} } \sum_{\substack{z_i \in \{0, 1\}\\ \Sigma \bm{z}_{\mathcal{N}_i} \in \{0, 1, \ldots, d_i\} }} m_i(z_i, \Sigma \bm{z}_{\mathcal{N}_i}, \bm{X}_i; \hat{\bm{\beta}}_{z, \alpha}^{\text{WLS}}) \pi(z_i, \Sigma{\bm{z}_{\mathcal{N}_i}};\alpha)\label{eq:drwls2},
\end{align}
where $\hat{\bm{\beta}}_{z, \alpha}^{\text{WLS}}$ corresponds to the vector of estimated coefficients in the weighted outcome regression model conditional on $Z_i = z$ and treatment saturation $\alpha$ with weights $\omega_i^{z,\alpha}$. 

Similar to \cite{liu2019doubly}, to see why this estimator has the DR property, first notice that $\hat{Y}^{\text{IP-WLS}}(z, \alpha) = m^{-1}\sum_{\nu=1}^m \hat{Y}^{\text{IP-WLS}}_{\nu}(z, \alpha),$
where 
$$\hat{Y}^{\text{IP-WLS}}_{\nu}(z, \alpha) = \frac{1}{N_{\nu}} \sum_{i\in C_{\nu}} \sum_{ \Sigma{\bm{z}_{\mathcal{N}_i}} = 0}^{d_i} m_i(z, \Sigma \bm{z}_{\mathcal{N}_i}, \bm{X}_i; \hat{\bm{\beta}}_{z, \alpha}^{\text{WLS}}) \pi(\Sigma{\bm{z}_{\mathcal{N}_i}};\alpha). $$
By definition, $\hat{\bm{\beta}}_{z, \alpha}^{\text{WLS}}$ is the solution of 
\begin{align}
\sum_{\nu = 1}^m \bm{\psi}^{\text{WLS}}_{z,\alpha}(\bm{O}_{\nu}; \bm{\beta}_{z,\alpha}, \hat{\bm{\gamma}}, \hat{\phi}_b) = 0,
\label{eqn:wls0}
\end{align}
where $\bm{\psi}^{\text{WLS}}_{z,\alpha}(\bm{O}_{\nu}; \bm{\beta}_{z,\alpha}, \bm{\gamma}, \phi_b)$ is the following set of likelihood equations
$$\bm{\psi}^{\text{WLS}}_{z,\alpha}(\bm{O}_{\nu}; \bm{\beta}_{z,\alpha}, \bm{\gamma}, \phi_b) = \sum_{i \in C_{\nu}} L_{ij} \omega_{i}^{z,\alpha}\left\{Y_i - \bm{L}_i^{\top} \bm{\beta}_{z,\alpha}\right\}, \quad j = 1, \ldots, p+2,$$
where $L_{ij}$ denotes the $j$-th element of $\bm{L}_i$. As argued by \cite{liu2019doubly}, we deduce from (\ref{eqn:wls0}) that the solution $\hat{\bm{\beta}}_{z, \alpha}^{\text{WLS}}$ satisfies the following equality:
\begin{align*}
   &\sum_{\nu=1}^m  \sum_{i \in C_{\nu}} \omega_{i}^{z, \alpha}\left\{Y_i - \bm{L}_i^{\top} \hat{\bm{\beta}}_{z, \alpha}^{\text{WLS}}\right\} \\
   &= \sum_{\nu=1}^m  \sum_{i \in C_{\nu}} \frac{\mathds{1}(Z_i = z)\pi(\Sigma \bm{Z}_{\mathcal{N}_i};\alpha)}{\binom{d_i}{\Sigma \bm{Z}_{\mathcal{N}_i}}f(z, \bm{Z}_{\mathcal{N}_i}| \bm{X}_i, \bm{X}_{\mathcal{N}_i}; \hat{\bm{\gamma}}, \hat{\phi}_b)}\left\{Y_i - m_i(z, \Sigma \bm{Z}_{\mathcal{N}_i}, \bm{X}_i, \hat{\bm{\beta}}_{z, \alpha}^{\text{WLS}})\right\}=0, 
\end{align*}
where $m_i(z, \Sigma \bm{Z}_{\mathcal{N}_i}, \bm{X}_i, \hat{\bm{\beta}}_{z, \alpha}^{\text{WLS}}) := \bm{L}_i^{\top} \hat{\bm{\beta}}_{z, \alpha}^{\text{WLS}}$. If components are balanced, i.e. $N_{\nu} = N/m$, $\forall \nu =1,\ldots,m$, it implies that we can write $\hat{Y}^{\text{IP-WLS}}(z, \alpha)$ as
\begin{multline}
\hat{Y}^{\text{IP-WLS}}(z, \alpha) = \frac{1}{N}\sum_{\nu=1}^m  \sum_{i\in C_{\nu}} \biggr\{\sum_{\Sigma \bm{z}_{\mathcal{N}_i} = 0}^{d_i}  m_i(z, \Sigma \bm{z}_{\mathcal{N}_i}, \bm{X}_i; \hat{\bm{\beta}}_{z, \alpha}^{\text{WLS}}) \pi(\Sigma \bm{z}_{\mathcal{N}_i};\alpha) +\\ \frac{\mathds{1}(Z_i = z)}{\binom{d_i}{\Sigma \bm{Z}_{\mathcal{N}_i}}f(Z_i, \bm{Z}_{\mathcal{N}_i} \ \rvert \ \bm{X}_i, \bm{X}_{\mathcal{N}_i}; \hat{\bm{\gamma}}, \hat{\phi}_b) } \left\{y_i(Z_i, \Sigma \bm{Z}_{\mathcal{N}_i}) - m_i(Z_i, \Sigma \bm{Z}_{\mathcal{N}_i}, \bm{X}_i; \hat{\bm{\beta}}_{z, \alpha}^{\text{WLS}})\right\}\pi(\Sigma \bm{Z}_{\mathcal{N}_i}; \alpha)\biggr\}\nonumber
\end{multline} 
 and that $\hat{Y}^{\text{IP-WLS}}(z, \alpha)$ has the same form as $\hat{Y}^{\text{DR-BC}}(z, \alpha)$. This equality carries over to estimators of the marginal average potential outcome, $\mu_{\alpha}$. Thus, it is clear that the regression estimators with inverse-propensity weighted coefficients in (\ref{eq:drwls1}) and (\ref{eq:drwls2}) also inherit the double robustness property when network components are balanced.

More generally, let $\psi_{z\alpha}^{\text{IP-WLS}}(\bm{O}_{\nu}; \mu_{z\alpha}, \bm{\beta}_{z\alpha}, \bm{\gamma}, \phi_b) = \hat{Y}^{\text{IP-WLS}}_{\nu}(z, \alpha) - \mu_{z\alpha}$ and\\ $\psi_{\beta_{0\alpha}}(\bm{O}_{\nu}; \bm{\beta}_{0\alpha}, \bm{\gamma}, \phi_b)$, $\psi_{\beta_{1\alpha}}(\bm{O}_{\nu}; \bm{\beta}_{1\alpha}, \bm{\gamma}, \phi_b)$, $\psi_{\gamma}(\bm{O}_{\nu}; \bm{\gamma}, \phi_b)$, and $\psi_{\phi_b}(\bm{O}_{\nu}; \bm{\gamma}, \phi_b)$ denote the estimating functions corresponding to $\hat{\bm{\beta}}_{0\alpha}$, $\hat{\bm{\beta}}_{1\alpha}$, $\hat{\bm{\gamma}}$, and $\hat{\phi}_b$ such that $$ \hat{\bm{\theta}}^{\text{IP-WLS}} = (\hat{Y}^{\text{IP-WLS}}(0,\alpha), \hat{Y}^{\text{IP-WLS}}(1,\alpha), \hat{\bm{\beta}}_{0\alpha}, \hat{\bm{\beta}}_{1\alpha}, \hat{\bm{\gamma}} , \hat{\phi}_b)^{\top}$$
 solves the estimating equation
$$\bm{\psi}_{\alpha}^{\text{IP-WLS}}(\bm{\theta}) = \sum_{\nu=1}^m \bm{\psi}_{\alpha}^{\text{IP-WLS}}(\bm{O}_{\nu}; \bm{\theta}) = \bm{0} $$
where $\bm{\theta}=(\mu_{0\alpha}, \mu_{1\alpha}, \bm{\beta}_{0\alpha},\bm{\beta}_{1\alpha},\bm{\gamma}, \phi_b)^{\top}$ and 
\begin{multline}
\bm{\psi}^{\text{IP-WLS}}_{\alpha}(\bm{O}; \bm{\theta})=( \psi_{0\alpha}^{\text{IP-WLS}}(\bm{O}; \mu_{0\alpha}, \bm{\beta}_{0\alpha}, \bm{\gamma}, \phi_b), \psi_{1\alpha}^{\text{IP-WLS}}(\bm{O}; \mu_{1\alpha}, \bm{\beta}_{1\alpha}, \bm{\gamma},\phi_b), \\\psi_{\beta_{0\alpha}}(\bm{O}; \bm{\beta}_{0\alpha},\bm{\gamma}, \phi_b),  \psi_{\beta_{1\alpha}}(\bm{O}; \bm{\beta}_{1\alpha},\bm{\gamma}, \phi_b), \psi_{\gamma}(\bm{O}; \bm{\gamma},\phi_b),\psi_{\phi_b}(\bm{O}; \bm{\gamma},\phi_b))^{\top}.
\end{multline}
The following proposition establishes the double robustness property and asymptotic normality of the regression estimator with inverse-propensity weighted coefficients whether or not components are balanced.
\begin{manualtheorem}{2}
\label{proposition:drwls}
If either $f(Z_i, \bm{Z}_{\mathcal{N}_i}| \bm{X}_i, \bm{X}_{\mathcal{N}_i}; \bm{\gamma}, \phi_b)$ or $m(Z_i, \Sigma \bm{Z}_{\mathcal{N}_i}, \bm{X}_i; \bm{\beta})$ is correctly specified, then $\hat{\bm{\theta}}^{\text{IP-WLS}} \xrightarrow{p} \bm{\theta}$ and $$\sqrt{m}(\hat{\bm{\theta}}^{\text{IP-WLS}} -  \bm{\theta}) \xrightarrow{d}  N(\bm{0}, \bm{U}(\bm{\theta})^{-1}\bm{V}(\bm{\theta})(\bm{U}(\bm{\theta})^{-1})^{\top}),$$
where $\bm{U}(\bm{\theta})=\mathbb{E}\left[-\partial \bm{\psi}^{\text{IP-WLS}}(\bm{O}_{\nu}; \bm{\theta})/ \partial \bm{\theta}^{\top} \right]$ and $\bm{V}(\bm{\theta}) = \mathbb{E}\left[\bm{\psi}^{\text{IP-WLS}}(\bm{O}_{\nu}; \bm{\theta}) \bm{\psi}^{\text{IP-WLS}}(\bm{O}_{\nu}; \bm{\theta})^{\top} \right].$
\end{manualtheorem}
The proof of Proposition \ref{proposition:drwls} can be found in Appendix \ref{A.C}. 
\clearpage
\section{Doubly Robust Estimation for Multilevel Outcomes}
\label{section:AppendixB}
 In the following section, we shift our focus to inferential methods that allow for a multilevel structure in the outcome data. We posit the following hierarchical linear model for the vector of outcomes in a component $C_{\nu}$ of size $N_{\nu}$, $\nu \in 1,\ldots, m$:
\begin{align}
\label{eqn:hierchical}
    \tilde{\bm{Y}}_{\nu} | \tilde{\bm{\Delta}}_{\nu}, \bm{\beta}, \sigma^2_{\varepsilon}, c_{\nu} \sim N(\tilde{\bm{\Delta}}_{\nu}\bm{\beta} + c_{\nu}, \sigma^2_{\varepsilon}),
    \end{align}
where $c_{\nu} \sim N(0, \sigma^2_c)$, $\bm{\Delta}_{\nu, i} $ is the observed covariate row-vector $(1, Z_{\nu, i}, h(\bm{Z}_{\mathcal{N}_{\nu,i}}), \bm{X}_{\nu,i})$ of unit $i$ in component $\nu$, $\tilde{\bm{\Delta}}_{\nu} = (\bm{\Delta}_{\nu, 1}, \ldots, \bm{\Delta}_{\nu, N_{\nu}})^{\top}$ is the corresponding component-level fixed effects matrix and $\bm{\beta} = (\beta_0, \beta_{Z}, \beta_{\bm{Z}_{\mathcal{N}}}, \beta_1, \ldots, \beta_p)$ as previously defined. If this assumption holds, using a conventional fixed effects approach would lead us to incorrectly model the marginal variance of the outcome. In the following, we review likelihood inference in the linear intercept model as a first step in formulating doubly robust estimators for multilevel network data.

\subsection{Likelihood Inference in the Random Intercept Linear Model}
\label{subsection:likelihood}
Inference for fixed effects and variance components is performed by exploiting the marginal likelihood \citep{wakefield2013bayesian}. By letting $\bm{Y} = (\tilde{\bm{Y}}_{1}, \ldots, \tilde{\bm{Y}}_{m})$ and $\bm{\Delta} = (\tilde{\bm{\Delta}}_{1}, \ldots, \tilde{\bm{\Delta}}_{m})$, it is easy to show that the marginal likelihood for parameter vector $(\bm{\beta}, \sigma^2_{\varepsilon}, \sigma^2_{c})$ under the random intercept linear model is given by
\begin{align}
\label{eqn:likelihoodrand}
\mathcal{L}(\bm{\beta}, \sigma^2_{\varepsilon}, \sigma^2_{c} |\bm{Y}, \bm{\Delta}) = \prod_{\nu=1}^m \mathrm{det}(2\pi \bm{V}_{\nu}(\sigma^2_{\varepsilon}, \sigma^2_{c}))^{-1/2} \exp\left[-\frac{1}{2}(\tilde{\bm{Y}}_{\nu} - \tilde{\bm{\Delta}}_{\nu}\bm{\beta})^{\top} \bm{V}_{\nu}(\sigma^2_{\varepsilon}, \sigma^2_{c})^{-1} (\tilde{\bm{Y}}_{\nu} - \tilde{\bm{\Delta}}_{\nu}\bm{\beta}) \right],  
\end{align}
where $\bm{V}_{\nu}(\sigma^2_{\varepsilon}, \sigma^2_{c})=  \sigma^2_{\varepsilon} I_{N_{\nu}} + \sigma^2_c \mathds{1}_{N_{\nu}} \mathds{1}^{\top}_{N_{\nu}}$ \citep{wakefield2013bayesian}. The maximum likelihood estimators for $\bm{\beta}$, $\sigma^2_{\varepsilon}$ and $\sigma^2_{c}$ are obtained by maximization of $\ell = \log\mathcal{L}$. One can easily verify that the score function for $\bm{\beta}$ has the closed-form expression 
\begin{align}
\label{eqn:scorebeta}
\frac{\partial \ell }{\partial \bm{\beta}}(\bm{\beta}, \sigma^2_{\varepsilon}, \sigma^2_c) = \sum_{\nu=1}^m \tilde{\bm{\Delta}}_{\nu}^{\top} \bm{V}_{\nu}^{-1}(\sigma^2_{\varepsilon}, \sigma^2_{c})(\tilde{\bm{Y}}_{\nu} - \tilde{\bm{\Delta}}_{\nu}\bm{\beta}),
\end{align}
yielding the MLE
$$\hat{\bm{\beta}} = \left(\sum_{\nu=1}^m  \tilde{\bm{\Delta}}_{\nu}^{\top} \bm{V}_{\nu}^{-1}(\hat{\sigma}^2_{\varepsilon}, \hat{\sigma}^2_{c})\tilde{\bm{\Delta}}_{\nu} \right)^{-1}\left(\sum_{\nu=1}^m  \tilde{\bm{\Delta}}_{\nu}^{\top} \bm{V}_{\nu}^{-1}(\hat{\sigma}^2_{\varepsilon}, \hat{\sigma}^2_{c})\tilde{\bm{Y}}_{\nu} \right), $$
where $\hat{\sigma}^2_{\varepsilon}$ and $\hat{\sigma}^2_c$ are the MLEs of $\sigma^2_{\varepsilon}$ and $\sigma^2_c$, respectively \citep{wakefield2013bayesian}. There is no closed-form expression for the score functions of $\sigma^2_{\varepsilon}$ and $\sigma^2_c$. Thus, numerical optimization is required to perform maximization of (\ref{eqn:likelihoodrand}) with respect to $\bm{\beta}$, ${\sigma}^2_{\varepsilon}$ and ${\sigma}^2_c$. This can be accomplished through the package \textbf{geex}, which finds point and variance estimates from any set of unbiased estimating equations \citep{saul2020calculus}. For the lack of analytic expressions for the score functions for ${\sigma}^2_{\varepsilon}$ and ${\sigma}^2_c$, the function \texttt{grad} from the package \textbf{numDeriv} can be used to program the corresponding unbiased estimating equations from the expression of the marginal likelihood in (\ref{eqn:likelihoodrand}).

Maximum likelihood estimation is known to yield biased estimators of the variance components in small to moderate samples, as it proceeds as if $\bm{\beta}$ were known \citep{wakefield2013bayesian}. An alternative approach, restricted MLE, allows for unbiased estimation of the variance components by taking into account the degrees of freedom used to estimate $\bm{\beta}$ \citep{wakefield2013bayesian}. However, since our primary interest here is to conduct inference about the marginal effects and not the variance components, we recommend the use of MLE. Indeed, restricted MLE introduces bias in the fixed effects estimates in small to moderate samples, which is not the case for MLE. 

\subsection{Regression Estimation of Causal Effects under the Random Intercept Linear Model}
\label{subsection:regestimator}
Causal effects presented in the main text can be identified from the hierarchical model in (\ref{eqn:hierchical}). To see this, consider the following argument. Under consistency (Assumption 1) and conditional exchangeability (Assumption 3), the expected potential outcome under treatment $z=1$ and treatment saturation $\alpha$ conditional on individual covariates $\bm{X}_{\nu, i}$ and component-specific intercept $c_{\nu}$ is given by
\begin{align*}
\mathbb{E}_{Y | \bm{X}, c}\left[ \bar{y}_{\nu, i} (1, \alpha) \rvert \bm{X}_{\nu, i}, c_{\nu}\right] &= \mathbb{E}_{Y | \bm{X}, c}\left[ \sum_{\Sigma \bm{z}_{\mathcal{N}_{\nu, i}} = 0}^{d_{\nu,i}} y_{\nu, i}(1, z_{\mathcal{N}_{\nu, i}}) \pi(z_{\mathcal{N}_{\nu, i}}, \alpha) \biggr \rvert \bm{X}_{\nu, i}, c_{\nu} \right]\\
&= \sum_{\Sigma \bm{z}_{\mathcal{N}_{\nu, i}} = 0}^{d_{\nu,i}} \mathbb{E}_{Y |\bm{X}, c}\left[Y_{\nu, i} \rvert Z_{\nu, i} = 1, \bm{Z}_{\mathcal{N}_{\nu,i}} =  z_{\mathcal{N}_{\nu, i}},\bm{X}_{\nu, i}, c_{\nu} \right]\pi(z_{\mathcal{N}_{\nu, i}}, \alpha)\\
&= \beta_0 + \beta_{Z} + \beta_{\bm{Z}_{\mathcal{N}}} \sum_{\Sigma \bm{z}_{\mathcal{N}_{\nu, i}} = 0}^{d_{\nu,i}} h(\bm{z}_{\mathcal{N}_{\nu, i}})\pi(\bm{z}_{\mathcal{N}_{\nu, i}}, \alpha) + \sum_{k=1}^p \beta_k X_{\nu, i, k} + c_{\nu}.
\end{align*}
Integration over $\bm{X}_{\nu, i}$ and $c_{\nu}$ results in the average potential outcome introduced in the main text, which corresponds to a population-average or marginal estimand in the context of multilevel data \citep{shardell2018joint}:
\begin{align*}
\mu_{1, \alpha} = \beta_0 + \beta_{Z} + \beta_{\bm{Z}_{\mathcal{N}}} \sum_{\Sigma \bm{z}_{\mathcal{N}_{\nu, i}} = 0}^{d_{\nu,i}} h(\bm{z}_{\mathcal{N}_{\nu, i}})\pi(\bm{z}_{\mathcal{N}_{\nu, i}}, \alpha) + \sum_{k=1}^p \beta_k \mathbb{E}[X_{\nu, i, k}].
\end{align*}
Similarly, we find that $\mu_{0, \alpha} = \mu_{1,\alpha} - \beta_Z$. Therefore, in this simple example in which there is no interaction between individual treatment and neighbourhood treatment, the direct effect of treatment under treatment saturation $\alpha$ is given by the constant
$$DE(\alpha) = \mu_{1,\alpha} - \mu_{0, \alpha} = \beta_{Z}, $$
which is consistently estimated by the fixed effect estimator $\hat{\beta}_Z$ obtained by maximizing the marginal likelihood in (\ref{eqn:likelihoodrand}) if the working hierarchical model is correctly specified. Note that since we are considering linear outcomes, marginal and component-specific causal estimands coincide.

We now proceed to formulate an extension of the regression estimator presented in the main text for multilevel outcome data. For notational simplicity, from this point on we will drop the component index $\nu$ for quantities defined on the elementary units, such that $i$ always denotes the index of a unit in component $\nu$. Define $m_i(z, \Sigma \bm{z}_{\mathcal{N}_i}, \bm{X}_i; \bm{\beta})=\beta_0 + \beta_{Z} + \beta_{\bm{Z}_{\mathcal{N}}} h(\bm{z}_{\mathcal{N}_{i}}) + \sum_{k=1}^p \beta_k X_{i, k}$. The regression estimator of $\mu_{z, \alpha}$ is given by
$\hat{Y}^{\text{REG}}(z,\alpha) = m^{-1} \sum_{\nu=1}^m \hat{Y}_{\nu}^{\text{REG}}(z, \alpha)$, where
\begin{align*}
\hat{Y}_{\nu}^{\text{REG}}(z, \alpha) = N_{\nu}^{-1} \sum_{i\in C_{\nu}} \sum_{\Sigma \bm{z}_{\mathcal{N}_i} = 0}^{d_i}  m_i(z, \Sigma \bm{z}_{\mathcal{N}_i}, \bm{X}_i; \hat{\bm{\beta}}) \pi(\Sigma \bm{z}_{\mathcal{N}_i};\alpha).
\end{align*}
Note that this expression of the regression estimator is identical to that found in the main text, with the exception that the marginal predictions $m_i(z, \Sigma \bm{z}_{\mathcal{N}_i}, \bm{X}_i; \hat{\bm{\beta}})$ are based on the hierarchical model in (\ref{eqn:hierchical}). We can leverage the M-estimation framework to obtain the point estimate along with an asymptotic variance estimate for $\hat{Y}_{z\alpha}^{\text{REG}}$. Let $\psi_{z\alpha}^{\text{REG}}(\bm{O}; \mu_{z\alpha}, \bm{\beta}, \sigma^2_{\varepsilon}, \sigma^2_c) = \hat{Y}^{\text{REG}}_{\nu}(z,\alpha) - \mu_{z\alpha}$ and $\psi_{\beta}(\bm{O}_{\nu}; \bm{\beta}, \sigma^2_{\varepsilon}, \sigma^2_c)$, $\psi_{\sigma^2_{\varepsilon}}(\bm{O}_{\nu}; \bm{\beta}, \sigma^2_{\varepsilon}, \sigma^2_c)$, and $\psi_{\sigma^2_{c}}(\bm{O}_{\nu}; \bm{\beta}, \sigma^2_{\varepsilon}, \sigma^2_c)$ denote the score equations for $\hat{\bm{\beta}}$ (given by (\ref{eqn:scorebeta})), $\hat{\sigma}^2_{\varepsilon}$, and $\hat{\sigma}^2_{c}$ such that $$\hat{\bm{\theta}}^{\text{REG}} = (\hat{Y}^{\text{REG}}(0, \alpha), \hat{Y}^{\text{REG}}(1, \alpha), \hat{\bm{\beta}},\hat{\sigma}^2_{\varepsilon}, \hat{\sigma}^2_{c})^{\top}$$
 solves the estimating equation
$$\bm{\psi}^{\text{REG}}(\bm{\theta}) = \sum_{\nu=1}^m \bm{\psi}^{\text{REG}}(\bm{O}_{\nu}; \bm{\theta}) = \bm{0},$$
where $\bm{\theta}=(\mu_{0\alpha}, \mu_{1\alpha}, \bm{\beta},\sigma^2_{\varepsilon},\sigma^2_c)^{\top}$ and \begin{align*}
\bm{\psi}^{\text{REG}}(\bm{O}; \bm{\theta})=\begin{pmatrix}\psi_{0\alpha}^{\text{REG}}(\bm{O}; \mu_{0\alpha}, \bm{\beta}, \sigma^2_{\varepsilon}, \sigma^2_c)\\
\psi_{1\alpha}^{\text{REG}}(\bm{O}; \mu_{1\alpha}, \bm{\beta}, \sigma^2_{\varepsilon}, \sigma^2_c)\\
\psi_{\beta}(\bm{O}; \bm{\beta}, \sigma^2_{\varepsilon}, \sigma^2_c)\\ \psi_{\sigma^2_{\varepsilon}}(\bm{O}; \bm{\beta}, \sigma^2_{\varepsilon}, \sigma^2_c)\\
\psi_{\sigma^2_{c}}(\bm{O}; \bm{\beta}, \sigma^2_{\varepsilon}, \sigma^2_c)\\\end{pmatrix}.
\end{align*} As mentioned in Appendix \ref{subsection:likelihood}, this M-estimation routine can be implemented in R with the package \textbf{geex}, utilizing the analytic expressions for $\psi_{z\alpha}^{\text{REG}}(\bm{O}_{\nu}; \mu_{z\alpha}, \bm{\beta}), \ z=0,1,$ and $\psi_{\beta}(\bm{O}; \bm{\beta}, \sigma^2_{\varepsilon}, \sigma^2_c)$ and defining $\psi_{\sigma^2_{\varepsilon}}(\bm{O}; \bm{\beta}, \sigma^2_{\varepsilon}, \sigma^2_c)$ and $\psi_{\sigma^2_{c}}(\bm{O}; \bm{\beta}, \sigma^2_{\varepsilon}, \sigma^2_c)$ numerically. Provided a correctly specified outcome model, consistency and asymptotic normality of $\hat{Y}^{\text{REG}}(z, \alpha)$ follows from standard estimating equation theory \citep{stefanski2002calculus}.
\subsection{Regression Estimation with Residual Bias Correction}
Building on the random-intercept treatment model developed in the main text and the regression estimator presented in Appendix \ref{subsection:regestimator}, it is straightforward to formulate an extension of the regression estimator with residual bias correction (DR-BC) for multilevel outcomes. The DR-BC estimator for multilevel network data can be written just as before, that is $\hat{Y}^{\text{DR-BC}}(z,\alpha) = m^{-1} \sum_{\nu=1}^m \hat{Y}_{\nu}^{\text{DR-BC}}(z, \alpha)$, where
\begin{multline}
\hat{Y}_{\nu}^{\text{DR-BC}}(z, \alpha) = N_{\nu}^{-1}  \sum_{i\in C_{\nu}} \biggr\{ \sum_{\Sigma \bm{z}_{\mathcal{N}_i} = 0}^{d_i}  m_i(z, \Sigma \bm{z}_{\mathcal{N}_i}, \bm{X}_i; \hat{\bm{\beta}}) \pi(\Sigma \bm{z}_{\mathcal{N}_i};\alpha) +\\ \frac{\mathds{1}(Z_i = z)}{\binom{d_i}{\Sigma \bm{Z}_{\mathcal{N}_i}}f(Z_i, \bm{Z}_{\mathcal{N}_i} \ \rvert \ \bm{X}_i, \bm{X}_{\mathcal{N}_i}; \hat{\bm{\gamma}}) } \left\{y_i(Z_i, \Sigma \bm{Z}_{\mathcal{N}_i}) - m_i(Z_i, \Sigma \bm{Z}_{\mathcal{N}_i}, \bm{X}_i; \hat{\bm{\beta}})\right\}\pi(\Sigma \bm{Z}_{\mathcal{N}_i}; \alpha)\biggr\}. \nonumber
\end{multline}
The DR-BC estimator can also be expressed as the solution to a vector of estimating equations. Let \\$\psi_{z\alpha}^{\text{DR-BC}}(\bm{O}_{\nu}; \mu_{z\alpha}, \bm{\beta}, \sigma^2_{\varepsilon}, \sigma^2_c, \bm{\gamma}, \phi_b) = \hat{Y}^{\text{DR-BC}}_{\nu}(z,\alpha) - \mu_{z\alpha}$ and $\psi_{\beta}(\bm{O}_{\nu}; \bm{\beta}, \sigma^2_{\varepsilon}, \sigma^2_c)$, $\psi_{\sigma^2_{\varepsilon}}(\bm{O}_{\nu}; \bm{\beta}, \sigma^2_{\varepsilon}, \sigma^2_c)$, \\$\psi_{\sigma^2_c}(\bm{O}_{\nu}; \bm{\beta}, \sigma^2_{\varepsilon}, \sigma^2_c)$, $\psi_{\gamma}(\bm{O}_{\nu}; \bm{\gamma}, \phi_b)$, and $\psi_{\phi_b}(\bm{O}_{\nu}; \bm{\gamma}, \phi_b)$ denote the estimating functions corresponding to $\hat{\bm{\beta}}$, $\hat{\sigma}^2_{\varepsilon}$, $\hat{\sigma}^2_c$, $\hat{\bm{\gamma}}$, and $\hat{\phi}_b$ such that $ \hat{\bm{\theta}}^{\text{DR-BC}} = (\hat{Y}^{\text{DR-BC}}(0, \alpha), \hat{Y}^{\text{DR-BC}}(1, \alpha), \hat{\bm{\beta}}, \hat{\sigma}^2_{\varepsilon},\hat{\sigma}^2_c, \hat{\bm{\gamma}}, \hat{\phi}_b)^{\top}$
 solves the estimating equation
$$\bm{\psi}^{\text{DR-BC}}(\bm{\theta}) = \sum_{\nu=1}^m \bm{\psi}^{\text{DR-BC}}(\bm{O}_{\nu}; \bm{\theta}) = \bm{0}, $$
where $\bm{\theta}=(\mu_{0\alpha}, \mu_{1\alpha}, \bm{\beta}, \sigma^2_{\varepsilon}, \sigma^2_c,\bm{\gamma}, {\phi}_b)^{\top}$ and \begin{align*}
\bm{\psi}^{\text{DR-BC}}(\bm{O}; \bm{\theta})&=\begin{pmatrix} \psi_{0\alpha}^{\text{DR-BC}}(\bm{O}; \mu_{0\alpha}, \bm{\beta}, \sigma^2_{\varepsilon}, \sigma^2_c, \bm{\gamma}, \phi_b)\\
\psi_{1\alpha}^{\text{DR-BC}}(\bm{O}; \mu_{1\alpha}, \bm{\beta}, \sigma^2_{\varepsilon}, \sigma^2_c, \bm{\gamma}, \phi_b)\\
\psi_{\beta}(\bm{O}; \bm{\beta},\sigma^2_{\varepsilon}, \sigma^2_c)\\
\psi_{\sigma^2_{\varepsilon}}(\bm{O}; \bm{\beta},\sigma^2_{\varepsilon}, \sigma^2_c)\\
\psi_{\sigma^2_c}(\bm{O}; \bm{\beta},\sigma^2_{\varepsilon}, \sigma^2_c)\\
\psi_{\gamma}(\bm{O}_{\nu}; \bm{\gamma},\phi_b)\\
\psi_{\phi_b}(\bm{O}_{\nu}; \bm{\gamma}, \phi_b)
\end{pmatrix}.
\end{align*}
Except for the inclusion of estimating equations for the variance components $\sigma^2_{\varepsilon}$ and $\sigma^2_c$, this M-estimation setup has the same structure as that of the DR-BC estimator in the main text. It is straightforward show that the DR-BC estimator for multilevel outcomes also inherits the double robustness property. The proof follows along the same lines as the proof for Proposition 1.
\subsection{Regression Estimation with Inverse-Propensity Weighted Coefficients}
\label{section:appendixb.4}
The idea behind inverse-propensity weighted regression estimation (IP-WLS) method is to move the estimated coefficients away from $\hat{\beta}$ to repair the bias in the regression estimators of average potential outcomes \citep{kang2007demystifying}. In the main text, we have shown that the IP-WLS estimator has the DR property in the case of independent outcomes. In the following, we present the implementation of this estimator under the hierarchical linear outcome model in (\ref{eqn:hierchical}). We also argue that the extension of IP-WLS to multilevel outcomes may not be doubly robust, as empirical results demonstrate.

Weighted hierarchical linear models are often used when analyzing complex survey data where weights are inverse selection probabilities. For the purpose of causal inference, we could instead set the weights to be inverse probability-of-treatment weights in a hierarchical linear model. Multiple packages in \texttt{R} can be used to fit such models. The package \textbf{lme4} fits mixed models when there are no weights or when there are weights only for the first-level units \citep{bates2014fitting}. The package \textbf{WeMix} implements methods for mixed models using weights at multiple levels for up to three levels \citep{bailey2020package}. If faced with multilevel outcome data, one could easily compute the IP-WLS estimator described the main text leveraging marginal model predictions given by \textbf{lme4} or \textbf{WeMix}. However, when propensity scores are unknown (i.e., $\bm{\gamma}$ and $\sigma^2_b$ are unknown), valid confidence intervals can be only be obtained by accounting for the estimation of $\bm{\gamma}$ and $\sigma^2_b$. In the following, we describe the M-estimation setup for this problem and in doing so, we formulate analytical expressions for the score equations for $\bm{\beta}_{0\alpha}$ and $\bm{\beta}_{1\alpha}$.

We restrict ourselves to the case where only first-level units are weighted. In other words, we assume that all components contribute equally to the likelihood. The approach implemented in \textbf{WeMix} was first described by \cite{rabe2006multilevel} and consists of weighting each individual contribution to the conditional log-likelihood within a cluster, yielding a log-pseudolikelihood. Recall that to compute the IP-WLS estimator of the average potential outcome under treatment $z$ and treatment saturation $\alpha$, one must stratify on individuals with observed treatment $Z=z$ upon fitting the weighted regression model. Under the hierarchical linear outcome model in (\ref{eqn:hierchical}), the complete log-likelihood for $(\bm{\beta}_{z\alpha}, \sigma^2_{\varepsilon, z, \alpha} , \sigma^2_{c,z, \alpha})$ under treatment $z$ and treatment saturation $\alpha$ (indices $z$ and $\alpha$ are only placed on the likelihood for the sake of notational simplicity) is given by
\begin{align}
\label{pseudologlik}
&\ell_{z,\alpha}(\bm{\beta}, \sigma^2_{\varepsilon} , \sigma^2_{c}\rvert \bm{Y}, \bm{L}) \nonumber \\
&= \sum_{\nu=1}^m \log \int_{c_{\nu}} \prod_{i\in C_{\nu}} \left[\frac{1}{\sqrt{2\pi \sigma^2_{\varepsilon}}^{\omega_{i}}} \exp\left(-\frac{\omega_{i}}{2\sigma^2_{\varepsilon}} (Y_{i} - \bm{L}_{i}^{\top} \bm{\beta} - c_{\nu})^2 \right) \right] \frac{1}{\sqrt{2\pi \sigma^2_c}} \exp\left(-\frac{1}{2\sigma^2_c} c_{\nu}^2 \right)dc_{\nu},
\end{align}
where $\omega_{i}  \equiv \omega_i^{z,\alpha} = \mathds{1}(Z_{i} = z) \pi(\Sigma \bm{Z}_{\mathcal{N}_{ i}};\alpha)/\left[\binom{d_i}{\Sigma \bm{Z}_{\mathcal{N}_i}}f(z, \bm{Z}_{\mathcal{N}_{i}}| \bm{X}_{i}, \bm{X}_{\mathcal{N}_{ i}}; \bm{\gamma})\right]$ and \\$\bm{L}_{i} = (1, \Sigma \bm{Z}_{\mathcal{N}_{i}},\bm{X}_{ i})^{\top}$, $i\in C_{\nu}$, $\nu = 1,\ldots,m$. Thus, only observations satisfying $Z_i = z$ will contribute to the conditional likelihood for a given random effect $c_{\nu}$. 
Let $N_{\nu}^{z}$ denote the number of observations within component $C_{\nu}$ that receive treatment $z$. It may be useful to note that the score equations for $\bm{\beta}_{z\alpha}$ have a closed form with this model specification. To see this, first note that the pseudo likelihood may be written as
\begin{multline*}
\mathcal{L}_{z,\alpha}(\bm{\beta}, \sigma^2_{\varepsilon} , \sigma^2_{c}\rvert \bm{Y}, \bm{L})
= \Bigg(\prod_{\substack{\nu \in \{1,\ldots, m\}\\ i \in C_{\nu}  : Z_i = z}} \frac{\sigma_{\varepsilon}^{1-\omega_i}}{\sqrt{\omega_i}}  \Bigg) \times \\\Bigg(\prod_{\nu=1}^m \mathrm{det}(2\pi \bm{V}^z_{\nu}(\sigma^2_{\varepsilon}, \sigma^2_{c}, \bm{W}))^{-1/2} \exp\left[-\frac{1}{2}(\tilde{\bm{Y}}_{\nu}^z - \tilde{\bm{L}}_{\nu}^z\bm{\beta})^{\top} \bm{V}_{\nu}^z(\sigma^2_{\varepsilon}, \sigma^2_{c}, \bm{W})^{-1} (\tilde{\bm{Y}}^z_{\nu} - \tilde{\bm{L}}^z_{\nu}\bm{\beta}) \right]\Bigg)
\end{multline*}
where the $z$ superscript indicates that we subset on observations $i \in C_{\nu}$ satisfying $Z_i = z$, the covariance matrix of the observations with $Z_i=z$ within a component is given by $\bm{V}_{\nu}^z(\sigma^2_{\varepsilon,z,\alpha}, \sigma^2_{c,z,\alpha}, \bm{W}_{z,\alpha}) = \sigma^2_{\varepsilon,z,\alpha} \bm{W}_{z,\alpha}^{-1} + \sigma^2_{c,z,\alpha} \mathds{1}_{N_{\nu}^z} \mathds{1}^{\top}_{N_{\nu}^z}$, $\bm{W} \equiv \bm{W}_{z,\alpha}$ is a $N_{\nu}^z \times N_{\nu}^z$ diagonal matrix with $\omega_i^{z,\alpha}$ as the $i$-th entry and $\tilde{\bm{L}}^z_{\nu}=(\bm{L}_1^{\top}, \ldots, \bm{L}_{N_{\nu}^z}^{\top})^{\top}$ is the component-level fixed effects matrix for observations with $Z_i=z$. Thus, this reduces to the unweighted case where $\bm{V}_{\nu}(\sigma^2_{\varepsilon}, \sigma^2_c)$ is replaced by $\bm{V}_{\nu}^z(\sigma^2_{\varepsilon,z,\alpha}, \sigma^2_{c,z,\alpha}, \bm{W}_{z,\alpha})$, so that the score function for $\bm{\beta}_{z\alpha}$ is given by
\begin{align}
\label{eqn:scorebetawls}
\frac{\partial \ell }{\partial \bm{\beta}_{z\alpha}}(\bm{\beta}_{z\alpha}, \sigma^2_{\varepsilon,z,\alpha}, \sigma^2_{c,z,\alpha}) = \sum_{\nu=1}^m \tilde{\bm{L}}_{\nu}^{z\top} \bm{V}_{\nu}^{z}(\sigma^2_{\varepsilon,z,\alpha}, \sigma^2_{c,z,\alpha}, \bm{W}_{z,\alpha})^{-1}(\tilde{\bm{Y}}_{\nu}^z - \tilde{\bm{L}}_{\nu}^z\bm{\beta}_{z\alpha}).
\end{align}
As in the previous case, there is no closed-form expression for the score functions for $\sigma^2_{\varepsilon,z,\alpha}$ and $\sigma^2_{c,z,\alpha}$ and one must resort to numerical optimization routines to compute the corresponding MLEs. 

The inverse-propensity weighted estimator of $\mu_{z, \alpha}$ for multilevel network data can be written just as in Appendix \ref{section:AppendixA}, that is $\hat{Y}^{\text{IP-WLS}}(z, \alpha) = m^{-1} \sum_{\nu=1}^m \hat{Y}_{\nu}^{\text{IP-WLS}}(z,\alpha)$, where
\begin{align} \hat{Y}_{\nu}^{\text{IP-WLS}}(z,\alpha) = \frac{1}{N_{\nu}} \sum_{i\in C_{\nu} } \sum_{ \Sigma{\bm{z}_{\mathcal{N}_i}} = 0}^{d_i} m_i(z, \Sigma \bm{z}_{\mathcal{N}_i}, \bm{X}_i; \hat{\bm{\beta}}_{z, \alpha}^{\text{WLS}}) \pi(\Sigma{\bm{z}_{\mathcal{N}_i}};\alpha),\nonumber
\end{align}
and $\hat{\bm{\beta}}_{z, \alpha}^{\text{WLS}}$ corresponds to the vector of estimated coefficients in the weighted hierarchical outcome regression model conditional on $Z_i = z$ and treatment saturation $\alpha$ with weights $\omega_i^{z,\alpha}$.
This extension of the IP-WLS estimator for multilevel outcomes can be expressed as the solution to the following vector of estimating equations.

Let $\psi_{z\alpha}^{\text{IP-WLS}}(\bm{O}_{\nu}; \mu_{z\alpha}, \bm{\beta}_{z\alpha}, \sigma^2_{\varepsilon,z,\alpha},\sigma^2_{c,z,\alpha}, \bm{\gamma}, \phi_b) = \hat{Y}^{\text{IP-WLS}}_{\nu}(z, \alpha) - \mu_{z\alpha}$ and \\$\psi_{\beta_{0\alpha}}(\bm{O}_{\nu}; \bm{\beta}_{0\alpha}, \sigma^2_{\varepsilon,0,\alpha},\sigma^2_{c,0,\alpha},\bm{\gamma}, \phi_b)$, $\psi_{\sigma^2_{\varepsilon,0,\alpha}}(\bm{O}_{\nu}; \bm{\beta}_{0\alpha}, \sigma^2_{\varepsilon,0,\alpha},\sigma^2_{c,0,\alpha},\bm{\gamma}, \phi_b)$, \\ $\psi_{\sigma^2_{c,0,\alpha}}(\bm{O}_{\nu}; \bm{\beta}_{0\alpha}, \sigma^2_{\varepsilon,0,\alpha},\sigma^2_{c,0,\alpha},\bm{\gamma}, \phi_b)$,
$\psi_{\beta_{1\alpha}}(\bm{O}_{\nu}; \bm{\beta}_{1\alpha}, \sigma^2_{\varepsilon,1,\alpha},\sigma^2_{c,1,\alpha},\bm{\gamma}, \phi_b)$, \\ $\psi_{\sigma^2_{\varepsilon,1,\alpha}}(\bm{O}_{\nu}; \bm{\beta}_{1\alpha}, \sigma^2_{\varepsilon,1,\alpha},\sigma^2_{c,1,\alpha},\bm{\gamma}, \phi_b)$, $\psi_{\sigma^2_{c,1,\alpha}}(\bm{O}_{\nu}; \bm{\beta}_{1\alpha}, \sigma^2_{\varepsilon,1,\alpha},\sigma^2_{c,1,\alpha},\bm{\gamma}, \phi_b)$, $\psi_{\gamma}(\bm{O}_{\nu}; \bm{\gamma}, \phi_b)$, and \\ $\psi_{\phi_b}(\bm{O}_{\nu}; \bm{\gamma}, \phi_b)$ denote the estimating functions corresponding to $\hat{\bm{\beta}}_{0\alpha}$, $\hat{\sigma}^2_{\varepsilon,0,\alpha}$, $\hat{\sigma}^2_{c,0,\alpha}$, $\hat{\bm{\beta}}_{1\alpha}$, $\hat{\sigma}^2_{\varepsilon,1,\alpha}$, $\hat{\sigma}^2_{c,1,\alpha}$, $\hat{\bm{\gamma}}$, and $\hat{\phi}_b$ such that $$ \hat{\bm{\theta}}^{\text{IP-WLS}} = (\hat{Y}^{\text{IP-WLS}}(0,\alpha), \hat{Y}^{\text{IP-WLS}}(1,\alpha), \hat{\bm{\beta}}_{0\alpha}, \hat{\sigma}^2_{\varepsilon,0,\alpha}, \hat{\sigma}^2_{c,0,\alpha}, \hat{\bm{\beta}}_{1\alpha}, \hat{\sigma}^2_{\varepsilon,1,\alpha}, \hat{\sigma}^2_{c,1,\alpha}, \hat{\bm{\gamma}} , \hat{\phi}_b)^{\top}$$
 solves the estimating equation
$$\bm{\psi}_{\alpha}^{\text{IP-WLS}}(\bm{\theta}) = \sum_{\nu=1}^m \bm{\psi}_{\alpha}^{\text{IP-WLS}}(\bm{O}_{\nu}; \bm{\theta}) = \bm{0}, $$
where $\bm{\theta}=(\mu_{0\alpha}, \mu_{1\alpha}, \bm{\beta}_{0\alpha}, {\sigma}^2_{\varepsilon,0,\alpha}, {\sigma}^2_{c,0,\alpha}, \bm{\beta}_{1\alpha}, {\sigma}^2_{\varepsilon,1,\alpha}, {\sigma}^2_{c,1,\alpha}, \bm{\gamma}, \phi_b)^{\top}$ and $$\bm{\psi}^{\text{IP-WLS}}(\bm{O}; \bm{\theta})=\begin{pmatrix}
\psi_{0\alpha}^{\text{IP-WLS}}(\bm{O}_{\nu}; \mu_{0\alpha}, \bm{\beta}_{0\alpha}, \sigma^2_{\varepsilon,0,\alpha},\sigma^2_{c,0,\alpha}, \bm{\gamma}, \phi_b) \\
\psi_{1\alpha}^{\text{IP-WLS}}(\bm{O}_{\nu}; \mu_{1\alpha}, \bm{\beta}_{1\alpha}, \sigma^2_{\varepsilon,1,\alpha},\sigma^2_{c,1,\alpha}, \bm{\gamma}, \phi_b)\\ \psi_{\beta_{0\alpha}}(\bm{O}_{\nu}; \bm{\beta}_{0\alpha}, \sigma^2_{\varepsilon,0,\alpha},\sigma^2_{c,0,\alpha},\bm{\gamma}, \phi_b)\\ \psi_{\sigma^2_{\varepsilon,0,\alpha}}(\bm{O}_{\nu}; \bm{\beta}_{0\alpha}, \sigma^2_{\varepsilon,0,\alpha},\sigma^2_{c,0,\alpha},\bm{\gamma}, \phi_b)\\ \psi_{\sigma^2_{c,0,\alpha}}(\bm{O}_{\nu}; \bm{\beta}_{0\alpha}, \sigma^2_{\varepsilon,0,\alpha},\sigma^2_{c,0,\alpha},\bm{\gamma}, \phi_b)\\
\psi_{\beta_{1\alpha}}(\bm{O}_{\nu}; \bm{\beta}_{1\alpha}, \sigma^2_{\varepsilon,1,\alpha},\sigma^2_{c,1,\alpha},\bm{\gamma}, \phi_b)\\ \psi_{\sigma^2_{\varepsilon,1,\alpha}}(\bm{O}_{\nu}; \bm{\beta}_{1\alpha}, \sigma^2_{\varepsilon,1,\alpha},\sigma^2_{c,1,\alpha},\bm{\gamma}, \phi_b)\\ \psi_{\sigma^2_{c,1,\alpha}}(\bm{O}_{\nu}; \bm{\beta}_{1\alpha}, \sigma^2_{\varepsilon,1,\alpha},\sigma^2_{c,1,\alpha},\bm{\gamma}, \phi_b) \\ \psi_{\gamma}(\bm{O}; \bm{\gamma},\phi_b) \\ \psi_{\phi_b}(\bm{O}; \bm{\gamma},\phi_b)
\end{pmatrix}.$$ 
As previously mentioned, $\psi_{\beta_{z\alpha}}(\bm{O}_{\nu}; \bm{\beta}_{z\alpha}, \sigma^2_{\varepsilon,z,\alpha},\sigma^2_{c,z,\alpha},\bm{\gamma}, \phi_b)$ corresponds to the score function in (\ref{eqn:scorebetawls}) whereas the score functions for $\sigma^2_{\varepsilon,z,\alpha}$ and $\sigma^2_{c,z,\alpha}$ must be programmed numerically using the expression of the log-pseudolikelihood in (\ref{pseudologlik}) and the function \texttt{grad} from the \textbf{numDeriv} package.

Despite its higher computational cost and complexity, the IP-WLS may not hold any advantage over the DR-BC estimator in the case of multilevel outcomes. As opposed to the simpler case of conditionally independent outcomes presented in the main text, there is no theoretical guarantee that this estimator is doubly robust, and empirical results suggest that it is biased unless the outcome model is correctly specified. One possible explanation lies in the fact the weighting matrix in the score function for $\bm{\beta}_{z\alpha}$ in (\ref{eqn:scorebetawls}) is no longer a simple diagonal matrix with inverse probability-of-treatment weights as diagonal elements, such that $\hat{\bm{\beta}}_{z, \alpha}^{\text{WLS}}$ no longer satisfies
\begin{align*}
   \mathbb{E}_F\left[ \frac{1}{N_{\nu}} \sum_{i \in C_{\nu}} \omega_{i}^{z, \alpha}\left\{Y_i - \bm{L}_i^{\top} \hat{\bm{\beta}}_{z, \alpha}^{\text{WLS}}\right\}\right] =0.
\end{align*}
This equality is used in the proof of Proposition 2 in Appendix \ref{A.C} to demonstrate the equivalence of the DR-BC and IP-WLS estimators in the case of conditionally independent outcomes. In Appendix \ref{section:B}, we present additional simulation results which suggest that in the case of a correctly specified treatment model and a misspecified outcome model, the bias of the IP-WLS estimator for multilevel outcomes is not vanishing with increasing number of network components. Thus, in its current form, this estimator is not doubly robust in the presence of multilevel outcomes.
\clearpage
\section{Additional Results for the First Simulation Scheme}
\label{section:AppendixC}
\subsection{Additional Simulation Results for the Main Simulation Scheme}
\begin{figure}[H]
    \includegraphics[scale=0.26]{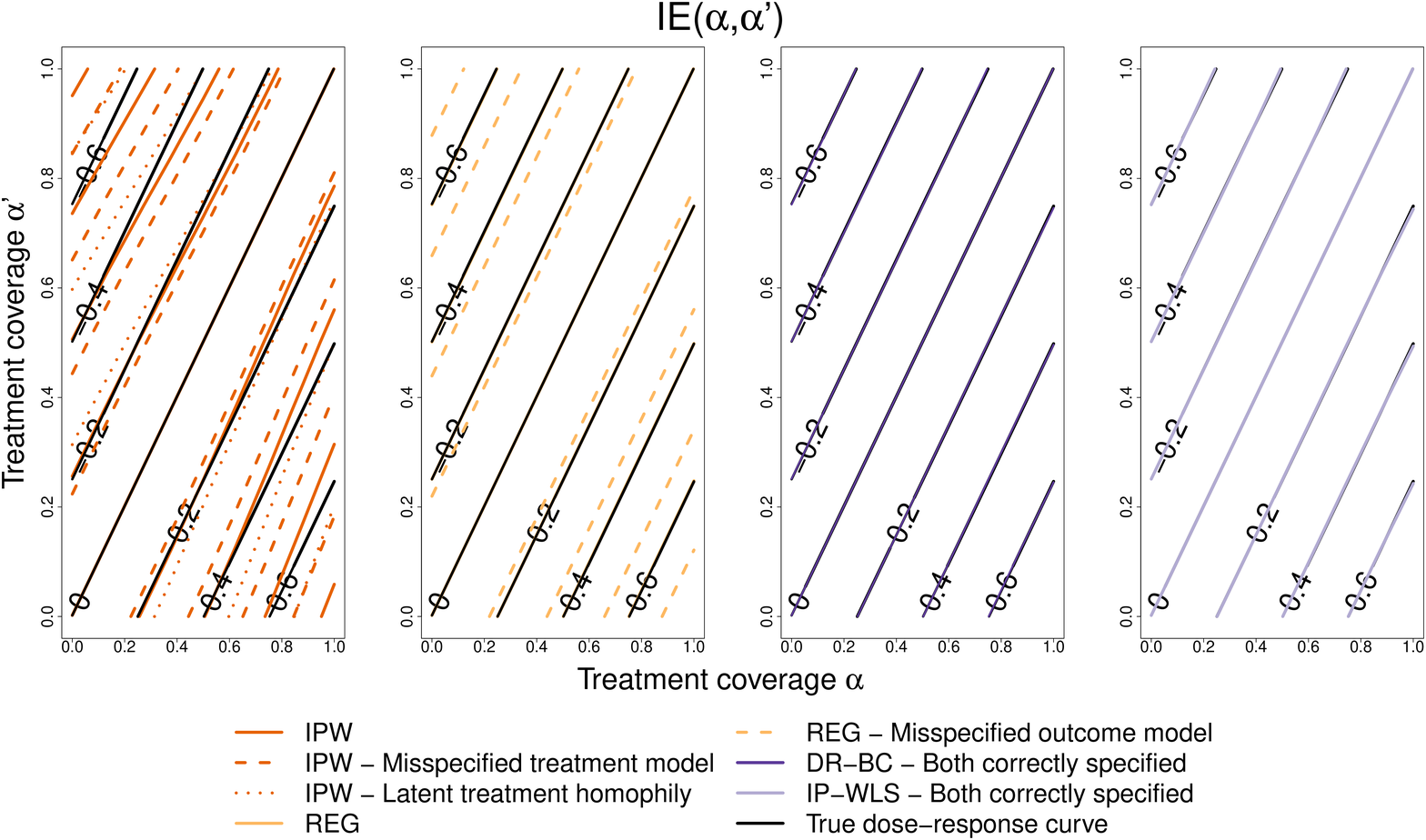}
    \caption{Estimates of the surface $IE(\alpha, \alpha')$, where each panel corresponds to a different estimator (IPW, REG, DR-BC, or IP-WLS). The black solid line represents the true effect. Notice the near overlap of the REG, DR-BC, and IP-WLS estimates based on correctly specified models with the true surface. }
    \label{fig:IEpreliminary}
\end{figure}

Figure \ref{fig:IEpreliminary} shows a different version of the right panel of Figure 2 in the main body of the paper, where each panel is dedicated to a different estimator (IPW, REG, DR-BC, or IP-WLS) to facilitate the reading of the plot.
Table \ref{tab:tab1} shows the bias and the mean squared error (MSE) of the IPW, REG, DR-BC, and IP-WLS estimators of the indirect/spillover effect using the correctly specified treatment model versus the treatment model deprived of the homophilous variable under the first simulation scheme presented in the main text. 
In accordance with the model used to generate the potential outcomes, the indirect treatment effect increases for larger gaps between coverage $\alpha'$ and the reference coverage $\alpha$. It is worth noting the greater stability as measured by MSE of the REG, DR-BC, and IP-WLS estimators, whether or not the correct treatment model is used. 

\begin{table}[ht]
\centering
\scriptsize
\caption{Bias and mean squared error (MSE) of the IPW, REG, DR-BC, and IP-WLS estimators using the correct outcome model and two different treatment models, the correctly specified model vs. the model deprived of the variable driving treatment homophily.}
\begin{tabular}{c c c c c c c c c c c}
\toprule
    & &  & \multicolumn{2}{c}{IPW} & \multicolumn{2}{c}{REG} & \multicolumn{2}{c}{DR-BC} & \multicolumn{2}{c}{IP-WLS}  \\
    \cmidrule(l){4-5}
    \cmidrule(l){6-7}
    \cmidrule(l){8-9} \cmidrule(l){10-11}
  Scenario & Estimand &Value & $\text{Bias}$ & MSE & $\text{Bias}$ & MSE & $\text{Bias}$ & MSE & $\text{Bias}$ & MSE\\
   \midrule
  No latent homophily  & $DE(0.2)$ & 2.200 & 0.004 & 0.112 & 0.001 & 0.009 & -0.001 & 0.015 & $<0.001$ & 0.015 \\ 
  ($Z_i\notind\bm{Z}_{\mathcal{N}_i} | \bm{X}_i$) & $DE(0.5)$ & 2.499 & -0.005 & 0.058 & -0.002 & 0.005 & -0.005 & 0.008 & -0.005 & 0.008 \\ 
 & $DE(0.8)$ & 2.797 & -0.046 & 0.319 & -0.005 & 0.011 & -0.007 & 0.020 & -0.006 & 0.019 \\ 
 & $IE(0.5,0.2)$ & 0.298 & -0.001 & 0.029 & 0.001 & 0.002 & $<0.001$ & 0.004 & $<0.001$ & 0.004 \\ 
  & $IE(0.8,0.2)$ & 0.597 & 0.034 & 0.132 & 0.001 & 0.009 & 0.001 & 0.015 & 0.001 & 0.015 \\ 
  & $IE(0.8,0.5)$ & 0.299 & 0.035 & 0.083 & $<0.001$ & 0.002 & 0.001 & 0.009 & $<0.001$ & 0.009 \\ 
  \addlinespace
 Latent homophily& $DE(0.2)$ & 2.200 & -0.140 & 0.106 & 0.001 & 0.009 & $<0.001$ & 0.012 & $<0.001$ & 0.013 \\ 
  ($Z_i \not\!\perp\!\!\!\perp \bm{Z}_{\mathcal{N}_i} | \bm{X}_i$) & $DE(0.5)$ & 2.499 & -0.007 & 0.029 & -0.002 & 0.005 & -0.003 & 0.006 & -0.003 & 0.006 \\ 
  & $DE(0.8)$ & 2.797 & 0.256 & 0.28 & -0.005 & 0.011 & -0.005 & 0.016 & -0.005 & 0.016 \\
  & $IE(0.5,0.2)$ & 0.298 & -0.048 & 0.025 & 0.001 & 0.002 & -0.001 & 0.003 & -0.001 & 0.003 \\ 
  & $IE(0.8,0.2)$ & 0.597 & -0.052 & 0.105 & 0.001 & 0.009 & $<0.001$ & 0.012 & $<0.001$ & 0.013 \\ 
  & $IE(0.8,0.5)$ & 0.299 & -0.004 & 0.062 & $<0.001$ & 0.002 & 0.001 & 0.007 & 0.001 & 0.008 \\
  \bottomrule
\end{tabular}
\label{tab:tab1}
\end{table}

\subsection{Simulations under Violations of the Stratified Interference Assumption}

\begin{figure}[H]
\includegraphics[scale=0.46]{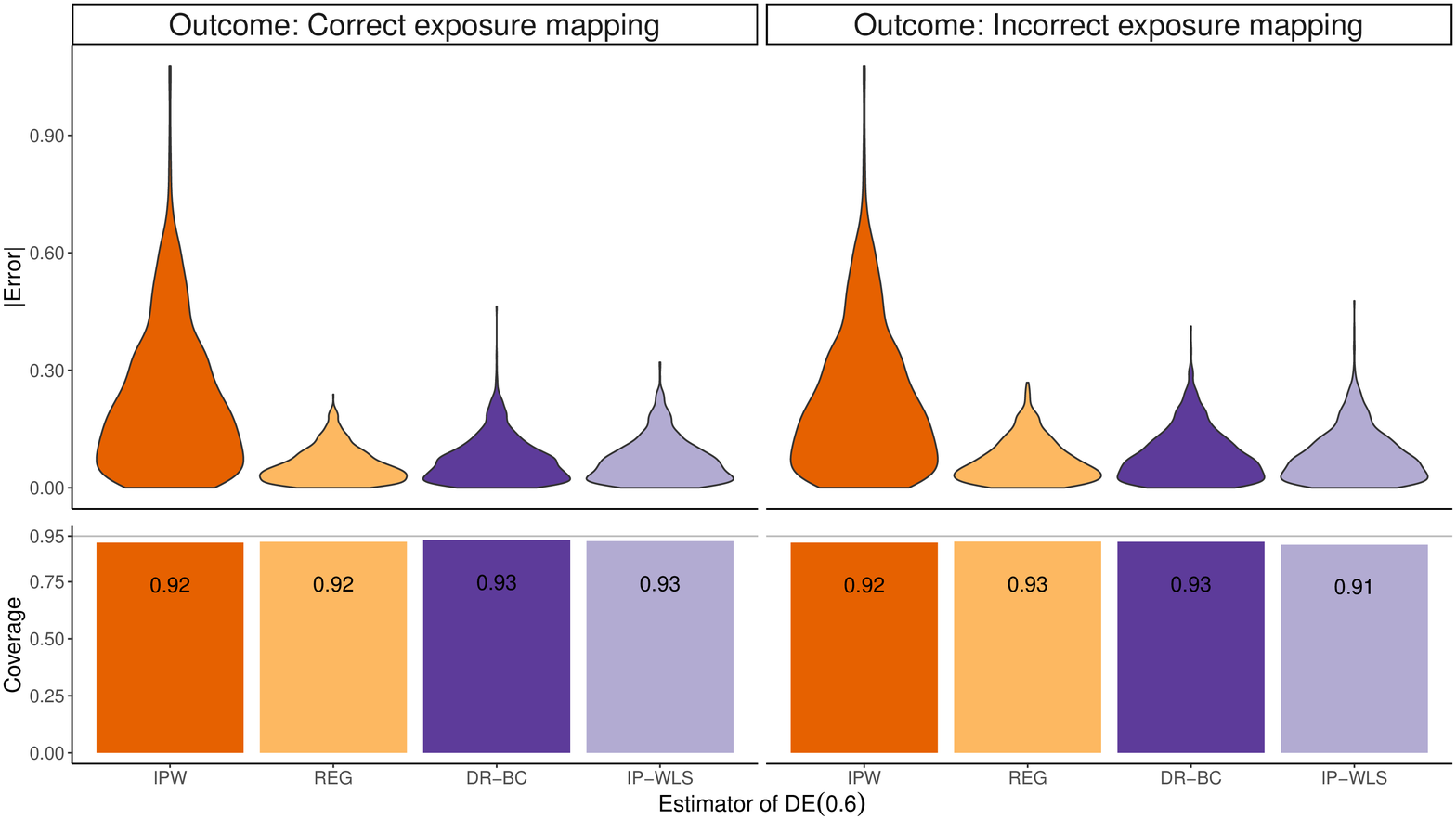}
\caption{Absolute bias and confidence interval coverage for the IPW, REG, DR-BC, and IP-WLS estimators of $DE(0.6)$ under the exposure mapping $\tilde{\phi}$. In the first scenario (correct exposure mapping), results for IPW, REG, DR-BC, and IP-WLS are based on 994, 1,000, 993, and 992 simulated datasets, respectively. In the second scenario, results for IPW, REG, DR-BC, and IP-WLS are based on 994, 1,000, 995, and 995 simulated datasets, respectively.} 
    \label{fig:absbiasviolation1}
\end{figure}

We also assessed the sensitivity of simulation results to deviations of the stratified interference assumption. In an additional scenario, we considered an outcome model where the stratified assumption is violated in the form of a misspecified exposure mapping. We used the potential outcome model, $y_i(z_i, \bm{z}_{\mathcal{N}_i})$, given by:
\begin{align*} y_i(z_i, \bm{z}_{\mathcal{N}_i})  = 2 + 2 z_i + \tilde{\phi}(\bm{z}_{\mathcal{N}_i})  + z_i \tilde{\phi}(\bm{z}_{\mathcal{N}_i}) -1.5 |X_{1i}| + 2X_{2i} - 3 |X_{1i}|X_{2i} + \varepsilon_i, \ \varepsilon_i \sim N(0,1) \end{align*}
where $\tilde{\phi}(\bm{z}_{\mathcal{N}_i})=\tilde{d}_i^{-1} \sum_{j \in \mathcal{N}_i} \mathds{1}(X_{2i}=X_{2j})z_j$ and $\tilde{d}_i = \sum_{j \in \mathcal{N}_i}  \mathds{1}(X_{2i} = X_{2j})$. In words, the potential outcome of node $i$ depends on the sum of treated neighbours who have the same value of $X_{2}$ as they do. Under this model, individual average potential outcomes are obtained using
$$\bar{y}_{i}(z;\alpha) = \sum_{\Sigma \tilde{\bm{z}}_{\mathcal{N}_i} = 0 }^{\tilde{d}_i} y_{i}(z, \bm{z}_{\mathcal{N}_i}) \tilde{\pi}( \Sigma \tilde{\bm{z}}_{\mathcal{N}_i}; \alpha),$$
where $\Sigma \tilde{\bm{z}}_{\mathcal{N}_i} = \sum_{j \in \mathcal{N}_i} \mathds{1}(X_{2i}=X_{2j})z_j$ and $$\tilde{\pi}(\Sigma \tilde{\bm{z}}_{\mathcal{N}_i}; \alpha) = \binom{\tilde{d}_i}{\Sigma \tilde{\bm{z}}_{\mathcal{N}_i}}\alpha^{\Sigma\tilde{\bm{z}}_{\mathcal{N}_i}}(1-\alpha)^{\tilde{d}_i -\Sigma \tilde{\bm{z}}_{\mathcal{N}_i}}. $$
We fitted the oracle outcome model which uses the correct exposure mapping
$$\mathbb{E}[Y_i|\bm{X}_i, Z_i, \bm{Z}_{\mathcal{N}_i}] = \beta_0 + \beta_1 Z_i + \beta_2 \tilde{\phi}(\bm{Z}_{\mathcal{N}_i}) + \beta_3 Z_i \tilde{\phi}(\bm{Z}_{\mathcal{N}_i}) + \beta_4 |X_{1i}| + \beta_5 X_{2i} + \beta_6 |X_{1i}|X_{2i}. $$
Additionally, we fitted the outcome model with an incorrect exposure mapping to the simulated data, that is to say, the exposure mapping introduced in the main body of the paper (the proportion of treated neighbours):
$$\mathbb{E}[Y_i|\bm{X}_i, Z_i, \bm{Z}_{\mathcal{N}_i}] = \beta_0 + \beta_1 Z_i + \beta_2 p(\bm{Z}_{\mathcal{N}_i}) + \beta_3 Z_i p(\bm{Z}_{\mathcal{N}_i}) + \beta_4 |X_{1i}| + \beta_5 X_{2i} + \beta_6 |X_{1i}|X_{2i}. $$
We used the correctly specified treatment model in the main body of the paper and computed the IPW, REG, DR-BC, and IP-WLS estimators. Figure \ref{fig:absbiasviolation1} displays the absolute bias and empirical coverage of Wald-type 95\% confidence intervals for the IPW, REG, DR-BC, and IP-WLS estimators of $DE(0.6)$ for the two outcome models (correct vs. incorrect exposure mapping). Using the correct exposure mapping, the bias of the REG, DR-BC, and IP-WLS estimators are -0.0003, 0.0013, and 0.0006 respectively, compared to respective values of 0.0009, -0.0012, and -0.0027 for the incorrect exposure mapping. For this particular data generating scheme, the bias and coverage of the estimators are only mildly affected by the misspecification of the exposure mapping.

The stratified interference assumption is also violated if potential outcomes depend on treatments of nodes outside of the first-order neighbourhood. We assessed the sensitivity of results to a second-order dependence (friends of friends). We define $\mathcal{N}'_i$ to be the set of nodes corresponding to second-order neighbours of node $i$, and $d'_i$ to be the cardinality of $\mathcal{N}'_i$. Let $\bm{z}_{\mathcal{N}'_i}$ denote the vector of counterfactual treatments assigned to the second-order neighbours of node $i$. We used the following model for the potential outcome $y_i(z_i, \bm{z}_{\mathcal{N}_i}, \bm{z}_{\mathcal{N}'_i})$:
$$y_i(z_i, \bm{z}_{\mathcal{N}_i}, \bm{z}_{\mathcal{N}'_i}) = 2 + 2 z_i + p'(\bm{z}_{\mathcal{N}_i}, \bm{z}_{\mathcal{N}'_i}) + z_i p'(\bm{z}_{\mathcal{N}_i}, \bm{z}_{\mathcal{N}'_i}) -1.5 |X_{1i}| + 2X_{2i} - 3 |X_{1i}|X_{2i} + \varepsilon_i,$$
where $\varepsilon_i \sim N(0,1)$ and $p'(\bm{z}_{\mathcal{N}_i}, \bm{z}_{\mathcal{N}'_i}) = (d_i + d_i')^{-1} \sum_{j \in \mathcal{N}_i \cup \mathcal{N}'_i} z_j$. Under this model, the potential outcome depends on the proportion of first-order and second-order neighbours who received the treatment, as opposed to only first-order, as was done in the primary simulation settings. For simplicity, treatments received by first-order or second-order neighbours are assigned equal weights.

For IPW estimation, we assessed the use of a joint propensity score for the first-order neighbours compared to using a joint propensity score for both first- and second-order neighbours:
\begin{align*}f'(Z_i, \bm{Z}_{\mathcal{N}_i}, \bm{Z}_{\mathcal{N}'_i}\ \rvert \ \bm{X}_i, \bm{X}_{\mathcal{N}_i}, \bm{X}_{\mathcal{N}'_i}; \bm{\gamma}) = \int_{-\infty}^{\infty} \prod_{j \in \{i, \mathcal{N}_i, \mathcal{N}'_i\}} p_j^{Z_j}(1-p_j)^{1-Z_j}f(b_{\nu}) db_{\nu},
\label{eqn:propensity2}
\end{align*}
where $p_j$ is defined and parameterised the same way as in the main body of the paper. We compared estimators based on a correctly specified outcome model
$$\mathbb{E}[Y_i|\bm{X}_i, Z_i, \bm{Z}_{\mathcal{N}_i}] = \beta_0 + \beta_1 Z_i + \beta_2 p'(\bm{Z}_{\mathcal{N}_i}, \bm{Z}_{\mathcal{N}'_i}) + \beta_3 Z_i p'(\bm{Z}_{\mathcal{N}_i}, \bm{Z}_{\mathcal{N}'_i}) + \beta_4 |X_{1i}| + \beta_5 X_{2i} + \beta_6 |X_{1i}|X_{2i}. $$
to those based on an outcome model that assumed dependence of the outcome on the first-order neighbourhood treatment vector
$$\mathbb{E}[Y_i|\bm{X}_i, Z_i, \bm{Z}_{\mathcal{N}_i}] = \beta_0 + \beta_1 Z_i + \beta_2 p(\bm{Z}_{\mathcal{N}_i}) + \beta_3 Z_i p(\bm{Z}_{\mathcal{N}_i}) + \beta_4 |X_{1i}| + \beta_5 X_{2i} + \beta_6 |X_{1i}|X_{2i}. $$

We assessed the performance of the IPW, REG, DR-BC, and IP-WLS estimators under different specification scenarios: (a) both the outcome and the propensity score assume first- and second-order dependence; (b) only the propensity score assumes second-order depencence; (c) only the outcome model assumes second-order dependence; and (d) neither component assumes second-order dependence.

\begin{figure}[H]
\includegraphics[scale=0.47]{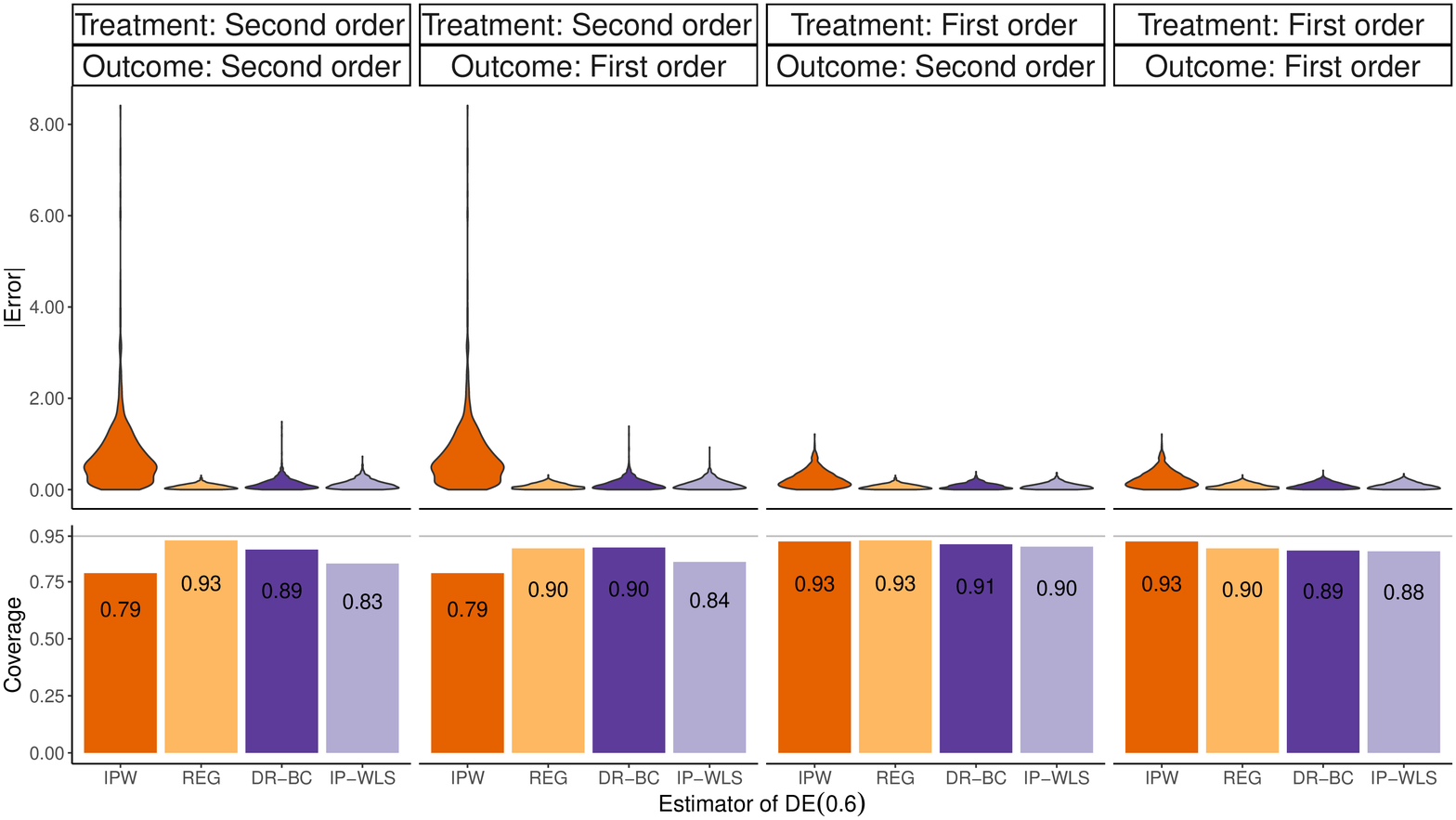}
\caption{Absolute bias and confidence interval coverage for the IPW, REG, DR-BC, and IP-WLS estimators of $DE(0.6)$ under a second-order dependence. In scenarios (a) and (b), results for IPW, REG, DR-BC, and IP-WLS are based on 997, 1,000, 994, and 993 simulated datasets, respectively. In scenarios (c) and (d), results for IPW, REG, DR-BC, and IP-WLS are based on 997, 1,000, 995, and 993 simulated datasets, respectively.}
\label{fig:absbiasviolation2}
\end{figure}

Absolute bias and empirical coverages of Wald-type 95\% confidence intervals for the IPW, REG, DR-BC, and IP-WLS estimators are shown in Figure \ref{fig:absbiasviolation2}. When the correct outcome model is used, the bias of the REG estimator is -0.0016 compared to 0.0403 when only the first-order neighbours are accounted for. The IPW estimator based on a joint propensity score for first- and second-order neighbours displays high instability (MSE of 1.14) compared to the strategy based on first-order neighbours (MSE of 0.09), albeit being less biased (absolute bias of 0.030 compared to 0.049). This is likely due to extreme weights that result from inverting the joint propensity score, which is often much smaller after widening the interference set as it is constructed by taking a product over a larger number of probabilities. A bias-variance trade-off is apparent in this situation, where the IPW estimator based on first-order interference, despite being incorrect, displays better finite sample properties. This in turn affects the performance of the DR-BC and DR-WLS, which seem to behave more optimally when the outcome model assumes second-order performance and the joint propensity score only includes first-order neighbours. For instance, for scenario (a), the bias of DR-BC and IP-WLS are 0.0080 and 0.040, respectively, whereas for scenario (c), these values are -0.0008 and -0.0004, respectively. The story is broadly the same when it comes to confidence interval coverage. However, the empirical coverage of the IP-WLS is more impacted by the use of the joint propensity score for first- and second-order neighbours compared to the usual joint propensity score introduced in the main body of the paper.

\clearpage
\section{Simulation Studies with Multilevel Outcomes}
\label{section:B}
In our main scenario, we simulated a network with 30 components each of size 30. In reality, it is \emph{highly} unlikely that network components would be found to be (even approximately) balanced. Furthermore, the simulation scheme presented in the main text confers a multilevel structure upon the treatment/exposure data with the inclusion of a component-level random intercept while assuming independent errors in the outcome model. Thus, in addition to the simulation scheme presented in the main text, we considered a more realistic network with unequal component sizes and incorporated a component-level random intercept into the model used to generate potential outcomes in order to investigate the performance of proposed estimators under conditions similar to those encountered in the analysis of Add Health. This scenario was carried out similarly to the first simulation scheme except that we considered $m=50$ network components and Step 1 was replaced with 
\begin{enumerate}
\item[1.\textsuperscript{*}] For $\nu \in \{1,\ldots, 30\}$, we sampled $N_{\nu}$ from the distribution $\mathrm{Poisson}(35)$, and for $\nu \in \{31,\ldots, 50\}$, we sampled $N_{\nu}$ from the distribution $\mathrm{Poisson}(12)$. For each component, we generated the random vector $\bm{H}_{\nu}$ of size $N_{\nu}$ whose elements are from a $\mathrm{Bernoulli}(0.5)$ distribution. The $m$ components were generated according to an exponential random graph model (ERGM) in which the conditional log-odds of two nodes having a tie while holding the rest of the network fixed was $1.5 \delta_1 - 2.5 \delta_2,$
where $\delta_1$ is the indicator for matching $H$ values and $\delta_2$ represents the increase in the number of ties \citep{hunter2008ergm}. The final network $G$ was taken as the union of the $m$ components. 
\end{enumerate}
and Step 2 was replaced with
\begin{enumerate}
\item[2.\textsuperscript{*}] For $i = 1, \ldots, N_{\nu}$, two baseline covariates were generated as $X_{1i} \sim \mathcal{N}(0,1)$ and $X_{2i} \sim \mathrm{Bernoulli}(0.5)$. We then generated potential outcomes under all possible individual treatments $z_i \in \{0,1\}$ and neighbourhood treatments $\Sigma \bm{z}_{\mathcal{N}_i} \in \{0,1,\ldots, d_i\}$ as
$$y_i(z_i, \bm{z}_{\mathcal{N}_i}) = 2 + 2 z_i + p(\bm{z}_{\mathcal{N}_i}) + z_i p(\bm{z}_{\mathcal{N}_i}) -1.5 |X_{1i}| + 2X_{2i} - 3 |X_{1i}|X_{2i} + c_{\nu} + \varepsilon_i, \ \varepsilon_i \sim N(0,1),$$
where $c_{\nu} \sim N(0,1)$ is independent of the random intercept $b_{\nu}$ in the treatment model generated in Step 3.
\end{enumerate}

\begin{figure}[!htb]
\centering
    \includegraphics[scale=0.46]{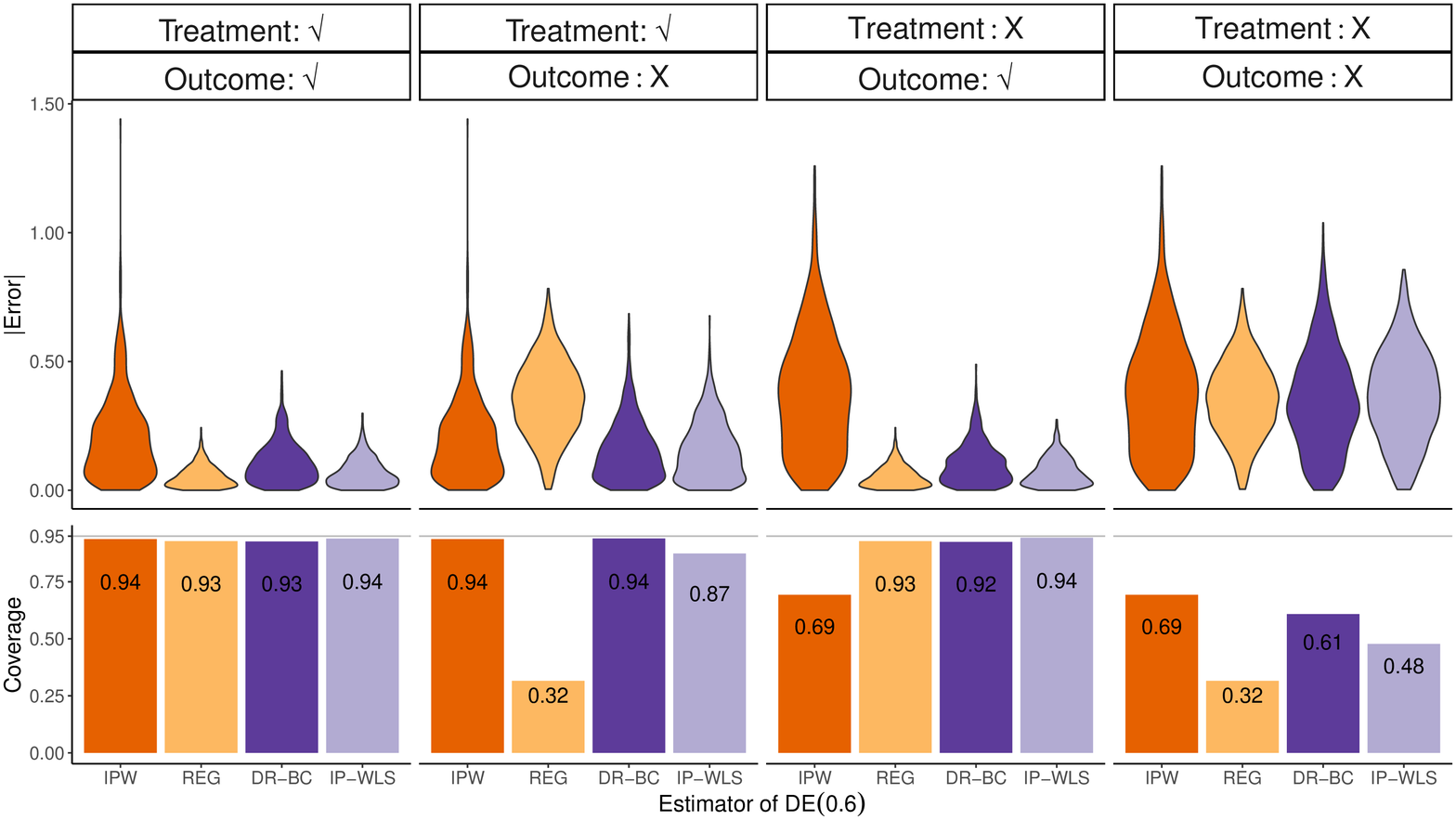}
    \caption{Absolute bias and confidence interval coverage for the IPW, REG, DR-BC, and IP-WLS estimators of $DE(0.6)$ under the unbalanced components simulation scheme. In scenario (a), results for the IPW, REG, DR-BC, and IP-WLS estimators are based on 996, 1,000, 995, and 645 simulated datasets, respectively. In scenario (b), these values are 996, 1,000, 995, and 769. In scenario (c), results for the IPW, REG, DR-BC, and IP-WLS estimators are based on 993, 1,000, 994, and 617 simulated datasets, respectively. In scenario (d), these values are 993, 1,000, 994, and 773.}
    \label{fig:absbias2}
\end{figure}

For this second simulation scheme, a combination of convergence issues and excess computation time led to the exclusion of between 227 and 383 simulated datasets for the assessment of the IP-WLS estimator. There was virtually no difference in the empirical distributions of population-specific parameters $DE(0.6)$ when comparing the included samples versus the excluded ones, which reassures about representativeness of the results. For the IPW estimator, 4 and 7 samples were excluded due to nonconvergence in scenarios (a)-(b) and scenarios (c)-(d), respectively. For the DR-BC estimator, between 5 and 7 samples had to be excluded across different scenarios. Finally, similar to the previous simulation scheme, no sample was excluded in the assessment of the REG estimator.

Figure \ref{fig:absbias2} shows the absolute bias of the IPW, REG, DR-BC, and IP-WLS estimators of $DE(0.6)$ under different misspecification scenarios for the second, unbalanced components data generation mechanism. As was the case for the previous simulation scheme, the bias of the DR-BC estimator is small whenever at least one of the two models is correctly specified. However, the same cannot be said about the IP-WLS estimator. For instance, for scenario (b), the bias of IPW, REG, DR-BC, and IP-WLS are -0.007, -0.353, -0.005, and -0.117, respectively, whereas for scenario (c), these values are -0.354, -0.002, -0.002, and -0.004, respectively. This suggests that IP-WLS sustains a bias whenever the outcome model is not correctly specified.

Empirical coverages of Wald-type 95\% confidence intervals for $DE(0.6)$ displayed at the bottom of Figure \ref{fig:absbias2} are approximately 95\% for singly robust methods when the working model is correctly specified and for the DR-BC method when either model is correct. In scenario (b), despite the average estimated standard error being close numerically to the empirical standard error (0.152 vs 0.153, results not shown), the empirical coverage associated to IP-WLS is only 87\%, which is well below 95\% even after accounting for finite-sampling error. The undercoverage of IP-WLS based confidence intervals can thus be attributed to the bias that was recorded in scenario (b).

In addition to the simulation scenario above, we also generated networks with varying number of network components, with $m \in \{100, 150, 200\}$. Given $m$, for $\nu \in \{1, \ldots, 0.6m\}$, we sampled $N_{\nu}$ from the distribution $\mathrm{Poisson}(35)$, and for $\nu \in \{0.6m + 1, m\}$, we sampled $N_{\nu}$ from the distribution $\mathrm{Poisson}(12)$. We generated 1,000 datasets in the same fashion as in the above simulation scheme with multilevel outcomes. For each scenario, we evaluated the absolute value of the bias, the empirical coverage probability of 95\% Wald-type confidence intervals, and the average estimated standard error (ASE) of the methods IPW, REG, and DR-BC using a correctly specified working model for the singly robust methods and at least one correctly specified model for DR-BC. 

\begin{figure}[!htb]
\centering
    \includegraphics[scale=0.47]{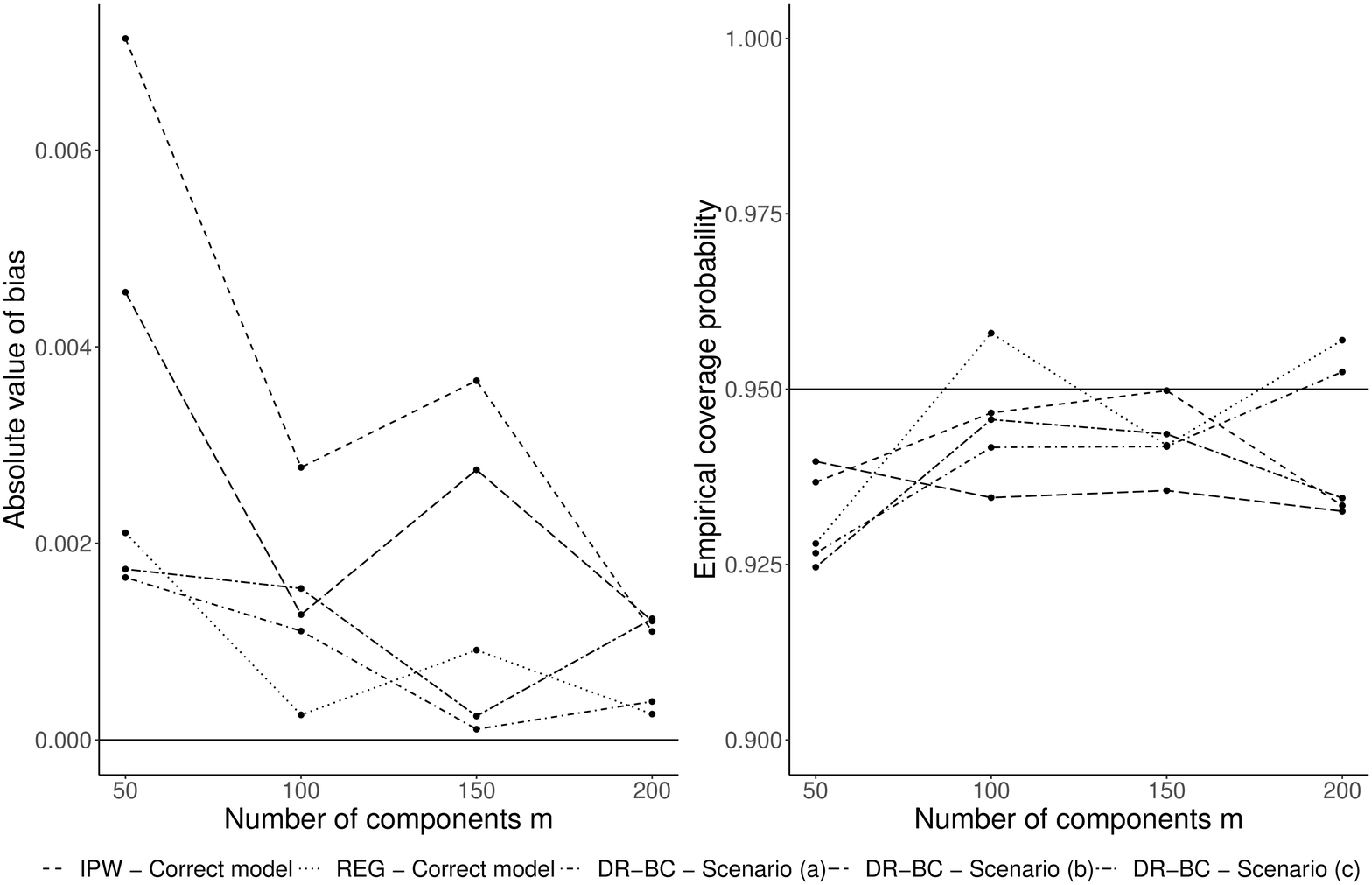}
    \caption{Absolute bias (left) of the IPW, REG, and DR-BC (scenarios (a), (b) and (c)) estimators of $DE(0.6)$ and corresponding Wald 95\% confidence intervals empirical coverage probability (right) with $m \in \{50, 100, 150, 200\}$. Scenario (a) corresponds to correctly specified treatment and outcome models, scenario (b) corresponds to a correct treatment model and a misspecified outcome model, and scenario (c) corresponds to a correct outcome model and a misspecified treatment model.} 
    \label{fig:absbiasecp}
\end{figure}

\begin{figure}[!htb]
\centering
    \includegraphics[scale=0.47]{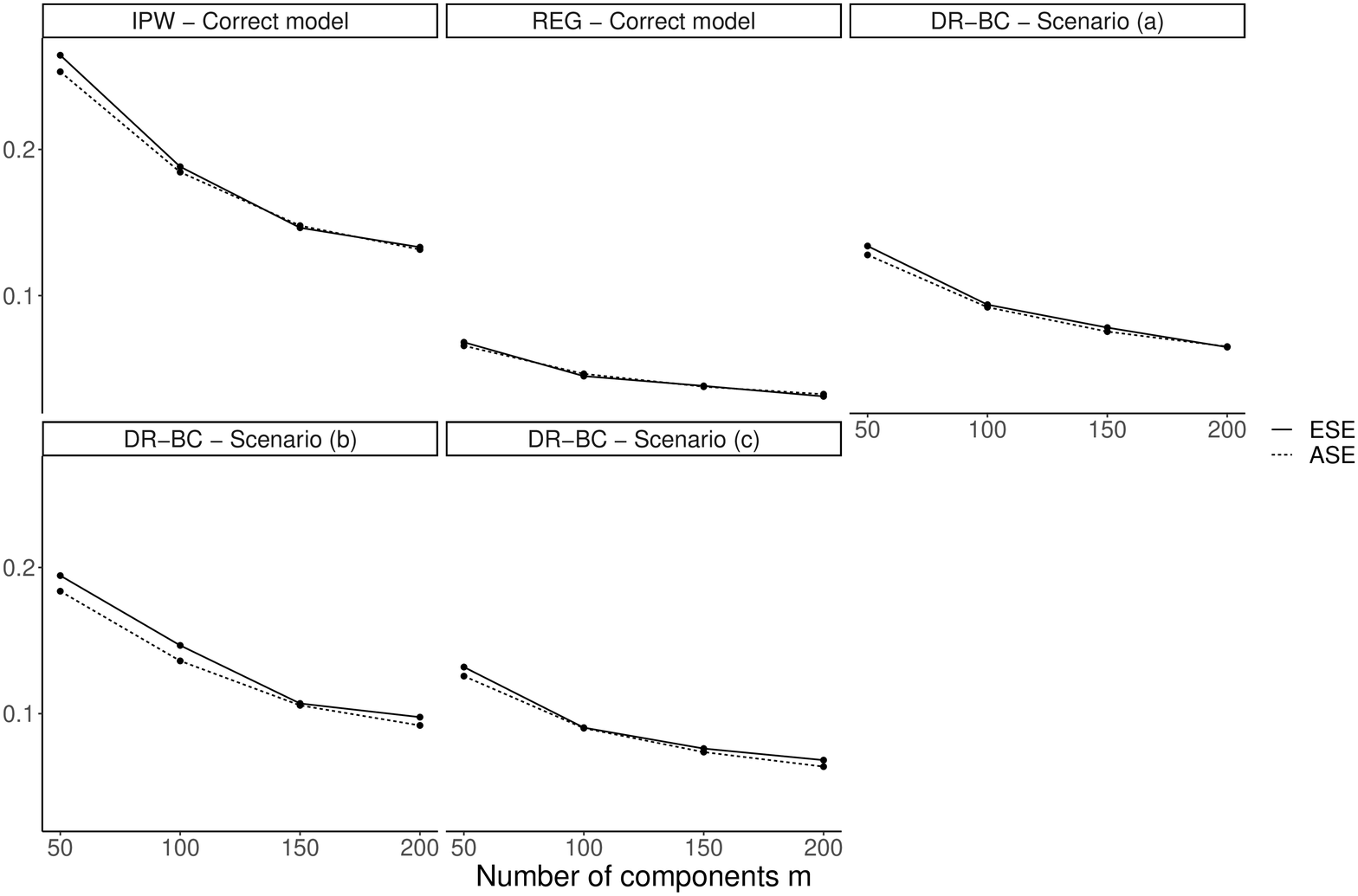}
    \caption{Comparison of the empirical standard error (ESE) and the average estimated standard errors (ASE) of the IPW, REG, and DR-BC (scenarios (a), (b) and (c)) estimators of $DE(0.6)$ with $m \in \{50, 100, 150, 200\}$. Scenario (a) corresponds to correctly specified treatment and outcome models, scenario (b) corresponds to a correct treatment model and a misspecified outcome model, and scenario (c) corresponds to a correct outcome model and a misspecified treatment model.} 
    \label{fig:ase}
\end{figure}

Figure \ref{fig:absbiasecp} displays the absolute value of bias on the left-hand side and the empirical coverage probability on the right-hand side for each of method and scenario combination. These results suggest the IPW estimator and the DR-BC estimator based on a correctly specified treatment model only (scenario (b)) converge more slowly than other methods, which is to be expected. The IPW estimator and DR-BC estimators based on only one correct model have an empirical coverage probability slightly under 95\% when $m=200$. Figure \ref{fig:ase} compares the empirical standard error (ESE) and the ASE for each method and scenario combination, which supports the use of the sandwich standard errors especially for large values of $m$.

\begin{figure}[!htb]
\centering
    \includegraphics[scale=0.45]{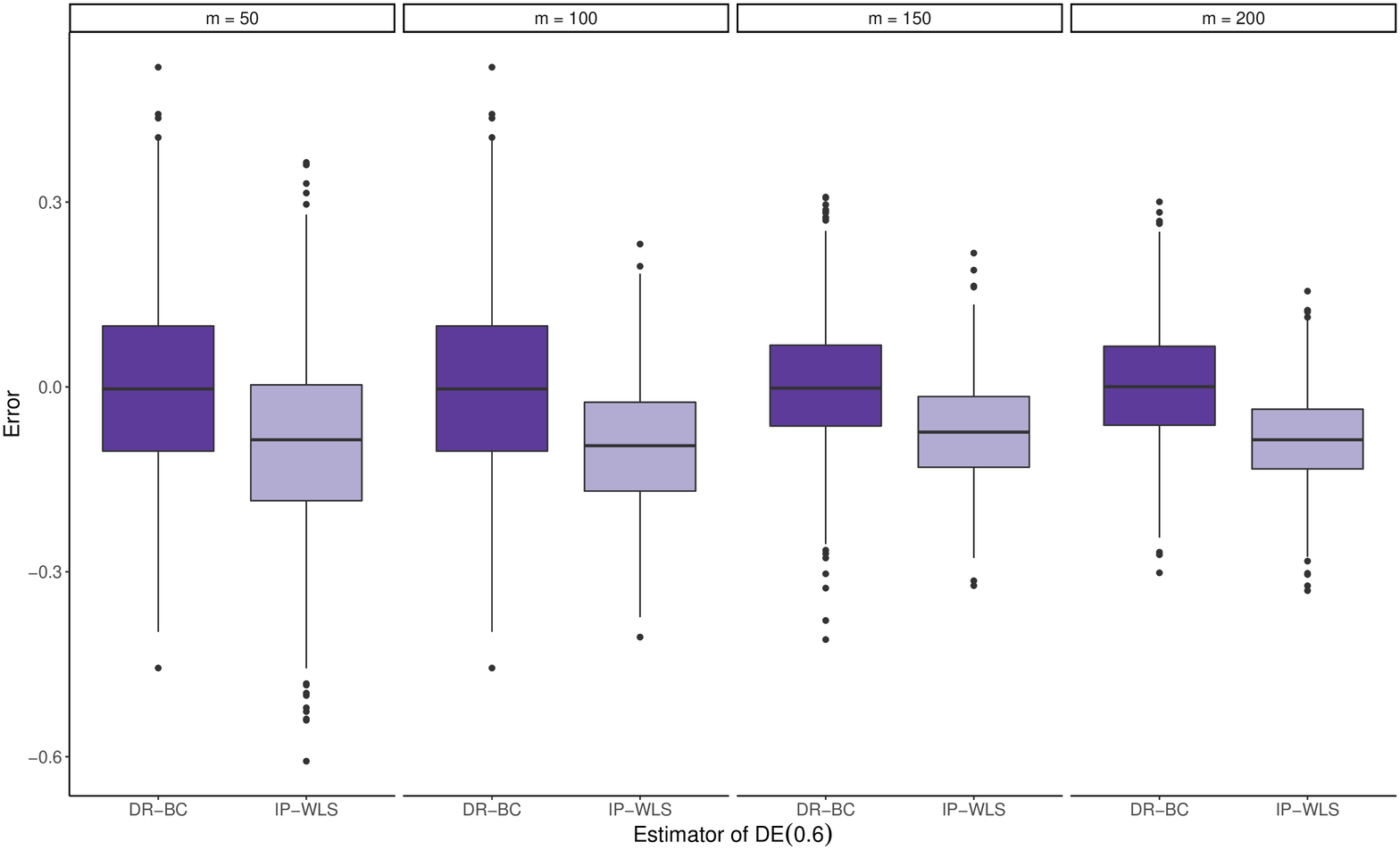}
    \caption{Bias for DR-BC and IP-WLS estimators of $DE(0.6)$ under the second simulation scheme when the outcome model is correctly specified but the treatment model is misspecified (scenario (b)) with $m \in \{50, 100, 150, 200\}$.} 
    \label{fig:biasdrwls}
\end{figure}

Figure \ref{fig:biasdrwls} compares the bias of the point estimates yielded by the DR-BC and DR-WLS methods under scenario (b), that is, when only the treatment model is correctly specified, for an increasing number of components. The bias values of the DR-BC estimator for 50, 100, 150, and 200 network components are -0.005, 0.001, 0.003, and 0.001, respectively, whereas for DR-WLS, these values are -0.090, -0.096, -0.073, and -0.085. Thus, the DR-WLS estimator based on a correctly specified treatment sustains a non-negligible bias even as the number of independent units increases, which suggests that it may not be doubly robust. This could also explain the low empirical coverage probability that was associated to DR-WLS under scenario (b) in Figure \ref{fig:absbias2}, since the ESE of the DR-WLS estimator was correctly estimated by the sandwich standard error (results not shown).

\clearpage
\section{Additional Information on the Add Health Study}
\label{section:AppendixE}
The National Longitudinal Study of Adolescent Health (Add
Health) is a longitudinal study of a nationally representative sample of 90,118 adolescent students in grades 7-12 in the
United States in 1994-95 who were followed through adolescence and the transition to adulthood. The survey collected data on socio-demographic characteristics, health
status, risk behaviors, academic achievement, and friendship nominations through in-school questionnaires
administered directly to students. In the friendships nominations questionnaire, each student was asked to nominate \emph{at most} five male and five female friends in any order. 

Leveraging the Add Health data, \cite{bifulco2011effect} adopted the peer effects model of \cite{manski1993identification} to investigate the causal effects of high school classmate characteristics on a wide range of post-secondary outcomes. They found that having a higher proportion of classmates with college-educated mothers was associated with a lower risk of dropping out of high school and a higher likelihood of attending
college. Recently, \cite{fletcher2020consequences} made direct use of the friendship nominations data to investigate the
effect of maternal education composition among adolescents’ social circles on their Grade Point Average
(GPA). \cite{fletcher2020consequences} regressed the outcome, GPA, on maternal education composition of each students’ friends using a high dimensional fixed effects model which included school and cohort fixed effects. They found that having a greater proportion of friends with college-educated mothers increases GPA
among female students, but not among male students, which suggests a spillover effect of maternal
education on GPA that disseminates through friendship ties, with potential effect modification by sex.

In our analysis reported in Section 6 of the main text, we examined an undirected network based on the data on friendship nominations from the first wave of data collection in 1994-95. Much like \cite{egami2021identification}, we defined a friendship as a symmetric relationship, meaning that there was an edge between students $i$ and $j$ if student $i$ listed $j$ as a friend in the in-school survey, or student $j$ listed $i$ as a friend, or both. This approach differs from that of \cite{fletcher2020consequences}, who analyzed mutual same-gender friendships. Unlike \cite{egami2021identification}, we did not restrict friendships to be within the same school. This allowed for ties to exist across schools, as shown in Figure 1 in the main text.

The full dataset presented some challenges, starting with the presence of partial actor non-response, which occurs when some, but not all, attribute information is available for one or more nodes in the network \citep{krause2018missing}. To prevent loss of network information, we performed a single imputation of the nodal attributes, which can be treated as ordinary missing data \citep{krause2018missing}. Imputation by chained equations using predictive mean matching with 20 cycles was used to this end. Because large network components impose a significant computational cost for the outcome regression estimator, schools with more than 400 students were excluded from the analysis. For the same reason, isolates were also excluded from the analysis sample. When subsampling schools with fewer than 400 students, we made sure to include a connected school (as illustrated in Figure \ref{fig:plot11}) so as to not break components apart.

\begin{figure}[ht]
\centering
\includegraphics[scale=0.5]{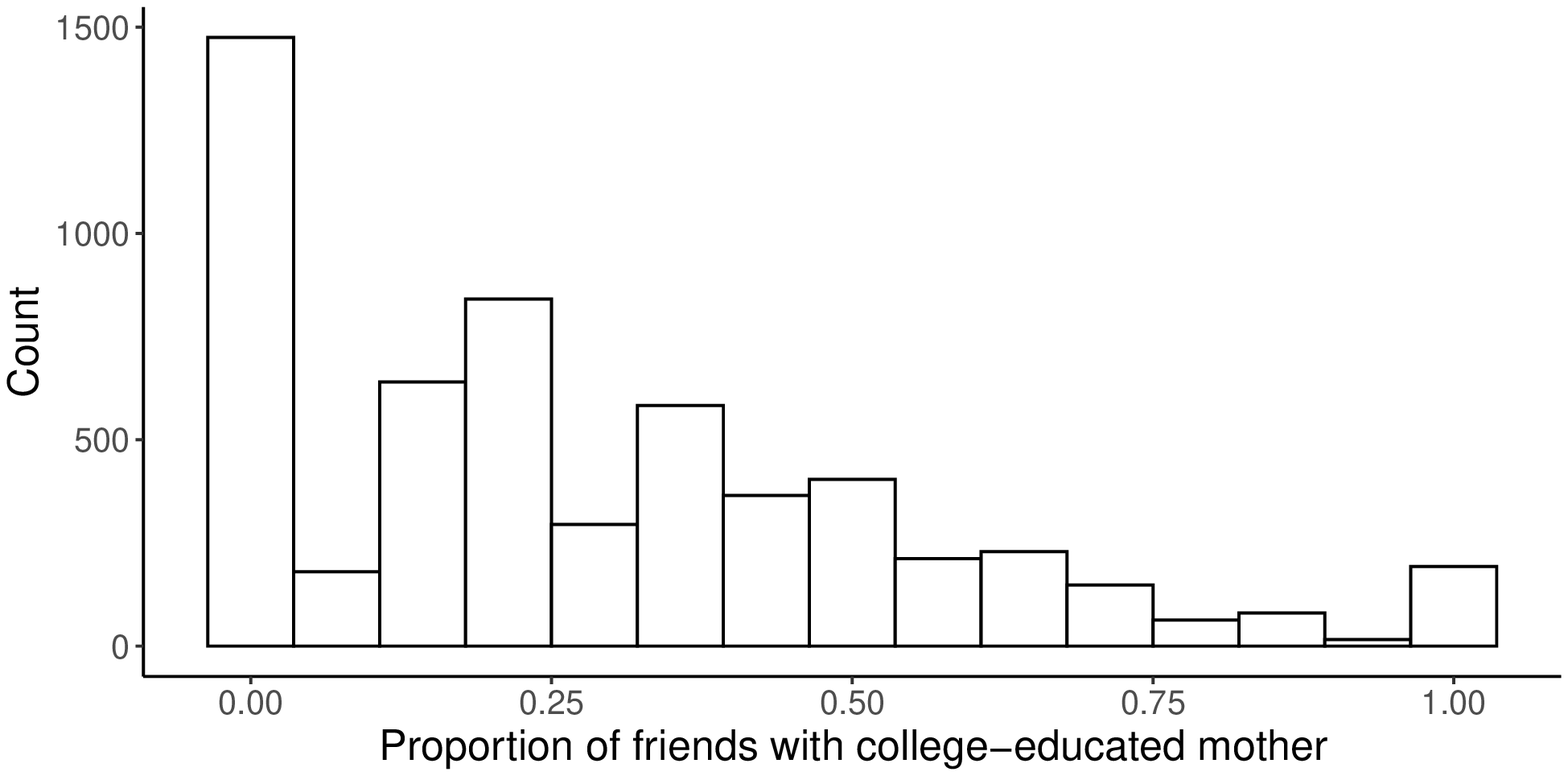}
\caption{Distribution of the observed proportion of friends with a college-educated mother in the Add Health analytical sample of 5,724 students.}
\label{fig:proptreated}
\end{figure}

Tables \ref{tab:table1} and \ref{tab:table12} exhibit network characteristics and descriptive statistics for the nodal attributes before and after the imputation, respectively. The average degree (i.e., average number of named friends per participant), defined as $N^{-1} \sum_{i \in \mathcal{N}} d_i$, is equal to 6.91 and the standard deviation of the degree distribution is 3.82. The edge density, defined as the ratio of the number of edges to the number of possible edges, has value 0.00131 in this network, indicating a very low density (weakly connected) social network. To define transitivity, we first need to define a triangle, which is a complete subgraph of order 3, and a connected triple, which is a subgraph of three nodes connected by two edges \citep{kolaczyk2009}. As shown in Table \ref{tab:table1}, the transitivity or clustering coefficient, defined as the ratio of the count of \emph{triangles} to the count of connected triples, is 0.255, which quantifies the extent to which egdges are clustered in the graph \citep{kolaczyk2009}. Finally, the assortativity coefficient based on maternal education (1 = mother has a four-year college degree, 0 = mother is a high school graduate or less) is defined as $$r = \frac{\sum_{i=0}^1 f_{ii} - \sum_{i=0}^1 f_{i+}f_{+i}}{1-\sum_{i=0}^1 f_{i+}f_{+i}},$$ where $f_{ij}$ is the fraction of edges in $G$ that join a node in the $i$-th category with a node in the $j$-th category, $f_{i+} = \sum_{j=0}^1 f_{ij}$ and $f_{+i} = \sum_{j=0}^1 f_{ji}$ \citep{igraph, kolaczyk2009}. The assortativity in this network is 0.170, which corresponds to the extent to which friends tend to share a similar values for maternal education. Figure \ref{fig:proptreated} displays a histogram for the observed proportion of friends with a college-educated in the analytical sample, which shows a right-skewed distribution.

\begin{table}[H]
\centering
\begin{threeparttable}
\caption{Characterization of the Add Health subnetwork of 5,724 students after excluding schools with more than 400 students and isolates}
    \small
    \begin{tabular}{l l c c }
    \toprule
       Global network  & Schools & \multicolumn{2}{c}{37} \\
      characteristics & Nodes & \multicolumn{2}{c}{5,724} \\
       & Components & \multicolumn{2}{c}{48}\\
         & Edges & \multicolumn{2}{c}{19,776}
\\
         & Average Degree (SD) & \multicolumn{2}{c}{6.91 (3.82)}\\
         & Edge density & \multicolumn{2}{c}{1.31$\cdot 10^{-3}$}\\
         & Transitivity & \multicolumn{2}{c}{0.255} \\
         & Assortativity\tnote{1}& \multicolumn{2}{c}{0.170} \\
         \midrule
         Between-school  & Nodes (Median, Q1-Q3) & \multicolumn{2}{c}{148 (71 - 237)}\\
         network characteristics & Edges (Median, Q1-Q3) & \multicolumn{2}{c}{451 (240 - 856)}\\
         & Average Degree (Mean, SD) & \multicolumn{2}{c}{6.62 (1.70)} \\
         & Edge density (Median, Q1-Q3) &  \multicolumn{2}{c}{0.045 (0.030 - 0.098)}\\
         & Transitivity (Median, Q1-Q3) &  \multicolumn{2}{c}{0.267 (0.221 - 0.361)}\\
         & Assortativity\tnote{1} (Median, Q1-Q3) & \multicolumn{2}{c}{0.027 (-0.025 - 0.052)}\\
         \bottomrule
    \end{tabular}
\begin{tablenotes}
\item[1] The assortativity coefficient is based on maternal education.
\end{tablenotes}
    \label{tab:table1}
\end{threeparttable}
\end{table}

\begin{table}[H]
\centering
\begin{threeparttable}
\caption{Descriptive statistics of nodal attributes in the Add Health subnetwork of 5,724 students after excluding schools with more than 400 students and isolates}
    \scriptsize
    \begin{tabular}{l l c c }
    \toprule
         & & Before imputation & After single imputation\\
         \cmidrule(l){3-3} \cmidrule(l){4-4}
                  Grade Point Average  & Mean (SD) &  2.92 (0.77) & 2.90 (0.77)\\
                  \cmidrule(l){2-2}  \cmidrule(l){3-3} \cmidrule(l){4-4}
         & Valid responses & 4,827 (84.3\%) & 5,724 (100.0\%)\\
         & Missing & 897 (15.7\%) 
 & 0 (0.0\%) \\
                  \midrule
         Maternal education & 4-year college degree &  1,464 (28.7\%)  & 1,638 (28.6\%)  \\
         & High school graduate or less &  3,644 (71.3\%) 
 & 4,086 (71.4\%)\\
  \cmidrule(l){2-2}  \cmidrule(l){3-3} \cmidrule(l){4-4}
         & Valid responses & 5,108 (89.2\%) & 5,724 (100.0\%)\\
         & Missing & 616 (10.8\%) & 0 (0.0\%) \\
         \midrule
         Age  & Mean (SD) &  14.54 (1.70)
 & 14.54 (1.69)\\
                  \cmidrule(l){2-2}  \cmidrule(l){3-3} \cmidrule(l){4-4}
         & Valid responses & 5,715 (99.8\%) & 5,724 (100.0\%)\\
         & Missing & 9 (0.2\%) 
 & 0 (0.0\%) \\
  \midrule
         Gender & Male &  2,594 (45.5\%) 
 &   2,604 (45.5\%) 

  \\
         & Female &   3,108 (54.5\%) 
 
 &  3,120 (54.5\%) 
\\
  \cmidrule(l){2-2}  \cmidrule(l){3-3} \cmidrule(l){4-4}
         & Valid responses & 5,702 (99.6\%) & 5,724 (100.0\%)\\
         & Missing & 22 (0.4\%) & 0 (0.0\%) \\
                  \midrule
         Race & White &  3,911 (68.3\%) & 3,911 (68.3 \%) \\
         & Black &   820 (14.3\%)  & 820 (14.3\%) \\
         & Asian &    97 ( 1.7\%)  & 97 (1.7\%) \\
         & Other &   896 (15.7\%) 
 & 896 (15.7\%) \\
   \cmidrule(l){2-2}  \cmidrule(l){3-3} \cmidrule(l){4-4}
 & Valid responses & 5,724 (100.0\%)& 5,724 (100.0\%)\\
         & Missing & 0 (0.0\%)& 0 (0.0\%) \\
         \midrule
         Lives with father & Yes &  4,505 (79.4\%)  & 5,542 (79.4\%) \\
         & No &  1,172 (20.6\%)  & 1,182 (20.6\%)\\
          \cmidrule(l){2-2}  \cmidrule(l){3-3} \cmidrule(l){4-4}
 & Valid responses & 5,677 (92.2\%)& 5,724  (100.0\%)\\
         & Missing & 47 (0.8\%)& 0 (0.0\%) \\
         \midrule 
         Mother was born in the US & Yes &  5,191 (92.1\%)  & 5,259 (91.9\%) \\
         & No &  443 (7.9\%)  & 465 (8.1\%) \\
         \cmidrule(l){2-2}  \cmidrule(l){3-3} \cmidrule(l){4-4}
 & Valid responses & 5,634 (98.4\%)& 5,724  (100.0\%)\\
         & Missing & 90 (1.6\%)& 0 (0.0\%) \\
         \bottomrule
    \end{tabular}
    \label{tab:table12}
\end{threeparttable}
\end{table}
\newpage
\section{Proofs}
\subsection{Unbiasedness of the IPW estimator}
\label{section:appendixA}
Define the collection of potential outcomes in the network as $\bm{y}(\cdot) = \left(y_1(\cdot), \ldots, y_N(\cdot)\right)^{\top}$ and the matrix of observed covariates as $\bm{X}=(\bm{X}_1^{\top}, \ldots, \bm{X}_N^{\top})^{\top}$. Treating the network as fixed, we have
\begin{multline*}
     \mathbb{E}_{Z, \bm{Z} | \bm{y}(\cdot), \bm{X}}\left[\hat{Y}^{\text{IPW}}(z, \alpha) \ |\ \bm{y}(\cdot), \bm{X}\right]\\
    =
    \frac{1}{m} \sum_{\nu = 1}^m \frac{1}{N_{\nu}} \sum_{i\in C_{\nu}}\mathbb{E}_{Z, \bm{Z} | \bm{y}(\cdot), \bm{X}}\left[\frac{y_i\left(Z_i, \Sigma {\bm{Z}_{\mathcal{N}_i}}\right) \mathds{1}(Z_i = z)\pi(\Sigma {\bm{Z}_{\mathcal{N}_i}};\alpha)}{  \binom{d_i}{\Sigma{\bm{Z}_{\mathcal{N}_i}}} f(Z_i, \bm{Z}_{\mathcal{N}_i} \ \rvert \ \bm{X}_i, \bm{X}_{\mathcal{N}_i})} \right]\\
   =
    \frac{1}{m} \sum_{\nu = 1}^m \frac{1}{N_{\nu}} \sum_{i\in C_{\nu}} \mathbb{E}_{Z, \bm{Z} | \bm{y}(\cdot), \bm{X}}\left[\frac{y_i\left(Z_i, \Sigma{\bm{Z}_{\mathcal{N}_i}}\right) \mathds{1}(Z_i = z)\pi(\Sigma{\bm{Z}_{\mathcal{N}_i}};\alpha)}{\binom{d_i}{\Sigma{\bm{Z}_{\mathcal{N}_i}}}P(Z_i = z_i, \bm{Z}_{\mathcal{N}_i} = \bm{z}_{\mathcal{N}_i} \ |\ \bm{X}_i, \bm{X}_{\mathcal{N}_i})}  \right]\\
    = \frac{1}{m} \sum_{\nu = 1}^m \frac{1}{N_{\nu}} \sum_{i\in C_{\nu}} \sum_{\substack{z_i \in \{0, 1\},\\ \bm{z}_{\mathcal{N}_i} \in \mathcal{Z}(d_i)}} \biggr( \frac{y_i\left(z_i, \Sigma{\bm{z}_{\mathcal{N}_i}}\right) \mathds{1}(z_i = z)\pi(\Sigma{\bm{z}_{\mathcal{N}_i}};\alpha)}{\binom{d_i}{\Sigma{\bm{z}_{\mathcal{N}_i}}}P(Z_i = z_i, \bm{Z}_{\mathcal{N}_i} = \bm{z}_{\mathcal{N}_i} \ |\ \bm{X}_i, \bm{X}_{\mathcal{N}_i})} \\ \times P(Z_i = z_i, \bm{Z}_{\mathcal{N}_i} = \bm{z}_{\mathcal{N}_i} \ |\ \bm{X}_i, \bm{X}_{\mathcal{N}_i}, y_i(\cdot)) \biggr).
    \end{multline*}
    By conditional exchangeability (Assumption 3), we have that $$P(Z_i = z_i, \bm{Z}_{\mathcal{N}_i} = \bm{z}_{\mathcal{N}_i} \ |\ \bm{X}_i, \bm{X}_{\mathcal{N}_i}, y_i(\cdot))  = P(Z_i = z_i, \bm{Z}_{\mathcal{N}_i} = \bm{z}_{\mathcal{N}_i} \ |\ \bm{X}_i, \bm{X}_{\mathcal{N}_i}),$$ so that
    \begin{align*}
    &\mathbb{E}_{Z, \bm{Z} | \bm{y}(\cdot), \bm{X}}\left[\hat{Y}^{\text{IPW}}(z, \alpha) \ |\ \bm{y}(\cdot), \bm{X}\right]\\&=  \frac{1}{m} \sum_{\nu = 1}^m \frac{1}{N_{\nu}} \sum_{i\in C_{\nu}}  \sum_{\substack{z_i \in \{0, 1\},\\ \bm{z}_{\mathcal{N}_i} \in \mathcal{Z}(d_i)}} \frac{y_i\left(z_i, \Sigma{\bm{z}_{\mathcal{N}_i}}\right) \mathds{1}(z_i = z)\pi(\Sigma{\bm{z}_{\mathcal{N}_i}};\alpha)}{\binom{d_i}{\Sigma{\bm{z}_{\mathcal{N}_i}}}}  \\
    &= \frac{1}{m} \sum_{\nu = 1}^m \frac{1}{N_{\nu}} \sum_{i\in C_{\nu}}  \sum_{ \bm{z}_{\mathcal{N}_i} \in \mathcal{Z}(d_i)} \frac{y_i\left(z, \Sigma{\bm{z}_{\mathcal{N}_i}}\right) \pi(\Sigma{\bm{z}_{\mathcal{N}_i}};\alpha)}{\binom{d_i}{\Sigma{\bm{z}_{\mathcal{N}_i}}}}  \quad \quad \tag*{(since $\mathds{1}(z_i = z) = 0$ if $z_i \neq z$)}\\
    &=\frac{1}{m} \sum_{\nu = 1}^m \frac{1}{N_{\nu}} \sum_{i\in C_{\nu}}  \sum_{ \Sigma{\bm{z}_{\mathcal{N}_i}} = 0}^{N_i} \frac{\binom{d_i}{\Sigma{\bm{z}_{\mathcal{N}_i}}} y_i\left(z, \Sigma{\bm{z}_{\mathcal{N}_i}}\right) \pi(\Sigma{\bm{z}_{\mathcal{N}_i}};\alpha)}{\binom{d_i}{\Sigma{\bm{z}_{\mathcal{N}_i}}}}\\
    &= \frac{1}{m} \sum_{\nu = 1}^m \frac{1}{N_{\nu}} \sum_{i\in C_{\nu}}  \sum_{ \Sigma{\bm{z}_{\mathcal{N}_i}} = 0}^{N_i} y_i\left(z, \Sigma{\bm{z}_{\mathcal{N}_i}}\right) \pi(\Sigma{\bm{z}_{\mathcal{N}_i}};\alpha)\\
    &= \bar{y}(z, \alpha).
\end{align*}
By the law of iterated expectation, it follows that 
$$\mathbb{E}\left[\hat{Y}^{\text{IPW}}(z, \alpha)\right] = \mathbb{E}_{\bm{y}(\cdot), \bm{X}}\left[ \mathbb{E}_{Z, \bm{Z} | \bm{y}(\cdot), \bm{X}}\left[\hat{Y}^{\text{IPW}}(z, \alpha) \ |\ \bm{y}(\cdot), \bm{X}\right] \right] = \mathbb{E}_{\bm{y}(\cdot), \bm{X}}[\bar{y}(z, \alpha)]
= \mu_{z, \alpha}.$$The unbiasedness of the estimator of the marginal average potential outcome, $\hat{Y}^{\text{IPW}}( \alpha)$, can be proved in a similar fashion.
\subsection{Proof of Proposition 1}
\label{A.B}
Let $\bm{\beta}_0$ and $(\bm{\gamma}_0, \phi_{b,0})$ denote the true values of the parameters in the outcome regression and the propensity score models, respectively. If the outcome regression model is correctly specified, then the parameter value $\bm{\beta}^*$ such that $\mathbb{E}_F\left[\psi_{\beta}(\bm{O}_{\nu}; \bm{\beta}^*)\right]= 0$ is equal to $\bm{\beta}_0$. Likewise, if the propensity score is correctly modeled, then the solution to $$\mathbb{E}_F\left[\left(\psi_{\gamma}(\bm{O}_{\nu}; \bm{\gamma}^*, \phi^*_b), \psi_{\phi_b}(\bm{O}_{\nu}; \bm{\gamma}^*, \phi^*_b)\right)\right]= 0$$ is such that $(\bm{\gamma}^*, \phi^*_b) = (\bm{\gamma}_0, \phi_{b,0})$. Note that the expectations are taken with respect to the true data generating mechanism.

If $(\bm{\gamma}^*, \phi^*_b) = (\bm{\gamma}_0, \phi_{b,0})$, then 
$f(Z_i, \bm{Z}_{\mathcal{N}_i} \ \rvert \ \bm{X}_i, \bm{X}_{\mathcal{N}_i}; \bm{\gamma}^*, \phi_b^*) = f(Z_i, \bm{Z}_{\mathcal{N}_i} \ \rvert \ \bm{X}_i, \bm{X}_{\mathcal{N}_i}; \bm{\gamma}_0, \phi_{b,0})$. Following Lee et al. (2021) and Liu et al. (2019), 
\begin{align*}
   &  \mathbb{E}_F\left[\frac{1}{N_{\nu}} \sum_{i \in C_{\nu}} \frac{y_i\left(Z_i, \sum{\bm{Z}_{\mathcal{N}_i}}\right) \mathds{1}(Z_i = z)\pi(\sum{\bm{Z}_{\mathcal{N}_i}};\alpha)}{  \binom{d_i}{\Sigma{\bm{Z}_{\mathcal{N}_i}}} f(Z_i, \bm{Z}_{\mathcal{N}_i} \ \rvert \ \bm{X}_i, \bm{X}_{\mathcal{N}_i}; \bm{\gamma}^*, \phi_b^*)}  \right] \\
   &= \mathbb{E}_{\bm{y}(\cdot), \bm{X}}\left[ \frac{1}{N_{\nu}}  \mathbb{E}_{Z, \bm{Z}|\bm{y}(\cdot), \bm{X}} \left[ \sum_{i \in C_{\nu}} \frac{y_i\left(Z_i, \sum{\bm{Z}_{\mathcal{N}_i}}\right) \mathds{1}(Z_i = z)\pi(\sum{\bm{Z}_{\mathcal{N}_i}};\alpha)}{  \binom{d_i}{\Sigma{\bm{Z}_{\mathcal{N}_i}}} f(Z_i, \bm{Z}_{\mathcal{N}_i} \ \rvert \ \bm{X}_i, \bm{X}_{\mathcal{N}_i}; \bm{\gamma}^*, \phi_b^*)}  \biggr \rvert y_i(\cdot), \bm{X}_i, \bm{X}_{\mathcal{N}_i}\right] \right]\\
   &= \mathbb{E}_{\bm{y}(\cdot), \bm{X}}\left[ \frac{1}{N_{\nu}}  \mathbb{E}_{Z, \bm{Z}| \bm{X}} \left[ \sum_{i \in C_{\nu}} \frac{y_i\left(Z_i, \sum{\bm{Z}_{\mathcal{N}_i}}\right) \mathds{1}(Z_i = z)\pi(\sum{\bm{Z}_{\mathcal{N}_i}};\alpha)}{  \binom{d_i}{\Sigma{\bm{Z}_{\mathcal{N}_i}}} f(Z_i, \bm{Z}_{\mathcal{N}_i} \ \rvert \ \bm{X}_i, \bm{X}_{\mathcal{N}_i};\bm{\gamma}^*, \phi_b^*)}  \biggr \rvert \bm{X}_i, \bm{X}_{\mathcal{N}_i}\right] \right] \quad \quad \tag*{(Conditional exchangeability)}\\
   &= \mathbb{E}_{\bm{y}(\cdot), \bm{X}}\left[ \frac{1}{N_{\nu}} \sum_{i\in C_{\nu}} \sum_{\substack{z_i \in \{0, 1\},\\ \bm{z}_{\mathcal{N}_i} \in \mathcal{Z}(d_i)}} \frac{y_i\left(z_i, \Sigma{\bm{z}_{\mathcal{N}_i}}\right) \mathds{1}(z_i = z)\pi(\Sigma{\bm{z}_{\mathcal{N}_i}};\alpha)}{\binom{d_i}{\Sigma{\bm{z}_{\mathcal{N}_i}}}f(z_i, \bm{z}_{\mathcal{N}_i} \ \rvert \ \bm{X}_i, \bm{X}_{\mathcal{N}_i}; \bm{\gamma}^*, \phi_b^*)}f(z_i, \bm{z}_{\mathcal{N}_i} \ \rvert \ \bm{X}_i, \bm{X}_{\mathcal{N}_i}; \bm{\gamma}^*, \phi_b^*) \right] \\
    &=  \mathbb{E}_{\bm{y}(\cdot), \bm{X}}\left[ \frac{1}{N_{\nu}} \sum_{i\in C_{\nu}} \sum_{ \Sigma\bm{z}_{\mathcal{N}_i} =0}^{d_i} y_i\left(z, \Sigma{\bm{z}_{\mathcal{N}_i}}\right) \pi(\Sigma{\bm{z}_{\mathcal{N}_i}};\alpha)\right] \\
   &= \mathbb{E}_{\bm{y}(\cdot), \bm{X}}\left[  \bar{y}^*(z, \alpha) \right] \\
   &= \mu_{z\alpha},
\end{align*}
where $N_{\nu}$ is treated as fixed and $\bar{y}^*(z, \alpha)$ denotes a component-level average potential outcome,
and \begin{multline*}
\mathbb{E}_F\Bigg[ \frac{1}{N_{\nu}}\sum_{i\in C_{\nu}}\biggr\{  \sum_{\Sigma \bm{z}_{\mathcal{N}_i} = 0}^{d_i}  m_i(z, \Sigma \bm{z}_{\mathcal{N}_i}, \bm{X}_i; \bm{\beta}^*) \pi(\Sigma \bm{z}_{\mathcal{N}_i};\alpha)\\ - \mathds{1}(Z_i = z)\frac{m_i(Z_i, \Sigma \bm{Z}_{\mathcal{N}_i}, \bm{X}_i; \bm{\beta}^*)}{\binom{d_i}{\Sigma{\bm{Z}_{\mathcal{N}_i}}}f(Z_i, \bm{Z}_{\mathcal{N}_i} \ \rvert \ \bm{X}_i, \bm{X}_{\mathcal{N}_i}; \bm{\gamma}^*, \phi_b^*) }\pi(\Sigma \bm{Z}_{\mathcal{N}_i}; \alpha)\biggr\} \Bigg] = 0,\\
\end{multline*}
which implies that $\mathbb{E}_F\left[\psi_{z\alpha}(\bm{O}; \mu_{z\alpha}, \bm{\beta}^*, \bm{\gamma}^*,\phi_b^*) \right]=0$. If $\bm{\beta}^*=\bm{\beta}_0$, then 
\begin{align*}
    &\mathbb{E}_F\left[\frac{1}{N_{\nu}}\sum_{i\in C_{\nu}}  \sum_{\Sigma \bm{z}_{\mathcal{N}_i} = 0}^{d_i}  m_i(z, \Sigma \bm{z}_{\mathcal{N}_i}, \bm{X}_i; \bm{\beta}^*) \pi(\Sigma \bm{z}_{\mathcal{N}_i};\alpha) \right] \\ &= \mathbb{E}_F\left[ \frac{1}{N_{\nu}}\sum_{i\in C_{\nu}} \sum_{\Sigma \bm{z}_{\mathcal{N}_i} = 0}^{d_i}  \mathbb{E}\left[y_i(z_i, \bm{z}_{\mathcal{N}_i})| \bm{X}_i \right] \pi(\Sigma \bm{z}_{\mathcal{N}_i};\alpha) \right]\\
   &= \mathbb{E}_F\left[  \bar{y}^*(z, \alpha) \right] \\
   &= \mu_{z\alpha},
\end{align*}
whereas 
\begin{multline*}
    \mathbb{E}_F\left[\frac{1}{N_{\nu}} \sum_{i \in C_{\nu}}   \frac{\mathds{1}(Z_i = z)\pi(\Sigma \bm{Z}_{\mathcal{N}_i}; \alpha)}{\binom{d_i}{\Sigma{\bm{Z}_{\mathcal{N}_i}}}f(Z_i, \bm{Z}_{\mathcal{N}_i} \ \rvert \ \bm{X}_i, \bm{X}_{\mathcal{N}_i}; \bm{\gamma}^*, \phi_b^*) } \left\{y_i(Z_i, \Sigma \bm{Z}_{\mathcal{N}_i}) - m_i(Z_i, \Sigma \bm{Z}_{\mathcal{N}_i}, \bm{X}_i; \bm{\beta}^*)\right\} \right]\\
    = \frac{1}{N_{\nu}} \mathbb{E}_F\Bigg[ \sum_{i\in C_{\nu}}  \sum_{\substack{z_i \in \{0, 1\},\\ \bm{z}_{\mathcal{N}_i} \in \mathcal{Z}(d_i)}} \frac{\mathds{1}(z_i = z)\pi(\Sigma \bm{z}_{\mathcal{N}_i}; \alpha)\left\{y_i(z_i, \Sigma \bm{z}_{\mathcal{N}_i}) - \mathbb{E}[y_i(z_i, \Sigma \bm{z}_{\mathcal{N}_i})|\bm{X}_i]\right\}}{\binom{d_i}{\Sigma{\bm{z}_{\mathcal{N}_i}}}f(z_i, \bm{z}_{\mathcal{N}_i} \ \rvert \ \bm{X}_i, \bm{X}_{\mathcal{N}_i}; \bm{\gamma}^*, \phi_b^*) }\\ \times f(z_i, \bm{z}_{\mathcal{N}_i} \ \rvert \ \bm{X}_i, \bm{X}_{\mathcal{N}_i}; \bm{\gamma}^*, \phi_b^*) \Bigg]=0.
\end{multline*}
It follows that $\mathbb{E}_F\left[\psi_{z\alpha}(\bm{O}; \mu_{z\alpha}, \bm{\beta}^*, \bm{\gamma}^*, \phi_b^*) \right]=0$ when either $(\bm{\gamma}^*, \phi^*_b) = (\bm{\gamma}_0, \phi_{b,0})$ or $\bm{\beta}^*= \bm{\beta}_0$, which establishes the double robustness property of the regression estimator with residual bias correction.

Under suitable regularity conditions \citep{stefanski2002calculus, van2000asymptotic}, by M-estimation theory it follows that $\hat{\bm{\theta}} \stackrel{p}{\longrightarrow}\bm{\theta}_0$ and $\sqrt{m} (\hat{\bm{\theta}}- \bm{\theta}_0)$ converges in distribution to a multivariate normal distribution $N(\bm{0}, \bm{U}^{-1}\bm{V}(\bm{U}^{-1})^{\top}),$
where $\bm{U}=\bm{U}(\bm{\theta})=\mathbb{E}_F\left[-\partial \bm{\psi}(\bm{O}_{\nu}; \bm{\theta})/ \partial \bm{\theta}^{\top} \right]$ and $\bm{V} = \bm{V}(\bm{\theta}) = \mathbb{E}_F\left[\bm{\psi}(\bm{O}_{\nu}; \bm{\theta}) \bm{\psi}(\bm{O}_{\nu}; \bm{\theta})^{\top} \right]$ and the true parameter vector $\bm{\theta}_0=(\mu_{0\alpha}, \mu_{1\alpha}, \bm{\beta}^*,\bm{\gamma}^*, \phi_b^*)^{\top}$ is defined as the solution to the equation 
$$\int \bm{\psi}(\bm{o}, \bm{\theta}_0)dF(\bm{o}) =0.$$
\subsection{Proof of Proposition 2}
\label{A.C}
Let $(\bm{\gamma}_0, \phi_{b,0})$ and $\bm{\beta}_{0,z\alpha}$ for $z=0,1$ denote the true values of the parameters in the propensity score and outcome regression models. Define $\bm{\beta}_{z\alpha}^*$ and $(\bm{\gamma}^*, \phi_{b}^*)$ to be such that $\mathbb{E}_F\left[ \psi_{\beta_{z\alpha}}(\bm{O}_{\nu}; \bm{\beta}_{z\alpha}^*, \bm{\gamma}^*, \phi_{b}^*) \right] = 0$ and $\mathbb{E}_F\left[\left(\psi_{\gamma}(\bm{O}_{\nu}; \bm{\gamma}^*, \phi^*_b), \psi_{\phi_b}(\bm{O}_{\nu}; \bm{\gamma}^*, \phi^*_b)\right)\right]= \bm{0}$, respectively. Treating $N_{\nu}$ as fixed, if $\mathbb{E}_F\left[\psi_{\beta_{z\alpha}}(\bm{O}_{\nu}; \bm{\beta}_{z\alpha}^*, \bm{\gamma}^*, \phi_{b}^*) \right] = 0$, then the following equality also holds:
$$\mathbb{E}_F\left[ \frac{1}{N_{\nu}}\sum_{i \in C_{\nu}} \frac{\mathds{1}(Z_i = z)\pi(\Sigma \bm{Z}_{\mathcal{N}_i};\alpha)}{\binom{d_i}{\Sigma \bm{Z}_{\mathcal{N}_i}}f(z, \bm{Z}_{\mathcal{N}_i}| \bm{X}_i, \bm{X}_{\mathcal{N}_i}; \bm{\gamma}^*, \phi_{b}^*)}\left\{y_i(Z_i, \Sigma \bm{Z}_{\mathcal{N}_i}) - m_i(z, \Sigma \bm{Z}_{\mathcal{N}_i}, \bm{X}_i, \bm{\beta}^*_{z, \alpha})\right\}\right]=0 .$$
Thus, if our goal is to show that $\mathbb{E}_F\left[\psi_{z\alpha}^{\text{IP-WLS}}(\bm{O}_{\nu}; \mu_{z\alpha}, \bm{\beta}_{z\alpha}^*, \bm{\gamma}^*, \phi_{b}^*)\right]=0$, we can proceed by showing that
\begin{align*} &\mathbb{E}_F[\hat{Y}^{\text{IP-WLS}}_{\nu}(z, \alpha)] - \mu_{z\alpha} = 0 \\
\iff  & \mathbb{E}_F\left[\frac{1}{N_{\nu}} \sum_{i\in C_{\nu}} \sum_{ \Sigma{\bm{z}_{\mathcal{N}_i}} = 0}^{d_i} m_i(z, \Sigma \bm{z}_{\mathcal{N}_i}, \bm{X}_i; \bm{\beta}_{z\alpha}^*) \pi(\Sigma{\bm{z}_{\mathcal{N}_i}};\alpha)  \right] -\mu_{z\alpha} =0
\end{align*}
\begin{multline}
 \iff \mathbb{E}_F\biggr[\frac{1}{N_{\nu}} \sum_{i\in C_{\nu}} \sum_{\Sigma \bm{z}_{\mathcal{N}_i} = 0}^{d_i} \biggr\{ m_i(z, \Sigma \bm{z}_{\mathcal{N}_i}, \bm{X}_i; \bm{\beta}_{z\alpha}^*) \pi(\Sigma \bm{z}_{\mathcal{N}_i};\alpha) +\\ \frac{\mathds{1}(Z_i = z)\pi(\Sigma \bm{Z}_{\mathcal{N}_i}; \alpha)}{\binom{d_i}{\Sigma \bm{Z}_{\mathcal{N}_i}}f(Z_i, \bm{Z}_{\mathcal{N}_i} \ \rvert \ \bm{X}_i, \bm{X}_{\mathcal{N}_i}; \bm{\gamma}^*, \phi_{b}^*) } \left\{y_i(Z_i, \Sigma \bm{Z}_{\mathcal{N}_i}) - m_i(Z_i, \Sigma \bm{Z}_{\mathcal{N}_i}, \bm{X}_i; \bm{\beta}_{z\alpha}^*)\right\}\biggr\}\biggr] - \mu_{z\alpha} =0 \nonumber 
\end{multline}
\begin{align*}
  \iff \hat{Y}_{\nu}^{\text{DR-BC}}(z, \alpha) - \mu_{z\alpha} = 0,   
\end{align*}
which we showed could hold if either $\bm{\beta}_{z\alpha}^* = \bm{\beta}_{0,z\alpha}$ or $(\bm{\gamma}^*, \phi_b^*) = (\bm{\gamma}_0, \phi_{b,0})$ in Proposition 1. Therefore, the regression estimator with inverse-propensity weighted coefficients has the double robustness property and the proof of asymptotic normality of $\hat{\bm{\theta}}$ follows along the same lines as the proof of Proposition 1 in Appendix \ref{A.B}.
\end{appendices}
\end{document}